\documentclass[12pt]{article}
\usepackage{cite}
\usepackage{amsmath, amsthm, amssymb, graphicx,slashed}

\parskip=\baselineskip
\def\Bbb{\mathbb}
\def\Tr{{\rm Tr}}
\def\16{{\bf 16}}
\def\1{{\bf 1}}
\def\2{{\bf 2}}
\def\4{{\bf 4}}
\def\bar{\overline}
\def\tilde{\widetilde}
\def\R{{\Bbb{R}}}\def\Z{{\Bbb{Z}}}
\def\N{{\mathcal N}}
\def\hat{\widehat}
\numberwithin{equation}{section}
\def\frak{\mathfrak}
\font\teneurm=eurm10 \font\seveneurm=eurm7 \font\fiveeurm=eurm5
\newfam\eurmfam
\textfont\eurmfam=\teneurm \scriptfont\eurmfam=\seveneurm
\scriptscriptfont\eurmfam=\fiveeurm

 \font\teneusm=eusm10 \font\seveneusm=eusm7 \font\fiveeusm=eusm5
\newfam\eusmfam
\textfont\eusmfam=\teneusm \scriptfont\eusmfam=\seveneusm
\scriptscriptfont\eusmfam=\fiveeusm

\font\tencmmib=cmmib10 \skewchar\tencmmib='177
\font\sevencmmib=cmmib7 \skewchar\sevencmmib='177
\font\fivecmmib=cmmib5 \skewchar\fivecmmib='177
\newfam\cmmibfam
\textfont\cmmibfam=\tencmmib \scriptfont\cmmibfam=\sevencmmib
\scriptscriptfont\cmmibfam=\fivecmmib
\def\cmmib#1{{\fam\cmmibfam\relax#1}}
\def\f{\cmmib f}

\parskip=\baselineskip
\def\Bbb{\mathbb}
\def\Tr{{\rm Tr}}

\def\1{{\bf 1}}
\def\2{{\bf 2}}
\def\3{{\bf 3}}
\def\4{{\bf 4}}
\def\bar{\overline}
\def\tilde{\widetilde}
\def\R{{\Bbb{R}}}\def\Z{{\Bbb{Z}}}
\def\N{{\mathcal N}}
\def\hat{\widehat}
\def\neg{\negthinspace}
\numberwithin{equation}{section}
\begin{document}

\begin{titlepage}
\begin{flushright}
hep-th/yymm.nnnn
\end{flushright}
\vskip 1.5in
\begin{center}
{\bf\Large{Supersymmetric Boundary Conditions }\vskip0cm
\bf\Large{in ${\cal N}=4$ Super Yang-Mills Theory} }\vskip 0.5cm
{Davide Gaiotto and Edward Witten} \vskip 0.3in {\small{
\textit{School of Natural Sciences, Institute for Advanced
Study}\vskip 0cm {\textit{Einstein Drive, Princeton, NJ 08540
USA}}}}

\end{center}
\vskip 0.5in

\baselineskip 16pt

\begin{abstract}
We study boundary conditions in ${\cal N}=4$ super Yang-Mills
theory that preserve one-half the supersymmetry.  The obvious
Dirichlet boundary conditions can be modified to allow some of the
scalar fields to have a ``pole'' at the boundary.  The obvious
Neumann boundary conditions can be modified by coupling to
additional fields supported at the boundary.  The obvious boundary
conditions associated with orientifolds can also be generalized.
In preparation for a separate study of how electric-magnetic
duality acts on these boundary conditions, we explore moduli
spaces of solutions of Nahm's equations that appear in the
presence of a boundary.  Though our main interest is in boundary
conditions that are Lorentz-invariant  (to the extent possible in
the presence of a boundary), we also explore non-Lorentz-invariant
but half-BPS deformations of Neumann boundary conditions.  We make
preliminary comments on the action of electric-magnetic duality,
deferring a more serious study to a later paper.
\end{abstract}
\end{titlepage}
\vfill\eject

\date{April, 2008}

\tableofcontents

\section{Introduction}\label{introduction}
Supersymmetric boundary conditions in two-dimensional
supersymmetric sigma models have been much studied, because of
their role in string theory and their importance in understanding
mirror symmetry. There has been comparatively very little study of
supersymmetric boundary conditions in four-dimensional
supersymmetric gauge theories.  In this paper, we begin such a
study, focusing on the case of boundary conditions in ${\cal N}=4$
super Yang-Mills theory that preserve one-half of the
supersymmetry.

One obvious choice comes from Neumann boundary conditions for
gauge fields, suitably extended to the rest of the supermultiplet;
another obvious choice comes from Dirichlet boundary conditions
for gauge fields.  A hybrid of the two can be constructed using an
involution (a symmetry of order two) of the gauge group.  Special
cases of these boundary conditions arise in string theory from
D3-NS5 systems, D3-D5 systems, and D3-branes interacting with an
orientifold five-plane.  All three of these constructions have
significant generalizations, as we explain in section
\ref{boundcon} of this paper, where we attempt a systematic survey
of superconformal boundary conditions that preserve one-half of
the supersymmetry.

Dirichlet boundary conditions, and its cousins, lead to an unusual
phenomenon.  A supersymmetric vacuum of ${\cal N}=4$ super
Yang-Mills theory on a half-space is not uniquely determined by
the boundary conditions and the values of the fields at infinity;
even after this data is fixed, the theory has a moduli space of
supersymmetric vacua that appear as solutions of Nahm's equations.
This phenomenon is explored in section \ref{longahm}.

The $S$-dual of this property of Dirichlet boundary conditions is
that gauge theory with gauge group $G$ and Neumann boundary
conditions can be coupled to a boundary superconformal field
theory with $G$ symmetry.  Here we make only a few preliminary
remarks about $S$-duality, deferring a more serious study to a
subsequent paper.

Though our main focus is on boundary conditions that preserve
Lorentz invariance (to the extent that this is possible in the
presence of a boundary) and even conformal symmetry, we also  in
section \ref{without} explore deformations of Neumann boundary
conditions that preserve one-half of the supersymmetry but violate
Lorentz invariance.

\def\cal{\mathcal}
Because Nahm's equations play an important role in this paper, we
mention a few references.  These equations were originally
introduced \cite{Nahm:1979yw} to study solutions of the Bogomolny
equation for monopoles.  See \cite{WeinbergYi} for a review in
that context. They were originally related to D-branes in
\cite{Diaconescu:1996rk}. Subsequent D-brane work
\cite{Tsimpis:1998zh, Kapustin:1998pb} uncovered some of the
issues involving D-branes, impurities, and discontinuities in
Nahm's equations that will be relevant in section \ref{longahm}.
As we explain most fully in section \ref{domainwalls}, the study
of supersymmetric boundary conditions is closely related to the
study of supersymmetric defects.  Early references on
supersymmetric defects via branes include
\cite{DeWolfe:2001pq,Erdmenger:2002ex,Constable:2002xt}.

A rough  analog of our problem in statistical mechanics is to
analyze Kramers-Wannier duality for the Ising model on a lattice
of finite spatial extent.  Kramers-Wannier duality exchanges order
and disorder, so it exchanges ordered and disordered boundary
conditions.  The four-dimensional problem we study is somewhat
similar.  One of the main differences is that as the boundary is
three-dimensional, the complexities of three-dimensional quantum
field theory can enter in the analysis of boundary conditions.

\section{Half-BPS Boundary Conditions}\label{boundcon}
Our goal is to describe supersymmetric boundary conditions -- and
more generally supersymmetric domain walls -- in four-dimensional
${\mathcal N}=4$ supersymmetric Yang-Mills theory. More
specifically, we will describe boundary conditions that are
maximally supersymmetric, which means that they preserve half of the
full underlying supersymmetry and in fact half of the superconformal
symmetry.  The full superconformal symmetry of ${\cal N}=4$ super
Yang-Mills is $PSU(4|4)$ (or $PSU(4|2,2)$, to be more precise about
the signature), and the unbroken subgroup will be $OSp(4|4)$.

 ${\mathcal N}=4$ super Yang-Mills theory
is conveniently obtained \cite{Brink:1976bc}
 by dimensional reduction from ten dimensions.
We begin in $\Bbb{R}^{1,9}$, with metric $g_{IJ}$, $I,J=0,\dots,9$
of signature $-++\dots +$. Gamma matrices $\Gamma_I$ obey
$\{\Gamma_I,\Gamma_J\}=2g_{IJ}$, and the supersymmetry generator
is a Majorana-Weyl spinor $\varepsilon$, obeying
$\bar\Gamma\varepsilon=\varepsilon$, where
$\bar\Gamma=\Gamma_0\Gamma_1\cdots\Gamma_9$.  The fields are a
gauge field $A_I$ and Majorana-Weyl fermion $\Psi$, also obeying
$\bar\Gamma\Psi=\Psi$. Thus, $\varepsilon$ and $\Psi$ both
transform in the $\bf{16}$ of $SO(1,9)$.  The supersymmetric
action is
\begin{equation}\label{supac}I=\frac{1}{e^2}\int \mathrm{d}^{10}x
\,\Tr\,\left(\frac{1}{2}F_{IJ}F^{IJ}-i\bar\Psi
\Gamma^ID_I\Psi\right).\end{equation} The conserved supercurrent is
\begin{equation}\label{upac}J^I =\frac{1}{2}\Tr\,\Gamma^{JK}F_{JK}\Gamma^I\Psi,\end{equation} and the
supersymmetry transformations are
\begin{align}\label{zupac}\delta A_I  &= i\bar\varepsilon\Gamma_I\Psi\\
\delta\Psi &=\frac{1}{2} \Gamma^{IJ}F_{IJ}\varepsilon.
\end{align}

We reduce to four dimensions by simply declaring that the fields
are allowed to depend only on the first four coordinates
$x^0,\dots,x^3$. This breaks the ten-dimensional Lorentz group
$SO(1,9)$ to $SO(1,3)\times SO(6)_R$, where $SO(1,3)$ is the
four-dimensional Lorentz group and $SO(6)_R$ is a group of
$R$-symmetries. Actually, the fermions transform as spinors of
$SO(6)_R$, and the $R$-symmetry group of the full theory is really
$Spin(6)_R$, which is the same as  $SU(4)_R$.  The ten-dimensional
gauge field splits as a four-dimensional gauge field $A_\mu$,
$\mu=0,\dots,3$, and six scalars fields $A_{3+i}$, $i=1,\dots,6$
that we rename as $\Phi_i$.  They transform in the fundamental
representation of $SO(6)_R$ . The supersymmetries $\varepsilon$
and fermions $\Psi$ transform under $SO(1,3)\times SO(6)_R$ as
$(\2,\1,\4)\oplus (\1,\2,\bar\4)$, where $(\2,\1)$ and $(\1,\2)$
are the two complex conjugate spinor representations of $SO(1,3)$
and $\4,\,\bar\4$ are the two complex conjugate spinor
representations of $SO(6)_R$.

Now we want to restrict to a half-space $x^3\geq 0$ and introduce
a supersymmetric boundary condition. We sometimes  write $y$ for
$x^3$. We will consider (until section \ref{without}) only
boundary conditions that are invariant under $SO(1,2)$ Lorentz
transformations that leave fixed the plane $y=0$, and moreover,
are also invariant under the larger group $SO(2,3)$ of conformal
transformations that preserve this plane. It is impossible to also
preserve the full $R$-symmetry group $SO(6)_R$, because, as we
explain momentarily, invariance under $SO(1,2)\times SO(6)_R$
would imply invariance under all of the supersymmetries, or none.
 Preserving all supersymmetries would imply preserving all
translation symmetries (since the commutator of two supersymmetries
is a translation generator), and this is incompatible with having a
boundary at $y=0$.

The problem with $SO(1,2)\times SO(6)_R$ as a symmetry of a
boundary condition is that under $SO(1,2)$, the two spinor
representations of $SO(1,3)$ are equivalent and real, and so under
$SO(1,2)\otimes SO(6)_R$, the supersymmetries transform as
$\2\otimes (\4\oplus\bar\4)$. Because the $\4$ and $\bar \4$ are
inequivalent complex representations, it follows that the space of
supersymmetries has no non-trivial invariant real subspace.  To
get such a subspace, we must reduce $SO(6)_R$ to a suitable
subgroup.

Actually, in order for a boundary condition to be conformally
invariant, the subgroup of $SO(6)_R$ must be $SO(3)\times SO(3)$,
embedded in $SO(6)_R$ in the obvious way. Indeed, the superconformal
group that contains the conformal group $SO(2,3)$ and has half of
the full superconformal symmetry of ${\mathcal N}=4$ super
Yang-Mills theory\footnote{We recall that this superconformal
symmetry is $PSU(4|4)$, with 32 supercharges, half of which are
preserved in $OSp(4|4)$.} is $OSp(4|4)$, whose bosonic part is
$SO(4)\times Sp(4,\R)$. Recall that $Sp(4,\R)$ is a double cover of
$SO(2,3)$, and that $SO(4)$ is a double cover of $SO(3)\times
SO(3)$. $SO(4)$ is the $R$-symmetry subgroup preserved by a boundary
condition with $OSp(4|4)$ symmetry, and that is why a conformally
invariant boundary condition must break $SO(6)_R$ to $SO(3)\times
SO(3)$ or $SU(4)_R$ to $SO(4)$.

Under $SO(4)_R$, the $\4$ and $\bar \4$ of $SU(4)_R$ are real and
equivalent, both transforming as $(\2,\2)$ under $SO(4)_R$, viewed
as a double cover of $SU(2)\times SU(2)$. So we can take any linear
combination $\4'$ of the $\4$ and $\bar \4$, and look for a boundary
condition that preserves a subspace $\2\times \4'$ of the global
supersymmetries. Our boundary conditions will also have manifest
conformal invariance, which will ensure the full $OSp(4|4)$.

Although, up to isomorphism, the unbroken supergroup does not depend
on which linear combination of the $\4$ and $\bar \4$ is chosen in
this construction, the boundary conditions that we can construct in
${\mathcal N}=4$ super Yang-Mills theory do depend very much on this
choice.  That leads to much of the richness of the theory.

$PSU(4|4)$ has a one-parameter group of outer automorphisms that
is responsible for the existence of a  family of inequivalent
embeddings of $OSp(4|4)$.  Represent an element $M$ of the
superalgebra $PSU(4|4)$ by a supermatrix
\begin{equation}\label{gorf}M=\begin{pmatrix} S&T\\
U&V\end{pmatrix}\,\end{equation} where $S$ and $V$ are bosonic
$4\times 4$ blocks and $U$ and $T$ are fermionic ones.  $M$ is
unitary and unimodular (in the $\Z_2$-graded sense), and in
$PSU(4|4)$, $M$ is equivalent to $\lambda M$ for any scalar
$\lambda$. Then $PSU(4|4)$ has a group $U(1)$ of outer
automorphisms, acting by $M\to VMV^{-1}$ with
\begin{equation}V=\begin{pmatrix}e^{i\beta}& 0 \\ 0 &
1\end{pmatrix},~{\beta\in\R}.\end{equation}  Conjugation by $U(1)$
generates the one-parameter family of embeddings of $OSp(4|4)$ in
$PSU(4|4)$.

\subsection{Basic Examples}\label{basicboundary}

It is convenient to  split the scalars $\Phi_i,\,i=1,\dots,6$ into
two groups acted on respectively by the two factors of
$SO(3)\times SO(3)\subset SO(6)_R$.   We take these two groups to
consist of the first three and last three $\Phi$'s; we rename
$(\Phi_1,\Phi_2,\Phi_3)$ as $\vec X=(X_1,X_2,X_3)$ and
$(\Phi_4,\Phi_5,\Phi_6)$ as $\vec Y=(Y_1,Y_2,Y_3)$. We sometimes
write $SO(3)_X$ and $SO(3)_Y$ for the two $SO(3)$ groups.

Though the $\bf{16}$ of $SO(1,9)$, in which the supersymmetries
transform, is irreducible, it is as already explained reducible as
a representation of $W=SO(1,2)\times SO(3)_X\times SO(3)_Y$.
Indeed, the action of $W$ commutes with the three operators
\begin{align}\label{gomot}\notag B_0 &=\Gamma_{456789}\\
             \notag B_1&=\Gamma_{3456}\\
             B_2&=\Gamma_{3789}.\end{align}
They obey $B_0^2=-1$, $B_1^2=B_2^2=1$, and $B_0B_1=-B_1B_0=B_2$,
etc., and  generate an action of $SL(2,\R)$.  We can decompose the
$\bf{16}$ of $SO(1,9)$ as $V_8\otimes V_2$, where $V_8$ transforms
in the real irreducible representation $(\2,\2,\2)$ of
$SO(1,2)\times SO(3)_X\times SO(3)_Y$, and $V_2$ is a
two-dimensional space in which the $B_i$ are represented by
\begin{align}\notag\label{dorf}B_0=&\begin{pmatrix}0 &1\\ -1
&0\end{pmatrix}\\
\notag B_1=&\begin{pmatrix}0 &1\\ 1
&0\end{pmatrix}\\
B_2=&\begin{pmatrix}1 &0\\
0&-1\end{pmatrix}.\end{align}

A boundary condition preserves supersymmetry if and only if it
ensures that the component of the supercurrent normal to the
boundary vanishes.  The supercurrent was written in eqn.
(\ref{upac}). For a supersymmetry generator $\varepsilon$, the
condition we need is that
\begin{equation}\label{zelf} \Tr\,\bar\varepsilon\,
\Gamma^{IJ}F_{IJ}\Gamma_3\Psi=0.\end{equation} For a half-BPS
boundary condition, we do not expect this to hold for all
$\varepsilon$, but only for $\varepsilon$ in a middle-dimensional
subspace of $V_8\otimes V_2$.  In fact, to achieve $OSp(4|4)$
invariance, the condition must hold precisely for
$\varepsilon=v\otimes\varepsilon_0$, where $\varepsilon_0$ is a
fixed element of $V_2$ and $v$ is an arbitrary element of $V_8$.
The choice of $\varepsilon_0$ is equivalent to a choice of
$OSp(4|4)$ embedding in $PSU(4|4)$.

The expression
$(\varepsilon,\tilde\varepsilon)=\bar\varepsilon\Gamma_3\tilde\varepsilon$
defines an $SO(1,2)\times SO(6)$-invariant quadratic form on the
$\bf{16}$ of $SO(1,9)$. For $\varepsilon=v\otimes \varepsilon_0$,
$\tilde\varepsilon=\tilde v\otimes \tilde\varepsilon_0$, we have
$(\varepsilon,\tilde\varepsilon)=\langle v,\tilde v\rangle \langle
\varepsilon_0,\tilde\varepsilon_0\rangle$, where the two factors
are antisymmetric inner products on $V_8$ and on $V_2$. If  we
think of $\varepsilon_0$ as a column vector $\begin{pmatrix} s \\
t
\end{pmatrix}$ and $\bar\varepsilon_0$ as the row vector
$\begin{pmatrix} t & - s \end{pmatrix}$, then we can
 write the inner product on $V_2$ as $\langle
\varepsilon_0,\tilde\varepsilon\rangle
=\bar\varepsilon_0\tilde\varepsilon_0$.

 What boundary conditions should we impose
on  $\Psi$ and the bosonic fields?  In general, a local  boundary
condition for fermions sets to zero half the components of the
fermions.  For invariance under $W=SO(1,2)\times SO(3)\times
SO(3)$, the boundary condition on $\Psi$ must be that
$\Gamma_3\Psi=\Psi'\otimes \vartheta$, where $\Psi'$ takes values
in\footnote{$\Psi$ takes values in the $\bf{16}$ of $SO(1,9)$ and
$\Gamma_3\Psi$ in the $\bf{16}'$.  Multiplication by
$\Gamma_{012}$ exchanges these spaces while commuting with
$SO(1,2)\times SO(3)_X\times SO(3)_Y$ and with the $B$'s.  So for
our purposes, we can identify them both as $V_8\otimes V_2$.}
$V_8\otimes \frak g$ ($\frak g$ is the Lie algebra of $G$) and
$\vartheta$ is a fixed vector in $V_2$. Note that, as $\Gamma_3$
reverses the ten-dimensional chirality, we have
\begin{equation}\label{tony}\bar\Gamma \Psi'=-\Psi'.\end{equation}

Eqn. (\ref{zelf}) is equivalent to
\begin{align}\label{longerlist}
\notag
0&=\bar\varepsilon\left(\Gamma^{\mu\nu}F_{\mu\nu}+2\Gamma^{3\mu}F_{3\mu}\right)
\Psi' \\
\notag 0&=\sum_{\mu=0,1,2}\bar\varepsilon\left(\Gamma^{\mu a}D_\mu
X_a\right)\Psi' \\
\notag 0 & = \sum_{\mu=0,1,2}\bar\varepsilon\left(\Gamma^{\mu
m}D_\mu Y_m\right)\Psi'\\
\notag 0 & = \bar\varepsilon \Gamma^{am}[X_a,Y_m]\Psi' \\
\notag 0 & = \bar\varepsilon\left(
2\Gamma^{3a}D_3X_a+\Gamma^{ab}[X_a,X_b]\right)\Psi'\\
0&
=\bar\varepsilon\left(2\Gamma^{3m}D_3Y_a+\Gamma^{mn}[Y_m,Y_n]\right)\Psi'.
\end{align}
Here Greek indices $\mu,\nu$ originate from ten-dimensional
indices $0,1,2$, while indices $a,b,c$ labeling $X$ and indices
$m,n,p$ labeling $Y$ originate from ten-dimensional indices
$4,5,6$ and $7,8,9$, respectively. Summations over all relevant
values are understood except where indicated. We must pick
$\varepsilon_0$ and $\vartheta$ as well as the boundary conditions
obeyed by the bosonic fields to ensure these equations.

Writing $\bar\varepsilon=\bar v\otimes \bar\varepsilon_0$,  we
want to eliminate $\bar v$ and $\Psi'$ and write these equations
just in terms of $\bar\varepsilon_0$ and $\vartheta$. To do this
in the first equation, we write $\Gamma^{3\mu}=-\frac{1}{2}
\epsilon^{\mu\nu\lambda}\Gamma_{\nu\lambda}\Gamma_{0123}$, where
$\epsilon^{\mu\nu\lambda}$ is the antisymmetric tensor in
$\R^{1,2}$ (with $\epsilon^{012}=1$).  Then, using (\ref{tony}),
we can replace $\Gamma_{0123}\Psi'$ by $B_0\Psi'$.  At this point,
the first equation in (\ref{longerlist}) reduces to
$\bar\varepsilon_0\left(F_{\mu\nu}-
\epsilon_{\mu\nu\lambda}F^{3\lambda}B_0\right) \vartheta =0$. To
similarly rewrite the second equation, we want to replace
$\Gamma^{\mu a}$ with the product of a matrix that acts in $V_8$
and one that acts in $V_2$.  We do this via $(\Gamma^{\mu
a})\Psi'=-\frac{1}{4}(\epsilon^{\mu\nu\lambda}\epsilon^{abc}\Gamma_{\nu\lambda}
\Gamma_{bc}B_2)\Psi', $ where eqn. (\ref{tony}) has been used.
With similar manipulations, we can write each equation just in
terms of $\bar\varepsilon_0$ and $\vartheta$:
\begin{align}\label{boho}
\notag &0= \bar\varepsilon_0\left(F_{\mu\nu}-
\epsilon_{\mu\nu\lambda}F^{3\lambda}B_0\right)\cdot \vartheta \\
\notag
&0=D_\mu X_a\cdot \bar\varepsilon_0 B_2\vartheta,~ \\
\notag &0=D_\mu Y_m \cdot \bar\varepsilon_0 B_1\vartheta, ~\\
\notag&0 =[X_a,Y_m]\cdot \bar\varepsilon_0B_0\vartheta\\
\notag & 0 =\bar\varepsilon_0\left([X_b,X_c]-\epsilon_{abc}D_3X_a
B_1\right)\vartheta\\
 & 0 =\bar\varepsilon_0\left([Y_m,Y_n]-\epsilon_{pmn}D_3Y_p
B_2\right)\vartheta.
\end{align}
(All expressions are to be evaluated at $y=0$.)  In analyzing these
equations, we will at first consider only boundary conditions that
preserve the full gauge symmetry.

To satisfy the first equation, we have to assume that the boundary
condition for the gauge fields is
\begin{equation}\label{elve}\epsilon_{\lambda\mu\nu}F^{3\lambda}+\gamma
F_{\mu\nu}=0,\end{equation} where $\gamma$ is a constant ($\gamma$
equals 0 for the usual Neumann boundary condition $F_{3\lambda}=0$
and $\infty$ for Dirichlet boundary conditions $F_{\mu\nu}=0,
\,\mu,\nu\not=3$).   Then in addition, we must choose
$\varepsilon_0$ and $\vartheta$ so that
\begin{equation}\label{guelve}\bar\varepsilon_0\left(1+\gamma
B_0\right)\vartheta=0.\end{equation}  The alternative of
satisfying the first equation in (\ref{boho}) by setting
$\bar\varepsilon_0 \vartheta=\bar\varepsilon_0 B_0\vartheta=0$ is
not viable, since it cannot be satisfied for real $\varepsilon_0$.

The nature of the remaining equations depends on whether $X$ or
$Y$ or both obeys Dirichlet boundary conditions or in other words
is required to vanish on the boundary.  If we place Dirichlet
boundary conditions on neither $X$ nor $Y$, then to obey the
second, third, and fourth equations we need $0=\bar\varepsilon_0
B_0\vartheta=\bar\varepsilon_0 B_1\vartheta=\bar\varepsilon_0
B_2\vartheta$.  But these conditions are overdetermined and force
$\vartheta=\varepsilon_0=0$.

If we place Dirichlet boundary conditions on both $X$ and $Y$,
then the second, third, and fourth equations become trivial.
However, the last two equations give $\bar\varepsilon_0
B_1\vartheta=\bar\varepsilon_0 B_2\vartheta=0$.  These equations
have no nonzero solution with real $\varepsilon_0$, so also this
case does not occur.

What remains is the case of Dirichlet boundary conditions on just
one of $X$ and $Y$.  Of course, the two cases are equivalent.  For
definiteness, we assume that $Y$ obeys Dirichlet boundary
conditions.   If we take the boundary condition on $X$ to be
\begin{equation}\label{oho}
D_3X_a+\frac{u}{2}\epsilon_{abc}[X_b,X_c]=0\end{equation} for some
constant $u$, then all equations are satisfied if
\begin{equation}\label{polo} 0=\bar\varepsilon_0 B_2\vartheta =
\bar\varepsilon_0\left(1+u B_1\right)\vartheta.\end{equation}

Eqns. (\ref{guelve}) and (\ref{polo}) enable us to determine
everything in terms of $\bar\varepsilon_0$, the assumed generator
of the unbroken supersymmetry. Let us write $\bar\varepsilon_0$ as
a row vector; by scaling we can put it in the form
$\bar\varepsilon_0=(1~a)$.  Then viewing $\vartheta$ as a column
vector, we find that up to scaling
\begin{equation}\label{varth}\vartheta=\begin{pmatrix}a\\
1\end{pmatrix}.\end{equation} Moreover,
\begin{equation}\label{params}
\gamma=-\frac{2a}{1-a^2}, ~~u=-\frac{2a}{1+a^2}.\end{equation}

Both $\gamma$ and $u$ change sign under $a\to-a$.  This results
from the action on the boundary conditions of a reflection
symmetry of the underlying super Yang-Mills theory. The symmetry
acts by a reflection of one of the spatial coordinates parallel to
the boundary, say $x^1$, and a sign change of $ X$. A reflection
of $x^1$ with a sign change of $ Y$ rather than $ X$ corresponds
to $a\to 1/a$, $\gamma\to-\gamma$, $u\to u$, which is also a
symmetry of the above formulas.

\subsubsection{Interpretation}\label{interp}

Let us now discuss the interpretation of some of these boundary
conditions.

{\it NS5-Like Boundary Condition} $~$ The first important case
arises if $\varepsilon$ is an eigenvector of $B_2$, or
equivalently if $a=0$ or $\infty$.  Then $\gamma$ and $u$ vanish,
meaning that the scalar fields $ X$ and the three-dimensional
gauge field $A_\mu$, $\mu=0,1,2$ obey Neumann boundary conditions.
They combine together from a three-dimensional point of view into
a vector multiplet. (This statement is explained  more fully in
section \ref{matter}.) Meanwhile, $Y$ and $A_3$ combine to a
hypermultiplet in the three-dimensional sense; it is subject to
Dirichlet boundary conditions.   In fact, for $G=U(N)$, these are
the boundary conditions that arise for parallel D3-branes ending
on a single NS5-brane whose world-volume is parametrized by
$x^0,x^1,x^2$ and $x^4,x^5,x^6$ (with the four-dimensional
$\theta$-angle vanishing).  We refer to boundary conditions that
preserve such supersymmetry as NS5-like.

{\it D5-Like Boundary Condition} $~$ A second important case is
that $\varepsilon$ is an eigenvector of $B_1$, or $a=\pm 1$.  Then
$\gamma$ is infinite, which means that the gauge field obeys
Dirichlet boundary conditions, with $F_{\mu\nu}$ vanishing on the
boundary for $\mu,\nu=0,1,2$.  $ Y$ also obeys Dirichlet boundary
conditions. Indeed, at $a=\pm 1$, $A_\mu$ and $ Y$ are a vector
multiplet from a three-dimensional point of view. The
hypermultiplet is described by $ X$ and $A_3$, and obeys modified
Neumann boundary conditions, with $u=\pm 1$ in (\ref{oho}).  These
rather simple boundary conditions preserve the same supersymmetry
of a system of D3-branes ending on a D5-brane (with the same
world-volume as the NS5-brane in the last paragraph), and we call
them D5-like. But as we discuss in section \ref{nahmdisc}, they do
not correspond to the case of D3-branes ending on a {\it single}
D5-brane.

One simple but important point is that the Dirichlet boundary
conditions for $\vec Y$ can be slightly generalized (in some
cases, this generalization can be realized in string theory by
displacing branes in the $\vec Y$ direction). Instead of taking
$\vec Y$ simply to vanish, we can pick any commuting triple $\vec
w\in \frak g$ (that is, any three elements $w_m\in \frak g$ such
that $[w_m,w_n]=0$) and take the boundary condition to be
\begin{equation}\label{takebo} \vec Y(0)=\vec w.\end{equation}
Because we take $\vec w$ to be {\it constant} (independent of the
spatial coordinates) and because the gauge field $A_\mu$ vanishes
on the boundary, this gives no contribution to the $D_\mu Y_m$
term in the boundary constraint (\ref{boho}). Because the
components of $\vec w$ commute, there is no contribution to the
$[Y_m,Y_n]$ term, and because $\bar\varepsilon B_0\vartheta=0$ for
D5-like supersymmetry, the $[X_a,Y_m]$ term is harmless.  This
establishes the supersymmetry of (\ref{takebo}).

{\it The $\theta$ Angle}$~$ Finally, let us consider the case of
generic $a$. The general conformally-invariant boundary condition
(\ref{elve}) for the gauge fields, which says that on the boundary
the normal part of the field strength is a prescribed multiple of
the tangential part, is the natural extension of Neumann boundary
conditions for gauge fields in the presence of a four-dimensional
$\theta$-angle. If one adds the $\theta$-term to the usual
Yang-Mills action, so that the combined action takes the form
\begin{equation}\label{takesf}I=\frac{1}{e^2}\int \mathrm{d}^{4}x \,\Tr\,\left(\frac{1}{
2}F_{\mu\nu}F^{\mu\nu}\right)+\frac{\theta}{ 8\pi^2}\int
\Tr\,F\wedge F,\end{equation} then upon varying $I$ with respect to
$A$, with no restriction on the variation of $A$ at the boundary,
one arrives at the boundary condition of eqn. (\ref{elve}) with
$\gamma=-\theta e^2/4\pi^2$.

\subsection{Boundary Conditions That Reduce The Gauge
Symmetry}\label{reduce}

The boundary conditions constructed in section \ref{basicboundary}
preserved the full gauge symmetry.  It turns out, however, that
there are also half-BPS boundary conditions that break part of the
gauge symmetry. Since this idea may seem strange at first, we
motivate it by starting with a natural special case, which arises
in string theory for D3-branes ending on an orientifold or
orbifold five-plane.  We will present this construction for
D5-type supersymmetry (which arises for an orientifold plane in
the 012456 directions or an orbifold that involves reflection of
directions 3789).  Or course, by exchanging $\vec X$ and $\vec Y$,
one can make a similar construction for NS5-like supersymmetry.

Instead of formulating the discussion in terms of a boundary
condition, we start with $\N=4$ super Yang-Mills theory on
$\R^{1,3}$, with no restriction on the sign of $x^3$.  However, we
require that all fields are invariant under a reflection $x^3\to
-x^3$, combined with a suitable automorphism.  Field theory on
$\R^{1,3}$ with this symmetry imposed is equivalent to field theory
on the half-space $x^3\geq 0$ with a suitable boundary condition.
The advantage of working on the covering space is that it makes it
more obvious how to reduce the gauge symmetry while preserving
supersymmetry.

To get a symmetry of $\N=4$ super Yang-Mills theory, a reflection
of space must be accompanied by a reflection of an odd number of
the scalar fields $\Phi^i$ (so as to preserve the orientation of
the underlying ten-dimensional spacetime $\R^{1,9}$).  To preserve
supersymmetry, it is necessary to reflect precisely
three\footnote{The total number of reflected coordinates,
including $x^3$, is then 4.  This is compatible with supersymmetry
since for instance $(\Gamma_{3789})^2=1$.} of the $\Phi^i$. To in
addition preserve the standard $SO(3)\times SO(3)$ $R$-symmetry
(rather than a group conjugate to this), we choose to reflect $
\vec X$ and not $ \vec Y$, or vice-versa.

In any event, we also accompany these reflections with an
automorphism $\tau$ of the gauge group $G$.  $\tau$ must obey
$\tau^2=1$ and may be either an inner automorphism or an outer
automorphism. Both cases can be realized in string theory with
D3-branes, by using certain orbifolds or orientifolds for inner or
outer automorphisms.  This will be discussed in detail elsewhere.
Here, we simply work in field theory.

It is convenient to decompose the Lie algebra $\frak g$ of $G$ as
$\frak g=\frak g^+\oplus \frak g^-$, where $\tau$ acts on $\frak
g^\pm$ by multiplication by $\pm 1$.  For any adjoint-valued field
$\Phi$, we write $\Phi=\Phi^++\Phi^-$, where $\Phi^\pm$ take
values in $\frak g^\pm$. We also write $\Phi^\tau$ for $\tau
\Phi\tau^{-1}$. We require that all fields should be invariant
under the action of $\tau$ combined with a reflection of
$x^3,x^7,x^8,x^9$:
\begin{align}\label{opsy}\notag A_\mu(x^3)&=A_\mu^\tau(-x^3),~
\mu=0,1,2\\ \notag A_3(x^3)&=-A_3^\tau(-x^3),\\ \notag \vec
X(x^3)&= -\vec X^\tau(-x^3)\\ \vec Y(x^3)&= \vec
Y^\tau(-x^3).\end{align}
 This implies certain conditions on the
behavior at the fixed plane $x^3=0$. Writing $\Phi|$ for the
restriction of a field $\Phi$ to $x^3=0$, we get
\begin{align}\label{topsy}
\notag F_{3\mu}^+|
&=F_{\mu\nu}^-|=0 \\
\notag D_3X^-|
&= X^+|=0 \\
 Y^-|
&=D_3Y^+|=0.\end{align}  To describe the boundary conditions on
the fermions, we write $\Psi'=\psi^+\otimes
\vartheta^++\psi^-\otimes \vartheta^-$, where $\psi^\pm$ is valued
in $V_8\otimes \frak g^\pm$, and $\vartheta^\pm$ is valued in
$V_2$.  By imitating the steps that led to eqns. (\ref{boho}),
(\ref{guelve}), and (\ref{polo}), one now finds that the condition
for maintaining one half of the supersymmetry is that
\begin{align}\label{olo}\notag \bar
\varepsilon_0\vartheta^+=\bar\varepsilon_0 B_1\vartheta^+&=0 \\
 \bar\varepsilon_0 B_0\vartheta^-=\bar\varepsilon_0 B_2\vartheta^-
 &=0.\end{align}
 These conditions  are
 equivalent to $\bar\varepsilon_0B_1=w \bar\varepsilon_0$,
 $B_1\vartheta^\pm=\mp w \vartheta^\pm$, where $w=\pm 1$; the two
 choices of $w$ are equivalent under a reflection  (say $x^1\to -x^1$)
that acts trivially on $x^3$ and
 reverses the sign of $\vec X$.  Since the eigenspaces of $B_1$ are
 one-dimensional, everything is determined up to scaling once $w$
 is chosen.

 The two choices of $w$ correspond to $a=0,\infty$;
equivalently, $\varepsilon_0$ is an eigenvector of $B_1$.  The
above boundary condition is D5-like in the sense of section
\ref{interp}. In fact, if $G=U(1)$ and  $\tau$ is the complex
conjugation operation that acts on the Lie algebra as
multiplication by $-1$ (thus, $\tau$ is ``charge conjugation''),
then the above is the standard Dirichlet or D5-like boundary
condition
 -- Dirichlet for $A_\mu$ and $\vec Y$, Neumann for $\vec X$ and $A_3$.
Since multiplication by $-1$ is not a symmetry of a nonabelian Lie
algebra, one might be puzzled what is the analog of this statement
for nonabelian $G$.  That will  become clear in section
\ref{generalization}.

Alternatively, if we set $\tau=1$ and exchange $\vec X$ and $\vec
Y$, we get the simplest NS5-like boundary condition of section
\ref{interp}.

\subsubsection{Generalization To Any $H$}\label{generalization}

The above construction has a generalization that may appear
surprising at first sight (but whose existence may become more
obvious in section \ref{angen}).

In the derivation, we decomposed $\frak g=\frak g^+\oplus \frak
g^-$, where $\frak g^+$ and $\frak g^-$ are even and odd under
$\tau$. Of course, $\frak g^+$ is a Lie algebra -- it is the Lie
algebra of the subgroup $H$ of $G$ that commutes with $\tau$.
Normally, $\frak g^-$ is not a Lie algebra.  In general, we have
\begin{equation}\label{ingen}[\frak g^+,\frak g^+]=\frak
g^+,~~[\frak g^+,\frak g^-]=\frak g^-,~~[\frak g^-,\frak g^-]=\frak
g^+,\end{equation} expressing the fact that $\frak g^+$ and $\frak
g^-$ are respectively even and odd  under $\tau$. The first equation
asserts that $\frak g^+$ is a Lie algebra. The second asserts that
$\frak g^-$ furnishes a representation of this Lie algebra. The
third equation asserts that $H$ is a very special type of subgroup
of $G$: the quotient $G/H$ is a symmetric space.

A close examination of the verification of the supersymmetry of the
boundary conditions of eqn. (\ref{topsy}) shows that while the first
two conditions in (\ref{ingen}) are needed, the third is not.
Therefore, we can generalize the above construction to the case of a
general
 subgroup $H\subset G$, not necessarily related
to a homogeneous space.  What we will get this way can no longer be
interpreted as the result of imposing reflection symmetry on gauge
theory on $\R^{1,3}$. But it will still give a half-BPS boundary
condition for gauge theory on the half-space.

In detail, we proceed as follows.  We pick an {\it arbitrary}
 subgroup\footnote{In most of this paper, our considerations are local and only
 the connected component of $H$ is relevant.}  $H$ of $G$,
and decompose the Lie algebra of $G$ as $\frak g= \frak h\oplus
\frak h^\perp$, where $\frak h$ is the Lie algebra of $H$, and
$\frak h^\perp$ is its orthocomplement. For any adjoint-valued
field $\Phi$, we write $\Phi=\Phi^++\Phi^-$, where $\Phi^+\in
\frak h$, $\Phi^-\in\frak h^\perp$.  Now we formulate $\N=4$ super
Yang-Mills theory on the half-space $x^3\geq 0$, restricting some
fields (or their normal derivatives) to $\frak h$ and some to
$\frak h^\perp$, according to eqn. (\ref{topsy}).  In this way, we
get a half-BPS boundary condition in which the gauge group is
reduced along the boundary from $G$ to $H$, for any $H\subset G$.
(In quantizing the theory, we divide by gauge transformations that
are $H$-valued along the boundary.)  Of course, by exchanging
$\vec X$ and $\vec Y$, we get a second such boundary condition. Of
these two boundary conditions, the first is D5-like and the second
is NS5-like.

An important special case is the case that $H$ is the trivial
subgroup of $G$, consisting only of the identity element.  Then
$\frak g^+=0$ and $\frak g^-=\frak g$; so for any field $\Phi$, we
have $\Phi^+=0, $ $\Phi^-=\Phi$. Then (\ref{topsy}) reduces to
standard Dirichlet boundary conditions (that is, Dirichlet for
$A_\mu$ and $\vec Y$, Neumann for $\vec X$ and $A_3$).

\subsubsection{Global Symmetries}\label{global}

An important property of boundary conditions with reduced gauge
symmetry is that they may admit global symmetries.  Let $K$ be the
subgroup of $G$ that commutes with $H$. The boundary conditions just
described, in which $G$ is reduced to $H$ along the boundary, admit
constant gauge transformations by an element of $K$. These behave as
global symmetries, since at the boundary they are not equivalent to
gauge transformations.  A local operator at $y\not=0$ is required to
be $G$-invariant, and so in particular $K$-invariant, but a local
operator at $y=0$ is only required to be $H$-invariant.  So in
particular,
 local operators that transform
non-trivially under $K$ exist at and only at $y=0$.  The $S$-dual
of this situation involves a construction that we will explain in
section \ref{matter}: for NS-like boundary conditions, it is
possible to introduce matter fields supported only at the
boundary.  These may carry global symmetries, and naturally local
operators that transform non-trivially under those symmetries
exist only on the boundary.

A special case is that if $H=1$ is the trivial group with only the
identity element, then $K$ is all of $G$.  In this case, $G$ acts by
global symmetries on the boundary.

The boundary condition with $H=1$ is actually not exotic at all.
It  coincides with the basic D5-like boundary conditions in which
the vector multiplet obeys Dirichlet boundary conditions and the
hypermultiplet obeys Neumann boundary conditions.  If $H=1$, then
for any field $\Phi$, we have $\Phi^+=0$ and $\Phi^-=\Phi$. As a
result, the boundary conditions (\ref{topsy}) are equivalent to
the D5-like boundary conditions summarized in section
\ref{interp}.

\subsubsection{Central Elements}\label{central}

In eqn. (\ref{topsy}), we have placed Dirichlet boundary
conditions on both $\vec X^+$ and $\vec Y^-$.   Just as in our
earlier treatment of (\ref{takebo}), these  conditions can be
slightly generalized\footnote{Eqn. (\ref{takebo}) is equivalent to
the special case of what follows in which $H$ is trivial, $\frak
g^+=0$ and $\frak g^-=\frak g$.}   so that the boundary values of
the fields in question are constant, but not zero. (This
generalization will typically break some of the global symmetries
that were just described.)

\def\ZZ{{\mathcal Z}}
First of all, we let $\ZZ(\frak g^+)$ denote the center of $\frak
g^+$, and we let $\ZZ(\frak g^-)$ denote the subspace of $\frak g^-$
that commutes with $\frak g^+$.  Let $\vec v$ and $\vec w$ be
triples of elements of $\ZZ(\frak g^+)$ and $\ZZ(\frak g^-)$,
respectively, such that the components of $\vec w$ commute with each
other. The components of $\vec v$ automatically commute with each
other since $\ZZ(\frak g^+)$ is abelian, and the components of $\vec
w$ commute with those of $\vec v$ since $\vec w$ commutes with
$\frak g^+$, which contains $\vec v$.  So in fact all components of
$\vec v$ and $\vec w$ commute.

Then without breaking supersymmetry, the simple Dirichlet boundary
conditions $\vec X^+(0)=\vec Y^-(0)=0$ can be replaced by
\begin{align}\label{replacex} \vec X^+(0) & = \vec v \notag\\
                              \vec Y^-(0) & = \vec w. \end{align}
Indeed, using (\ref{olo}) and the fact that all components of
$\vec v$ and $\vec w$ commute with each other and with $A_\mu(0)$,
one can verify the vanishing of all contributions to (\ref{boho})
that depend on $\vec v$ or $\vec w$.

\subsection{Coupling The NS System To Matter}\label{matter}

We have constructed quite a few half-BPS boundary conditions, but
nonetheless an attempt to understand the action of
electric-magnetic duality on the boundary conditions we have seen
so far would fail. Generically, duality maps boundary conditions
that we have described to ones that we have not yet described. We
explain an important extension for the NS5 case here and an
important extension for the D5 case in section \ref{otherone}.  It
will turn out that these two extensions make it possible to
describe the action of $S$-duality (though in this paper we take
only preliminary steps in that direction).

We begin with the NS5-like boundary condition summarized in
section \ref{interp}, in which $A_\mu$ and three scalars obey
Neumann boundary conditions, while $A_3$ and the other three
scalars obey Dirichlet boundary conditions.  However, we will make
a small change of notation from section \ref{basicboundary}. In
that section, we considered a one-parameter family of possible
choices of the unbroken supersymmetry, always denoting as $\vec Y$
the scalars that obey Dirichlet boundary conditions.  The
parameter that enters the choice of supersymmetry is important,
and we further explore its role elsewhere \cite{GaWi}. But in the
rest of the present paper, we will consider only boundary
conditions that have the same supersymmetry as the D3-D5 system,
or equivalently, if we exchange $\vec X$ and $\vec Y$, the same
supersymmetry as the D3-NS5 system.

We will describe several different constructions, and will want to
combine them together.  This is more straightforward if they all
preserve the same supersymmetry.  So in the rest of this paper, we
always assume that the generator $\bar\varepsilon$ of the unbroken
supersymmetry is an eigenvector of $B_1$.  A related statement is
that, in a sense that will become clear, though we will consider
many different boundary conditions for vector multiplets and
hypermultiplets, in the rest of this paper, $\vec X$ will always
transform in a hypermultiplet and $\vec Y$ will always be part of
a vector multiplet.

To put in this framework the simplest NS5-like boundary
conditions, we make a change of notation relative to section
\ref{basicboundary}, and exchange $\vec X$ and $\vec Y$. Thus, the
boundary conditions that we will generalize, without changing the
unbroken supersymmetry, are Neumann boundary conditions for
$A_\mu$ and $\vec Y$, together with Dirichlet boundary conditions
for $\vec X$, suitably extended to the rest of the supermultiplet.

\subsubsection{Three-Dimensional Theory With Infinite-Dimensional Gauge
Group}\label{freedom}

In particular, in our starting point, at the boundary $y=0$ there
are gauge fields of the full $G$ symmetry. This being so, one can
introduce additional degrees of freedom that carry the $G$
symmetry and are supported at the boundary. These additional
degrees of freedom must have $\N=4$ superconformal symmetry if the
combined system is to have that property, but otherwise they are
arbitrary.

Of course, we should ask here whether a bulk system with $\N=4$
supersymmetry (in the four-dimensional sense) can be coupled to a
boundary system with $\N=4$ supersymmetry (in the
three-dimensional sense), in such a way as to preserve the full
supersymmetry of the boundary theory.  A rather similar question,
involving defects instead of boundaries, was addressed in
reference \cite{DeWolfe:2001pq}.  Rather than performing a similar
calculation, we will take a short-cut, first of all to show that
the supersymmetric coupling exists at the classical level.  We
will assume to begin with that the boundary theory is described by
hypermultiplets that parametrize a hyper-Kahler manifold $Z$ with
$G$ symmetry.

The first step will be to describe the gauge theory on the
half-space $y\geq 0$ as a three-dimensional theory with an
infinite-dimensional gauge group.  We let $L$ be the half-line
$y\geq 0$, and we think of the half-space $y\geq 0$ as
$\R^{1,2}\times L$. We let $\hat G$ be the group of maps from $L$ to
$G$.  The Lie algebra of $\hat G$ is spanned by $\frak g$-valued
functions on $L$.  On this Lie algebra, there is a natural positive
definite inner product; if $a$ and $b$ are two such functions, we
define $\langle a,b\rangle=-\int \mathrm{d}y\,\Tr\,ab$, where
$-\Tr\,ab$ is a positive definite invariant inner product on $L$.
So formally we can write down in the usual way a supersymmetric
gauge theory action on $\R^{1,2}$, with ${\cal N}=4$ supersymmetry
in the three-dimensional sense, for a vector multiplet with gauge
group $\hat G$. The fields in this theory are the three-dimensional
gauge field $A_\mu$, $\mu=0,1,2$ (but not $A_3$), plus the scalars
$\vec Y$ (but not $\vec X$), and half of the fermions of $\N=4$
super Yang-Mills theory.

This theory, though formally supersymmetric, is not really
well-behaved unless we also add suitable hypermultiplets.  The
reason is that the kinetic energy contains no derivatives in the $y$
direction.  For example, the gauge theory part of the action is
\begin{equation}\label{lofus} \frac{1}{2e^2}\int_{\R^{1,2}}\mathrm{d}^3x \int_L \mathrm{d}y
\sum_{\mu,\nu=0,1,2}\Tr\,F_{\mu\nu}F^{\mu\nu}.\end{equation} Here
the integral over $\R^{1,2}$ is part of the definition of
three-dimensional gauge theory, and the integral over $L$ arises
because it is part of the definition of the quadratic form on the
Lie algebra.  Clearly, (\ref{lofus}) is part of the usual Yang-Mills
action in four dimensions, but the terms involving $F_{3\mu}$ and
containing derivatives in the $y$ direction are missing.

To complete the theory, we need hypermultiplets, namely the
additional fields $A_3$ and $\vec X$. They parametrize an
infinite-dimensional flat hyper-Kahler manifold. The hyper-Kahler
metric is
\begin{equation}\label{molfo}ds^2=-\int_L \mathrm{d}y \,\,\Tr\,\left(\delta
A_3^2+\sum_i\delta X_i^2\right).\end{equation} The three
hyper-Kahler forms are
\begin{equation}\label{golfo}\omega_i=\int_L \mathrm{d}y\,\,\Tr\,\left(\delta
A_3\wedge \delta X_i+\delta X_{i+1}\wedge \delta
X_{i-1}\right),~i=1,2,3,\end{equation}
 where we set $X_{i+3}=X_i$.
 This formula is covariant under $SO(3)$ rotations of $X_i$ and $\omega_i$, though
 not written so as to make this manifest.

 These equations
describe an infinite-dimensional flat hyper-Kahler manifold on
which $\hat G$ acts by gauge transformations. One point to mention
here is that the fields $X_a$ transform in the adjoint
representation of $\hat G$, but $A_3$,  because of its
inhomogeneous gauge transformation law $\delta
A_3=-D_3u=[u,A_3]-\partial_3u$ (where $u$ is the generator of a
gauge transformation), transforms in what one might call an
``affine deformation'' of the adjoint representation. This has no
close analog for finite-dimensional groups.

Nonetheless, the pair $(\vec X,A_3)$ form a hypermultiplet, that
is, they parametrize a hyper-Kahler manifold with $\hat G$ action.
 So following the
standard recipe, we can formally write down the three-dimensional
supersymmetric action for the coupling of this hyper-Kahler
manifold to the vector multiplet of $\hat G$.  The sum of this
action with the vector multiplet action described earlier is the
action of four-dimensional $\N=4$ super Yang-Mills theory on the
half-space. For example, the kinetic energy of the hypermultiplet
gives the $F_{3\mu}^2$ term that was missing in (\ref{lofus}).

{}From this point of view, there is no problem to add additional
hypermultiplets, with $G$ symmetry, that are supported at $y=0$.
First of all, there is a natural homomorphism from $\hat G$ to $G$
by evaluation at $y=0$.  Thus, if $g(y):L\to G$ is an element of
$\hat G$, we simply map $g(y)$ to its boundary value $g(0)$. So if
$Z$ is any space with $G$ symmetry, we can regard it as a space
with $\hat G$ symmetry: an element $g(y)\in \hat G$ acts on $Z$
via the given action of $g(0)$.  If therefore $Z$ is a
hyper-Kahler manifold with $G$ action, we can view it as a
hyper-Kahler manifold with $\hat G$ action.  Then we just write
down the standard $\N=4$ theory in the three-dimensional sense,
with the vector multiplets being those of the group $\hat G$, and
the hypermultiplets being $(A_3,\vec X)$ and the fields
parametrizing $Z$.

This construction gives a four-dimensional theory with a boundary
hypermultiplet.  The theory is conformally invariant at the
classical level if and only if the purely three-dimensional theory
with target $Z$ is conformally invariant.  In turn, that is so
precisely if the hyper-Kahler manifold $Z$ is conical, for example
if $Z$ is a linear manifold $\R^{4n}$ for some $n$.

It is also possible to modify this construction by taking the metric
on the Lie algebra of $\hat G$ to be $\langle a,b\rangle=-\int
\mathrm{d}y \,e(y)^{-2}\,\Tr\,ab$, with an arbitrary positive
definite  function $e(y)^2$.  This gives a construction of the
half-BPS Janus configuration, first described in field theory in
\cite{DEG}, for the case that the gauge coupling $e$ is a function
of $y$ but the angle $\theta$ is constant.  For the generalization
to varying $\theta$, see \cite{GaWi}.

\subsubsection{Shifted Boundary Condition of $\vec
Y$}\label{shiftboun}

By computing the hyper-Kahler moment map of $(A_3,\vec X)$, we can
get a new understanding of some known results about coupling of
bulk gauge fields to localized hypermultiplets
\cite{DeWolfe:2001pq}.
 To compute the hyper-Kahler moment
map, we must contract $\omega_i$ with the vector fields $\delta
A_3=-D_3\alpha$, $\delta X_i=[\alpha,X_i]$ that generate the
action of the gauge group. We call this vector field $V(\alpha)$.
Its contraction with $\omega_i$ is
\begin{equation}\iota_{V(\alpha)}\omega_i=\int
\mathrm{d}y\,\Tr\,\left(-D_3\alpha \delta X_i-\delta A_3
[\alpha,X_i]+\alpha [X_{i+1},\delta X_{i-1}]-\alpha [\delta
X_{i+1},X_{i-1}]\right).\end{equation} The definition of the
hyper-Kahler moment map $\mu_i(\alpha)$ is that
$\delta\mu_i(\alpha)=\iota_{V(\alpha)}\omega_i$.  A short
calculation, with some integration by parts, shows that
\begin{equation}\label{zota}\mu_i(\alpha)=\int \mathrm{d}y\,\,\Tr\,\left(
\alpha\left(\frac{DX_i}{ Dy}+[X_{i+1},X_{i-1}]\right)\right)+\Tr\,
\alpha X_i(0).\end{equation} In integrating by parts, we have
included a surface term at $y=0$, but a possible surface term at
$y=\infty$ vanishes if the energy is finite and will not be
important.

The consequences of this formula may be clearer if instead of
writing the pairing of the moment map $\vec\mu$ with an arbitrary
element $\alpha$ of the Lie algebra of $\hat G$, we write out
$\vec\mu$ as a $\frak g$-valued function on $L$:
\begin{equation}\label{hypmap}\vec\mu(y)=\frac{D\vec X}{Dy}+\vec X\times
\vec X(y)+\delta(y) \vec X(0).\end{equation} Now we can get a
somewhat better understanding of the NS boundary condition
summarized in section \ref{interp}.  In general, for coupling to any
hypermultiplets, the action contains a term $\int \mathrm{d}^3x
\,(\vec\mu,\vec\mu)$. In the present context, this means
$-\int_{\R^{2,1}} \mathrm{d}^3x\int_L \mathrm{d}y\, \Tr\,\vec\mu^2$.
Because of the delta function in $\vec\mu$, the action is finite
only if $\vec X(0)=0$, which (modulo the exchange of $\vec Y$ and
$\vec X$) is the boundary condition that we found in section
\ref{basicboundary}.

Now we can generalize this to the case that a boundary
hypermultiplet is present, parametrizing a hyper-Kahler manifold
$Z$.  $Z$ has its own hyper-Kahler moment map $\vec \mu^Z$, and the
hyper-Kahler moment map of the combined system is obtained by adding
this to eqn. (\ref{hypmap}):
\begin{equation}\label{nypmap}\vec\mu(y)=\frac{D\vec X}{Dy}+\vec X\times
\vec X(y)+\delta(y) (\vec X(0)+\vec\mu^Z).\end{equation} To keep the
action finite, it now must be that in the presence of the boundary
hypermultiplet, the boundary condition on $\vec X$ is shifted from
$\vec X(0)=0$ to
\begin{equation}\label{ympap}\vec X(0)+\vec \mu^Z=0.\end{equation}
This closely parallels a result in \cite{DeWolfe:2001pq}.

\subsubsection{Analog For General $H$}\label{angen}

We can now get a new understanding of the boundary conditions found
in section \ref{generalization} with the gauge symmetry reduced from
$G$ to $H$ along the boundary.

For any subgroup $H$ of $G$, we define a subgroup $\hat G_H$ of
$\hat G$ that consists of maps $g:L\to G$ such that $g(0)\in H$.
We take $(A_\mu,\vec Y)$ to be the vector multiplets of a
three-dimensional theory with gauge group $\hat G_H$.  And we
interpret $(A_3,\vec X)$ as hypermultiplets of this symmetry,
valued in the adjoint representation but with the boundary
condition that  $\vec X(0)$ is valued in $\frak h^\perp$. As
above, the  condition on $\vec X(0)$ can be explained by computing
the delta function contribution to the moment map, which turns out
to be the projection of $\vec X(0)$ to $\frak  h$ (the projection
arises simply because the Lie algebra of $\hat G_H$ is spanned by
functions $\alpha:L\to \frak g$ with $\alpha(0)\in \frak h$).

The $\N=4$ supersymmetric theory with this vector multiplet and
hypermultiplet is one that we have already constructed.  It arises
from gauge theory on a half-space $\R^{1,2}\times L$ with the
boundary condition constructed in section \ref{generalization} in
which the gauge symmetry is reduced from $G$ to $H$ on the
boundary.

Moreover, it should be clear now that this system can be coupled to
any boundary hypermultiplets that parametrize a hyper-Kahler
manifold $Z$ with $H$ action.  The group $\widehat G_H$ has a
homomorphism to $H$ by mapping a function $g(y)$ representing an
element of $\hat G_H$ to its boundary value $g(0)$.  So $Z$ can be
regarded as a hyper-Kahler manifold with $\hat G_H$ symmetry. Hence,
we can simply borrow the standard formulas for coupling vector
multiplets and hypermultiplets in three dimensions.

Eqn. (\ref{ympap}) still holds and shows that in the presence of
the boundary hypermultiplet, the boundary condition on $\vec X$
becomes \begin{equation}\label{lofty}\vec
X^+(0)+\vec\mu^Z=0,\end{equation} where $\vec X^+(0)$ is the
projection of $\vec X(0)$ to $\frak h$.

\subsubsection{Coupling To A More General Boundary Theory}\label{abstract}

Hopefully, we have given a fairly clear recipe for coupling $\N=4$
super Yang-Mills in bulk to boundary hypermultiplets.  One can
also, without any difficulty, add vector multiplets that are
supported on the boundary and couple to the same hypermultiplets.
One simply replaces the group $\hat G$ in the above by $\hat
G\times J$, where $J$ is a finite-dimensional compact gauge group
that ``lives'' at $y=0$.  The boundary hypermultiplets can then be
coupled to $J$ as well as $\hat G$.  Therefore, this recipe
extends to the coupling of the bulk theory to any boundary theory
of hypermultiplets and vector multiplets. The recipe is also
useful for understanding the coupling to a more general CFT if
that theory arises by renormalization group flow from a weakly
coupled theory of vector multiplets and hypermultiplets with $G$
action. Many interesting three-dimensional CFT's arise in this
way.

To understand the coupling of $\N=4$ super Yang-Mills theory in bulk
to a completely general CFT would require a more abstract approach
that we will not develop here.  One simple comment is that if this
CFT has a Higgs branch, the description we have given is valid for
describing the low energy coupling of the bulk $\N=4$ theory to that
Higgs branch.  (A full understanding that is not just valid at low
energy would require returning to the underlying CFT.)  Another
useful point is that eqn. (\ref{ympap}) holds in general, provided
$\vec\mu^Z$ is understood as a suitable CFT operator (whose
expectation value on the Higgs branch coincides with the classical
hyper-Kahler moment map).

\def\O{\CO}

 Going back to the simple case that $Z$ parametrizes $\R^{4n}$
with a linear action of $G$, we would like to know that the
coupling is conformally invariant quantum mechanically and not
just classically. For a detailed treatment of a similar problem
(involving bulk rather than boundary impurities), see
\cite{DeWolfe:2001pq}. A partial shortcut is to observe that
global ${\mathcal N}=4$ supersymmetry in this situation actually
implies superconformal symmetry.  A collection of free
hypermultiplets supported on a hyperplane or a boundary  (and
coupled to gauge fields in bulk) simply does not admit any
possible counterterm of scaling dimension 3 or less that preserves
global ${\mathcal N}=4$ supersymmetry.

\subsubsection{Shifting The Boundary Conditions}\label{shifting}

Finally, we want to describe from the present point of view the
possibility, explained in section \ref{central}, to shift the
boundary conditions on $\vec X$ and $\vec Y$ by constants.

In general, in coupling a vector multiplet to hypermultiplets, one
is free to add a constant to the moment map, as long as this
preserves gauge invariance. The resulting parameters are usually
called Fayet-Iliopoulos (FI) parameters.  In the present context,
this means that we can pick any triple $\vec v$ valued in the center
of $\frak h$, and shift the moment map by a boundary term
proportional to $\vec v$. Eqn. (\ref{nypmap}) then becomes
\begin{equation}\label{gelf}\vec\mu(y)=\frac{D\vec X}{Dy}+\vec X\times
\vec X(y)+\delta(y) (\vec X(0)+\vec\mu^Z-\vec v),\end{equation} and
the boundary condition (\ref{lofty}) on $\vec X$ becomes
\begin{equation}\label{ofty}\vec X^+(0)+\vec\mu^Z=\vec
v.\end{equation}  This is the boundary condition of section
\ref{central}, or more precisely the generalization of it to include
the coupling to a boundary matter system with moment map
$\vec\mu^Z$.

Now let us discuss the other term in eqn. (\ref{replacex}), the
shift in the boundary value of $\vec Y^-$ by elements $\vec w\in
\frak g^-$ that commute with each other and with $\frak h$.  As
they commute with $H$, the components of $\vec w$ are elements of
the Lie algebra of the global symmetry group $K$ described in
section \ref{global}. As they commute with each other, the
components of $\vec w$ can be conjugated to a maximal torus $T_K$
of $K$.  Thus, they lie in an abelian group of global symmetries.

In three-dimensional $\N=4$ supersymmetry with a finite dimensional
gauge group coupled to hypermultiplets, an abelian group $F$ of
global symmetries leads to parameters -- often called mass terms --
that can be incorporated in the theory.  The standard way to
describe these parameters is to weakly gauge $F$, give expectation
values to the scalar fields in the vector multiplet of $F$, and then
turn off the gauge coupling of $F$.

It is not clear to us whether, in our situation with an
infinite-dimensional gauge group, one can introduce the mass
parameters in precisely this way.\footnote{One can gauge the global
symmetry $T_K$, which means the following.  Let $H'=H\times T_K$.
Then repeating the analysis of section \ref{angen} with $H'$
replacing $H$, we arrive at a theory in which $T_K$ has been gauged.
But it does not seem to be natural to vary the $T_K$ gauge coupling
independently of the bulk $G$ gauge coupling. This problem has no
analog for finite-dimensional gauge groups.} We therefore offer the
following alternative for introducing the mass parameters $\vec w$
in our situation.

We recall first that the Lie algebra of $\hat G_H$ consists of
functions $\phi:L\to \frak g$ such that $\phi(0)\in \frak h$, or
equivalently $\phi^-(0)=0$.  For any element $c\in {\cal Z}(\frak
g^-)$ (the subspace of $\frak g^-$ that commutes with $\frak h=\frak
g^+$), we can deform the adjoint representation of $\frak g$ to the
space of functions $\phi:L\to \frak g$ that obey $\phi^-(0)=c$. Such
a continuous deformation of a representation has no analog for a
finite-dimensional compact group.

Now we modify the $\hat G_H$ vector multiplet as follows.  We make
no change in the three-dimensional $\hat G_H$ gauge fields $A_\mu$,
or in the fermions.  But instead of interpreting $\vec Y$ as three
scalar fields valued in the adjoint representation of $\hat G_H$,
and thus obeying the boundary condition $\vec Y^-(0)=0$, we consider
each component $Y_m$, $m=1,2,3$ to take values in a deformed adjoint
representation with $c=w_m$.

Though the fields $Y_m$ are not quite adjoint-valued, their
commutators with each other or with adjoint-valued fields such as
the other fields in the vector multiplet are adjoint-valued. And
their commutators with hypermultiplet fields take values in the same
spaces as at $w_m=0$.  To verify these statements, one uses the fact
that the $w_m$ commute with each other and with $H$, so that their
presence does not affect the relevant properties of commutators.
Given these facts, the three-dimensional supersymmetric action with
gauge group $\hat G_H$ can be defined, and supersymmetry verified,
in the usual way, despite the deformation of the adjoint
representation.

\subsection{The D5 System And Nahm's Equations}\label{nahm}

A vector multiplet with Neumann boundary conditions can be coupled
to boundary degrees of freedom, as described in section
\ref{matter}.  What can be the dual of this for a vector multiplet
with Dirichlet boundary conditions?  This question may seem
puzzling, because if a gauge field is required to vanish on the
boundary, there is no obviously natural way to couple it to boundary
degrees of freedom.  The answer to this question turns out to be
that half-BPS boundary conditions with Dirichlet boundary conditions
on gauge fields are automatically coupled, in effect, to certain
boundary degrees of freedom.

$\N=4$ super Yang-Mills theory on $\R^{1,3}$ has supersymmetric
vacua parametrized by expectation values of $\vec X$ and $\vec Y$.
To ensure supersymmetry, these expectation values must all
commute. What happens on a half-space?  It no longer makes sense,
of course, to look for vacua with unbroken {\it four}-dimensional
Poincar\'e supersymmetry, but we can look for vacua with {\it
three}-dimensional  Poincar\'e supersymmetry. Three-dimensional
Poincar\'e invariance requires that $F_{\mu\nu}$ and $F_{3\mu}$
should vanish.  It allows $\vec X$ and $\vec Y$ to have
expectation values, depending only on $y$.  We want to determine
the condition on $\vec X(y)$ and $\vec Y(y)$ that ensures
supersymmetry.

The supersymmetry variation of the fermion fields $\Psi$ of $\N=4$
super Yang-Mills theory is conveniently written
\begin{equation}\label{supvar}\delta\bar\Psi=\frac{1}{2}\bar\varepsilon
\Gamma^{IJ}F_{IJ}.\end{equation}  The condition for supersymmetry
is simply that the right hand side must vanish:
\begin{equation}\label{upvar}  \bar\varepsilon
\Gamma^{IJ}F_{IJ}=0.              \end{equation} This is the same
as the condition (\ref{zelf}) for a supersymmetric boundary
condition, except that the factor $\Gamma_3\Psi$ is missing.
Consequently, the equations resulting from (\ref{supvar}) are the
same as  eqns. (\ref{boho}) that characterize supersymmetric
boundary conditions, with the very important difference that the
factor of $\vartheta$ should be omitted -- so that in effect we
must satisfy eqn. (\ref{boho}) for all choices of $\vartheta$.

After imposing three-dimensional Poincar\'e invariance, we are left
with three equations:
\begin{align}\label{noho}
\notag&0 =[X_a,Y_m]\cdot \bar\varepsilon_0B_0\\
\notag & 0 =\bar\varepsilon_0\left([X_b,X_c]-\epsilon_{abc}D_3X_a
B_1\right)\\
 & 0 =\bar\varepsilon_0\left([Y_m,Y_n]-\epsilon_{pmn}D_3Y_p
B_2\right). \end{align} The first tells us that all components of
$\vec X$ and $\vec Y$ commute. The second tells us that unless
$\bar\varepsilon_0$ is an eigenvector of $B_1$, we have $D\vec
X/Dy=[\vec X,\vec X]=0$.  As a result, $\vec X$ coincides
everywhere with its value at spatial infinity (up to a gauge
transformation), and the different components of $\vec X$ must
commute. The third equation similarly tells us that unless
$\bar\varepsilon_0$ is an eigenvector of $B_2$, $\vec Y$ is a
commuting constant and coincides with its value at spatial
infinity.  Thus, for generic $\bar\varepsilon_0$, all components
of $\vec X$ and $\vec Y$ commute with each other and are
covariantly constant.

Something interesting happens only if $\bar\varepsilon_0$ is an
eigenvector of $B_1$ or $B_2$.  We will take $\bar\varepsilon_0$
to be an eigenvector of $B_1$. (As usual, the case that
$\bar\varepsilon_0$ is an eigenvector of $B_2$ simply differs by
exchanging $\vec X$ and $\vec Y$.)  If $\bar\varepsilon_0B_1=\pm
\bar\varepsilon_0$, then the condition for supersymmetry gives
\begin{equation}\label{zildo} \frac{DX^1}{Dy}=\pm
[X^2,X^3],\end{equation} and cyclic permutations. It also implies
that $\vec Y$ is a covariant constant whose components commute with
each other and with $\vec X$:
\begin{equation}\label{ildo}\frac{D\vec Y}{Dy}=[\vec Y,\vec
Y]=[\vec Y,\vec X]=0.\end{equation} More briefly, the components of
$\vec Y$ generate unbroken gauge symmetries.

The equations (\ref{zildo}) are known as Nahm's equations
\cite{Nahm:1979yw}, and arise frequently as conditions for
supersymmetry.  Even after specifying the behavior of $\vec X$ at
infinity, Nahm's equations have an interesting moduli space of
solutions, which we will explore in section \ref{longahm}.  The
existence of this moduli space means that, when vector multiplets
obey Dirichlet boundary conditions, as happens in the D5-like
case, there are in a sense boundary degrees of freedom already
present in the theory. The dual of this for gauge fields with
Neumann boundary conditions is that in that case, boundary degrees
of freedom can be naturally added, as in section \ref{matter}.

\subsubsection{Poles}\label{otherone}

Nahm's equations have another important consequence.  Poles in the
solutions of Nahm's equations can be used to generate new half-BPS
boundary conditions.  Though it may sound exotic, this idea is not
new; it reflects the familiar fact \cite{Diaconescu:1996rk},
\cite{Constable:1999ac} that D3-branes ending on D5-branes can be
described by solutions of Nahm's equations with poles.  For
related reasons, such poles played a crucial role in Nahm's
original use of his equation \cite{Nahm:1979yw}.  Defining a new
boundary condition by requiring a pole of a specified type is
somewhat analogous to defining 't Hooft operators in gauge theory
(or disorder operators in statistical mechanics) by requiring a
singularity of a prescribed type.

The basic singular solution of Nahm's equation is simple to
describe.  With one choice of sign, Nahm's equations can be written
\begin{equation}\label{morn} \frac{\mathrm{d}X^1}{\mathrm{d}y}+[X^2,X^3]=0,\end{equation}
and cyclic permutations.  Now let $t^1,t^2,t^3$ be any elements of
$\frak g$ that obey the $\frak{su}(2)$ commutation relations
$[t^1,t^2]=t^3$, and cyclic permutations.  Thus, specifying the
$t^i$ amounts to specifying a homomorphism of Lie algebras
$\rho:\frak{su}(2)\to\frak g$.  Having made such a choice, we obtain
a solution of Nahm's equations with a pole at the origin:
\begin{equation}\label{singsol} X^i(y)=\frac{t^i}{y}.\end{equation}

So far, when we have discussed gauge theory on the half-space $y\geq
0$, we have considered fields that are regular on this half-space,
including its boundary at $y=0$, and the question has been what
types of boundary values are allowed.  Somewhat as in the definition
of 't Hooft operators, we can introduce a new type of boundary
condition by requiring a singularity of a prescribed type at $y=0$.
If we wish in this way to get a supersymmetric boundary condition,
we must select a singularity that is compatible with supersymmetry.
The singularity $X^i\sim t^i/y$ clearly has this property, since it
is compatible with Nahm's equations.

So for every choice of a non-zero homomorphism
$\rho:\frak{su}(2)\to\frak g$, we get a new half-BPS boundary
condition as follows.  Setting $t^i$ to be the images of a
standard set of $\frak{su}(2)$ generators, we require that the
behavior of $X^i$ near $y=0$ is $X^i\sim t^i/y$.

This preserves the same supersymmetry that is preserved by Dirichlet
boundary conditions on gauge fields, since that is the supersymmetry
that is preserved by Nahm's equations.  A boundary condition of this
type breaks the gauge symmetry from $G$ to the subgroup $G'$ that
commutes with $\rho$.  This gives a different type of half-BPS
boundary condition with reduced gauge symmetry from what was
described in section \ref{generalization}. For the same reason as in
that case, there is a group of global symmetries. This group is $F$,
the commutant of $\rho$ in $G$ (that is, the subgroup of $G$ that
commutes with $\rho$).

As we explain next, the two constructions can be combined, roughly
speaking by gauging a subgroup of $F$.

\subsection{Combining The Constructions}\label{combining}

We have described a significant generalization of each of the most
obvious half-BPS boundary conditions.  Neumann boundary conditions
were generalized in section \ref{matter} by including a boundary
CFT.  Dirichlet boundary conditions were generalized in section
\ref{nahm} using a  homomorphism $\rho:\frak{su}(2)\to\frak g$.
And orbifold boundary conditions were generalized in section
\ref{generalization} to depend on a choice of an arbitrary
subgroup $H$ of the gauge group. It is possible to combine all
three constructions, preserving the same supersymmetry, which we
take to be of D5-type.

We will make the construction in three steps.  Choosing an
$\frak{su}(2)$ embedding $\rho$, we require that $\vec X$ should
have the familiar pole $\vec X\sim \vec t/y$.

Fields that do not commute with $\rho$ will all vanish at the
boundary, because of terms in the Hamiltonian that involve
commutators with $\vec X$.  Denoting therefore as $\f$ the Lie
algebra of $F$ (the commutant of $\rho$), what remains is to
describe supersymmetric boundary conditions for the $ \f$-valued
parts of all fields. For this, in brief, we can use any
supersymmetric boundary condition in $F$ gauge theory. We pick any
subgroup $H$ of $F$ and decompose $\f=\f^+\oplus \f^-$, where
$\f^+=\frak h$ and $\f^-$ is the orthocomplement. Then as in
section \ref{generalization}, we expand any field $\Phi$ as
$\Phi^++\Phi^-$, with $\Phi^\pm\in\f^\pm$. We impose the boundary
conditions described in section \ref{generalization}:
\begin{align}\label{lopsy}
\notag F_{3\mu}^+|
&=F_{\mu\nu}^-|=0 \\
D_3X^-|&= X^+| =0 \notag  \\ Y^-|&= D_3Y^+| =0 .\end{align} The
condition $F_{\mu\nu}^-=0$ means that the curvature restricted to
the boundary is $\frak h$-valued, so that the gauge group along
the boundary is $H$.

If we take $H$ to be trivial, so that for every field $\Phi$,
$\Phi^+=0$ and $\Phi=\Phi^-$, this reduces to the boundary
condition of section \ref{otherone}. Whatever $H$ may be, since
the gauge symmetry along the boundary is $H$, we can introduce
boundary hypermultiplets (or more general boundary variables) with
$H$ symmetry and couple them to the bulk gauge fields.  When we do
this, the boundary condition on $\vec X$ shifts from $\vec X^+|=0$
to $\vec X^+|+\vec\mu^Z=0$, where $\vec\mu^Z$ is the moment map
for the boundary variables.

A unified way to describe the whole construction is to follow the
logic of section \ref{matter}.  We construct a three-dimensional
supersymmetric gauge theory with an infinite-dimensional gauge group
$\hat G_H$ consisting of maps $g:L\to G$ such that $g(0)\in H$. The
bulk vector multiplets are $(A_\mu,\vec Y)$.  We couple to
hypermultiplets $(\vec X,A_3)$ that are adjoint-valued but such that
$\vec X$ is required to have the pole $\vec X\sim \vec t/y$
determined by $\rho$. We add additional boundary hypermultiplets
(and possibly vector multiplets) as desired.  The supersymmetric
action we want then arises from the standard construction of a
three-dimensional supersymmetric gauge theory with vector multiplets
and hypermultiplets.

 At this stage, we can follow the logic of section \ref{shifting}
and introduce some additional parameters. These parameters are a
triple $\vec v$ of elements of the center of $\frak h$, and a
triple $\vec w$ of elements of $\f^-$ that commute with each other
and with $\frak h$.  The parameters are introduced by shifting the
boundary conditions, which become
\begin{align}\label{newpars} \vec X^+|+\vec\mu^Z& = \vec v\notag\\
                             \vec Y^-|          & =\vec
w.\end{align}

 The general maximally supersymmetric boundary
condition that we know of\footnote{Some of these boundary
conditions can be generalized to include the $\theta$ angle
\cite{GaWi}.  The unbroken supersymmetry is then not of D5-type,
but rotated by an outer automorphism of $PSU(4|4)$.} thus involves
a triple $(\rho,H,B)$, where $\rho$ is a homomorphism from
$\frak{su}(2)$ to $\frak g$, $H$ is a subgroup of $G$ that
commutes with $\rho$, and $B$ is an ${\cal N}=4$ supersymmetric
field theory with $H$ symmetry.  The parameters that such a
boundary condition depends upon (after fixing the parameters of
the bulk theory) are a triple $\vec v$ in the center of $\frak h$,
a triple $\vec w\in \f^-$ whose components commute with each other
and with $\frak h$, and the parameters of the theory $B$.

\begin{figure}
  \begin{center}
    \includegraphics[width=4in]{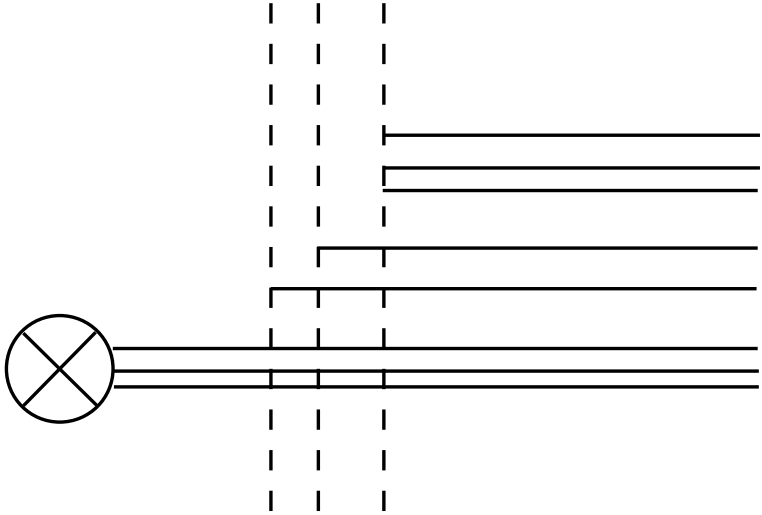}
  \end{center}
  \caption{\small A brane configuration whose purpose is to illustrate the general
half-BPS boundary condition.  A collection of semi-infinite
D3-branes with worldvolume in the 0123 directions (portrayed by
horizontal solid lines) ends on a collection of D5-branes that run
in the 012456 directions (portrayed by vertical dotted lines) and
one or more coincident NS5-branes that run in the 012789
directions (portrayed by the symbol $\bigotimes$).  In this and
subsequent pictures, the horizontal direction parametrizes $x^3$
and the vertical direction represents the 456 directions in
spacetime.}
  \label{fig0}
\end{figure}

\subsubsection{ A Brane Construction}

Since this general construction may seem rather elaborate, we
illustrate it with a brane configuration (fig. \ref{fig0}).
However, the reader may find the description of this brane
configuration clearer after reading section \ref{longahm}.

In the figure, we consider a $U(n)$ gauge theory associated to $n$
parallel  D3-branes, whose worldvolumes extend in directions 0123.
These D3-branes extend to infinity in $y=x^3$ in one direction.
They terminate in the other direction on D5-branes that extend in
the 012456 directions and NS5-branes that extend in the 012789
directions.
 Reading the figure from right to left, first several D3-branes end
on the same D5-brane. This gives a pole in Nahm's equations with a
non-trivial embedding $\rho:\frak{su}(2)\to \frak u(n)$. Then,
several D3-branes end one each on its own D5-branes. This gives a
subalgebra of $\frak u(n)$ in which $\vec X$ (which represents
motion in the 456 directions) obeys Neumann boundary conditions
and $\vec Y$ (which represents motion in the 789 directions) obeys
Dirichlet boundary conditions.  Finally, several D3-branes end on
an NS5-brane, giving a subalgebra of $\frak u(n)$ in which $\vec
X$ obeys Dirichlet boundary conditions and $\vec Y$ obeys Neumann
boundary conditions.

The figure is drawn for $n=7$, so the gauge group is $G=U(7)$. The
embedding $\rho:\frak{su}(2)\to \frak u(7)$ is of rank 3 and
reduces the gauge symmetry to $F=U(4)\times U(1)$, and as the
number of D3-branes ending on the NS5-brane is 2, the group $H$
that remains as a gauge group at the boundary is $H=U(2)$.  If the
number of NS5-branes is greater than 1, the $H$ gauge theory is
coupled to a non-trivial boundary conformal field theory.

The parameters $\vec v$ by which the boundary conditions on $\vec
X$ can be shifted arise from displacing the NS5-brane (or branes)
in the 456 directions.  The parameters $\vec w$ by which the
boundary conditions on $\vec Y$ can be shifted arise from
displacing the D5-branes in the 789 directions.

In the figure, to make the physics easier to describe, the various
fivebranes have been displaced from each other in the $y$
direction.  To reduce to the case of gauge theory on a half-space
with a boundary condition, one must take the limit in which all
fivebranes become coincident in $y$.

This example thus illustrates all of the ideas that are used in
constructing boundary conditions.

\subsection{Domain Walls}\label{domainwalls}

A close cousin of the problem of supersymmetric boundary conditions
is the problem of supersymmetric domain walls.   The theory of
half-BPS domain walls in $\N=4$ super Yang-Mills theory is known to
be quite rich; many examples have been constructed in the string
theory literature.

In fact, we do not really need anything new to describe such domain
walls in field theory, since the problem of domain walls can be
reduced to the problem that we have already considered of boundary
conditions.  Suppose that we want $\N=4$ super Yang-Mills theory
with one gauge group $G_1$ in the half-space $x^3<0$, and another
gauge group $G_2$ in the half-space $x^3>0$. What sort of half-BPS
domain walls can interpolate between these two theories?

We can reduce this question to one that we have already studied by
a simple ``folding'' trick.  Instead of saying that there is one
gauge theory to the left of the domain wall and one to the right,
we can flip the ``left'' theory over to the right and say that the
theory is trivial for $x^3<0$, and has gauge group $G_1\times G_2$
for $x^3>0$.

In folding or unfolding, we also must reverse the sign of three of
the scalar fields in the gauge theory factor that is flipped
between $x^3>0$ and $x^3<0$; merely changing the sign of $x^3$ is
not a symmetry of the theory.  To preserve D5-type supersymmetry,
we should reverse the sign of $\vec X$.

So the problem of finding a domain wall that interpolates between
$G_1 $ and $G_2$ is equivalent to describing boundary conditions
in the theory with gauge group $G_1\times G_2$.  For this, we can
use any of the constructions that we have seen above, all of which
are applicable to a general compact gauge group, not necessarily
simple.

\subsubsection{First Example}\label{firstex}

Let us give a few illustrative examples, in which we assume that
$\varepsilon_0$ is an eigenvector of $B_1$ or $B_2$.  Take
$G_1=G_2=G$, so that the gauge group away from the boundary is
$G\times G$. Let $H$ be a copy of $G$ diagonally embedded in
$G\times G$.  As in section \ref{generalization}, we can find in
$G\times G$ gauge theory a half-BPS boundary condition that breaks
the $G\times G$ gauge symmetry in bulk down to $H$ on the
boundary.  In fact, we do not really need the arguments of section
\ref{generalization} for this particular example; since we have
taken $G_1=G_2=G$, the unfolded theory simply has gauge group $G$
everywhere and is ordinary $\N=4$ super Yang-Mills with that gauge
group.  (One can verify that the arguments of section
\ref{generalization} give the same result as ``folding''
${\mathcal N}=4$ super Yang-Mills to a theory with gauge group
$G\times G$ on a half-space.) According to section \ref{angen}, we
can furthermore modify the folded theory by coupling to boundary
hypermultiplets that parametrize any hyper-Kahler manifold $Z$
with $H$ action. In the unfolded theory, what we have done is to
couple $\N=4$ super Yang-Mills theory with gauge group $G$ to
hypermultiplets that are supported on the hyperplane $x^3=0$. An
example coming from the D3-D5 system has been treated in detail in
\cite{DeWolfe:2001pq}.

\subsubsection{Generalization}\label{secondex}

To generalize this, take any group $G$ and subgroup $G'$, with an
embedding $i:G'\to G$.  Let $H$ be a copy of $G'$, regarded as a
subgroup of $G\times G'$ via the diagonal embedding $i\times
1:H\to G\times G'$. Consider $G\times G'$ gauge theory on a
half-space, with the half-BPS boundary conditions constructed in
section \ref{generalization} that break $G\times G'$ down to $H$
on the boundary. In the unfolded theory, this corresponds to a
supersymmetric domain wall with gauge group $G'$ on one side and
$G$ on the other. Various examples have been constructed in string
theory via branes and fluxes. The model can be modified to include
hypermultiplets with an arbitrary action of $H$ supported on the
domain wall.

This example can also be generalized to allow $\vec X$ to have a
pole at $y=0$, along the lines of eqn. (\ref{singsol}). (The pole
is in $\vec X$ rather than $\vec Y$ because of our choice of the
unbroken supersymmetry.)

In this example, it is not necessary to assume that $G'$ is a
subgroup of $G$.  We can take an arbitrary pair of gauge groups
$G$ and $G'$, and a third group $H$ with two embeddings $i:H\to G$
and $i':H\to G'$.  We regard $H$ as a subgroup of $G\times G'$ via
the diagonal embedding $i\times i':H\to G\times G'$, and consider
a half-BPS boundary condition with $G\times G'$ gauge symmetry in
a half-space reduced to $H$ on the boundary, possibly coupled to
boundary hypermultiplets with $H$ action. In the unfolded theory,
this sort of construction gives half-BPS domain walls
interpolating between gauge group $G$ on one side and $G'$ on the
other.  The subgroup of $G\times G'$ that commutes with $H$ acts
as global symmetries at the boundary.

\section{Moduli Spaces Of Solutions Of Nahm's Equations}\label{longahm}

\def\H{\mathcal H}
\def\M{\mathcal M}
As we explained in section \ref{nahm}, in $\N=4$ super Yang-Mills
theory on a half-space with suitable D5-like boundary conditions,
supersymmetric vacua arise from  solutions of Nahm's equations
\begin{equation}\label{rnahm}\frac{\mathrm{d}X_i}{\mathrm{d}y}+[X_{i+1},X_{i-1}]=0,~~i=1,2,3
\end{equation} on the half-line $L:y\geq 0$.
  $\vec X$ must also commute with the constant value of $\vec Y$ at
$y=\infty$; until section \ref{incy}, we assume that this constant
value vanishes.

What we really want to define is the moduli space of vacua of the
half-space theory for a given choice of the vacuum at infinity.
The vacuum at infinity is specified by a choice (up to conjugation
by a constant gauge transformation) of  the value of $\vec X$ at
$y=\infty$.  We write $\vec
X_\infty=(X_{1,\infty},X_{2,\infty},X_{3,\infty})$ for this
limiting value; the components of $\vec X_\infty$ must commute. It
is convenient to first consider the case that $\vec X_\infty$ is
regular, in the sense that the subgroup of $G$ that commutes with
all components of $\vec X_\infty$ is precisely a maximal torus
$T$. We let $\M$  denote the moduli space of solutions of Nahm's
equations with some appropriate condition at $y=0$, and with $\vec
X(y)\to\vec X_\infty$ (up to conjugation) for $y\to\infty$.

For the relevant boundary conditions, $\vec X$ is part of a
hypermultiplet, and therefore it is natural to think of $\M$ as a
Higgs branch of vacua. On general grounds, $\M$ is a hyper-Kahler
manifold. In fact, the relevant spaces of solutions of Nahm's
equations were used by Kronheimer \cite{Kronheimer,Kronheimer2} to
define hyper-Kahler metrics on certain spaces that arise in
representation theory.  For reviews and some later refinements,
see \cite{Biel,Abiel}.  We will try to give a fairly
self-contained explanation of the facts we need about Nahm's
equations, but essentially everything we explain is contained in
the above-cited references, or in the literature on Nahm's
equations applied to BPS monopoles in three dimensions (where
those equations originally arose \cite{Nahm:1979yw}). For a recent
survey of the extensive literature on Nahm's equations and
monopoles, see \cite{WeinbergYi}.  For previous results from a
D-brane perspective, see
\cite{Diaconescu:1996rk,Kapustin:1998pb,Tsimpis:1998zh}.

\subsection{The Hyper-Kahler Quotient}\label{hypquo}

\def\H{{\mathcal H}}
The proof that $\M$ is hyper-Kahler (see \cite{Kronheimer2}, section
3) uses the fact that it can be interpreted as a hyper-Kahler
quotient. We follow the logic of section \ref{nahm}. We complete
$\vec X$ to a hypermultiplet by adding $A=A_3$, the component of the
gauge field in the $y$ direction.  We pick a maximal torus $T$ with
Lie algebra $\frak t$, and we pick a regular triple $\vec
X_\infty\in \frak t$.  We require that $\vec X\to \vec X_\infty$ for
$y\to\infty$. (For the moment, we place no restriction on $\vec
X(0)$ except that it should be non-singular.) And we require that
$A$ is $\frak t$-valued at infinity.
 $\vec X$
and $A$ together parametrize a flat hyper-Kahler manifold $\mathcal
W$. The three symplectic forms of $\mathcal W$ are
\begin{equation}\label{ogolfox}\omega_i=\int_L \mathrm{d}y\,\,\Tr\,\left(\delta
A\wedge \delta X_i+\delta X_{i+1}\wedge \delta
X_{i-1}\right),~i=1,2,3.\end{equation}  We let $\hat G$ be the group
of gauge transformations $g:L\to G$ such that $g(0)=1$, and $g$ is
$T$-valued for $y\to\infty$.  $\hat G$ acts on $\mathcal W$ with a
hyper-Kahler moment map
\begin{equation}\label{hypmomt}\mu_i= \frac{DX_i}{Dy}+[ X_{i+1},
X_{i-1}],~~i=1,2,3,\end{equation} as in eqn. (\ref{hypmap}).
(Because $g(0)=1$, there is no delta function in the moment map at
$y=0$.) On general grounds, the hyper-Kahler quotient of $\mathcal
W$ by $\hat G$ is a hyper-Kahler manifold $\M$. The hyper-Kahler
quotient is obtained by setting to zero the moment map and dividing
by $\hat G$.

A convenient way to describe $\M$ is to eliminate $A$.  There is
always a unique map $g:L\to G$, with $ g(0)=1$, such that a gauge
transformation by $g$ sets $A$ to zero. After setting $A=0$, the
condition $\vec \mu=0$ becomes Nahm's equations. However, $g$ is
not necessarily an element of $\hat G$, since it may not be
$T$-valued for $y\to\infty$.  So after eliminating $A$, we can no
longer claim that $\vec X(y)\to \vec X_\infty$ for $y\to \infty$.
Rather, $\vec X(y)$ approaches a limit for $y\to\infty$ and this
limit is conjugate to $\vec X_\infty$ by a constant gauge
transformation.

The hyper-Kahler manifold obtained this way depends on $\vec
X_\infty$, of course, so we sometimes denote it as $\M(\vec
X_\infty)$.  $\M(\vec X_\infty)$ is smooth as long as $\vec
X_\infty$ is regular (which is needed for the above construction
to make sense as stated) because the condition that $g(0)=1$
ensures that the gauge group acts freely on $\mathcal W$.
Smoothness of $\M(\vec X_\infty)$ for regular $\vec X_\infty$ will
also be clear in section \ref{compman} when we describe $\M$ as a
complex manifold.  $\M(\vec X_\infty)$ can be continued to
non-regular values -- for instance, $\vec X_\infty=0$ -- but as
will also be clear in section \ref{compman}, it then develops
singularities.

The original finite-dimensional group $G$ acts on ${\M}(\vec
X_\infty)$, by gauge transformations at $y=0$.  To compute the
moment map for the $G$ action, we just repeat the computation of
eqn. (\ref{hypmap}), and then define $\vec \mu=\int
\mathrm{d}y\,\vec\mu(y)$. Now we pick up a delta function
contribution at $y=0$, since we do not require $g(0)=1$; indeed,
since we are imposing Nahm's equations, the delta function is all we
get.  So the hyper-Kahler moment map for the $G$ action is
\begin{equation}\label{hypgac}\vec \mu=\vec X(0).\end{equation}

\subsubsection{Including A Pole}\label{incpole}

As in section \ref{otherone}, we can construct a more general
boundary condition by choosing
 a homomorphism $\rho:\frak{su}(2)\to
\frak g$ and requiring that for $y\to 0$
\begin{equation}\label{dolfin}X_i(y)=\frac{
t_i}{y}+\dots.\end{equation} Here  $t_i$ are the images under $\rho$
of a standard basis of $\frak{su}(2)$; the ellipses refer to terms
regular at $y=0$.

We denote the subgroup of $G$ that commutes with $\rho$ as $H$, and
we call its Lie algebra $\frak h$.  We modify the above construction
by requiring that $A$ is $\frak h$-valued at $y=0$, so that $A(0)$
commutes with the polar part of $\vec X$.

The hyper-Kahler quotient now gives a hyper-Kahler manifold
$\M_\rho(\vec X)$ that (after gauging away $A$) parametrizes
solutions of Nahm's equations with the behavior of eqn.
(\ref{dolfin}) for $y\to 0$, and with $\vec X\to \vec X_\infty$ up
to conjugation for $y\to\infty$.

$\M_\rho(\vec X)$ admits an action of $H$.  The hyper-Kahler
moment map is
\begin{equation}\label{golfox}\vec \mu=\vec X_{\frak
h}(0).\end{equation} Here $\vec X_{\frak h}$ is the orthogonal
projection of $\vec X$ from $\frak g$ to $\frak h$.  Of course,
$\vec X_{\frak h}$ is regular at $y=0$, even though $\vec X$ has a
pole.

\subsection{The Complex Manifold}\label{compman}

For our purposes, the most useful way to understand the hyper-Kahler
manifold $\mathcal H$ is to describe it as a complex manifold in one
of its complex structures.  (We continue to follow
\cite{Kronheimer2}, section 3.) We first consider the case that
$\vec X$ has no pole at $y=0$.

\def\CX{{\mathcal X}}
\def\CA{{\mathcal A}}
\def\CD{{\mathcal D}}
\def\C{{\Bbb{C}}}
We let $\CX=X_1+iX_2$, $\CA=A+iX_3$.  In one complex structure on
the infinite-dimensional hyper-Kahler manifold $\mathcal W$, the
fields $\CX$ and $\CA$ are complex coordinates.  In this complex
structure, two of Nahm's equations combine to a single holomorphic
equation
\begin{equation}\label{compnahm}\frac{\CD \CX}{\CD
y}=0.\end{equation} Here $\CD\CX/\CD
y=\mathrm{d}\CX/\mathrm{d}y+[\CA,\CX]$ is the covariant derivative
of $\CX$ with respect to the complex-valued connection $\CA$.  In
solving eqn. (\ref{compnahm}), we require that $\CX(y)\to
\CX_\infty$ for $y\to\infty$, where
$\CX_\infty=X_{1,\infty}+iX_{2,\infty}$.

Eqn. (\ref{compnahm}) is invariant under complex-valued gauge
transformations, acting in the usual way $\CX\to g\CX g^{-1}$,
$\CD\to g\CD g^{-1}$, where now $g(y):L\to G_\C$ takes values in
the complexification $G_\C$ of $G$.  We also require that
$g(0)=1$, and that $g$ for large $y$ is valued in $T_\C$, the
complexification of $T$; these conditions mean that $g(y)$ is an
element of $\hat G_\C$, the complexification of the group $\hat G$
that was used in the construction of $\M$ as a hyper-Kahler
quotient.

By a standard type of argument,\footnote{Stability is not an issue
if $\CX_\infty$ is regular semi-simple, which for regular $\vec X$
is true for a generic choice of the coordinate axes in $\vec X$
space.} imposing the third Nahm equation and dividing by $\hat G$
is equivalent to simply dividing by $\hat G_\C$.  But dividing by
$\hat G_\C$ is a very simple operation.  If we relax the
requirement that $g(y)$ is $T_\C$-valued at infinity, then there
is a unique $G_\C$-valued gauge transformation, with $g(0)=1$,
that sets $\CA=0$.  Since $g(\infty)$ may not commute with
$\CX_\infty$, after we make this gauge transformation $\CX(y)$  is
conjugate for $y\to\infty$ to $\CX_\infty$ but need not equal
$\CX_\infty$.

In the gauge $\CA=0$, the complex Nahm equation (\ref{compnahm})
reduces to $\mathrm{d}\CX/\mathrm{d}y=0$, telling us that $\CX$ is a
constant. The boundary condition at $y=0$ (which just says that
$\CX$ is finite there) puts no restriction on the constant, and the
boundary condition at infinity simply tells us that $\CX$ is
conjugate to $\CX_\infty$.

So as a complex manifold, $\M$ is isomorphic to the conjugacy
class of $\CX_\infty$ in the complex Lie algebra $\frak g_\C$.  In
particular, this implies that if $\CX_\infty$ is regular
semi-simple (diagonalizable with distinct eigenvalues) then $\M$
is smooth.  If $\vec X_\infty$ is regular, then $\CX_\infty$ is
regular semi-simple for a generic choice of coordinate axes (that
is, a generic choice of which components of $\vec X$ we identify
as  $X_1+iX_2$).

\subsubsection{Conjugacy Classes In Complex Lie
Algebras}\label{conjclass}

\def\CO{{\mathcal O}}
Because of this result and related results that will soon appear, we
need a few simple results on conjugacy classes in complex Lie
groups.

Let $G$ be a compact Lie group of dimension $d$, and let $G_\C$ be
its complexification.  Then the complex dimension of $G_\C$ is also
$d$.  Let $x$ be an element of $\frak g_\C$, and $S$ the subgroup of
$G_\C$ that commutes with $x$.  Let $s$ be the complex dimension of
$S$.  The orbit $\CO_x$ of $x$ in $\frak g_\C$ is a complex manifold
of complex dimension $d-s$.

The smallest possible value of $s$ is $r$, the rank of $G$.  For
example, suppose that  $x$  can be conjugated to the Lie algebra
$\frak t_\C$ of a (complex) maximal torus $T_\C$.  Then $x$ at least
commutes with $T_\C$, of dimension $r$, so $s\geq r$.  If $x$ is a
generic element of $T_\C$, then $S=T_\C$ and $s=r$. $x$ is said to
be semisimple if it can be conjugated to a maximal torus, and
regular if $s=r$.

The starting point in our analysis was an assumption that $\vec
X_\infty$ is regular, meaning that the value of $\vec X$ at
infinity breaks the gauge group $G$ to its maximal torus $T$.
($\vec X_\infty$ is automatically semisimple; indeed,
supersymmetry requires that the components of $\vec X_\infty$
commute and so can be simultaneously conjugated to a maximal
torus.)  Then to avoid some technicalities we oriented the
coordinate axes in a generic fashion, so that
$\CX_\infty=X_{1,\infty}+iX_{2,\infty}$ is also regular
semi-simple.  This means that the gauge symmetry breaking is fully
reflected in $\CX_\infty$.

\subsubsection{Turning Off The Symmetry Breaking}\label{turnoff}

It is also of interest to ask what happens when we turn off the
gauge symmetry breaking at infinity.  An important subtlety will
arise, so to get our bearings we start with the example of
$G=SU(2)$.  If $\CX_\infty$ is regular semi-simple, then it is
conjugate to ${\rm diag}(w,-w)$ for some $w\in\C$.  As  a result,
the quadratic Casimir invariant $u=\Tr\,\CX^2$ is nonzero; in fact,
$u=2w^2$.  $u$ is a natural gauge-invariant measure of the symmetry
breaking.

What happens if we take $u\to 0$?  One might think that that means
that $\CX$ goes to zero and its orbit collapses to a point. That
is actually not the case.  The following nonzero element of
$\frak{sl}(2,\C)$ has $u=0$:
\begin{equation}\label{helfo}x=\begin{pmatrix} 0 & 1\\ 0 & 0
\end{pmatrix}.\end{equation}
$x$ is regular, since the subgroup $S$ of $SL(2,\C)$ that commutes
with $x$ is one-dimensional, being generated by $x$ itself.  In
general, for every value of the Casimir invariants of a complex
Lie group, there is a unique regular orbit.  For $SL(2,\C)$, $u$
is the only independent Casimir orbit;  the orbit $x$ is the
regular orbit with $u=0$. For every $u$, a regular element $w_u$
of $\frak{sl}(2,\C)$ with $\Tr\,w_u^2=u$ can be written as
follows:
\begin{equation}\label{elfo}w_u=\begin{pmatrix} 0 & 1\\ u/2& 0
\end{pmatrix}.\end{equation}
This family contains every regular conjugacy class precisely once.

The orbit $\CO_x$ of $x=w_0$ can easily be described explicitly.
Any element
\begin{equation}\label{anyem}\begin{pmatrix}a & b \\  c &
d\end{pmatrix}\end{equation} of $\frak{sl}(2,\C)$ is conjugate to
$x$ if and only if
\begin{equation}\label{horn}ad-bc=0\end{equation}
and $a,b,c,d$ are not all zero.

\def\N{{\mathcal N}}
Obviously, the orbit $\CO_x$ is not closed in $\frak{sl}(2,\C)$.  To
take its closure, we must relax the condition that $a,b,c$, and $d$
are not all zero.  If we do relax this condition, we get a subspace
of $\frak{sl}(2,\C)$ that is known as the nilpotent cone $\N$. It
parametrizes all nilpotent elements of the Lie algebra, conjugate to
$x$ or not. For our example of $SL(2,\C)$, $\N$ is the union of two
orbits; one orbit is $\CO_x$, and the second orbit is a single
point, the orbit $\CO_0$ of the zero element of $\frak{sl}(2,\C)$
with $a=b=c=d=0$. In fact, $\CO_0$ is a singularity of $\N$.  The
equation $ad-bc=0$ that defines $\mathcal N$ is a standard
description of the $A_1$ singularity. Topologically, for $SL(2,\C)$,
$\mathcal N$ is $\C^2/\Z_2$ or equivalently $\R^4/\Z_2$.

We can describe explicitly the family of solutions of the original
real Nahm equations (\ref{rnahm})  that is parametrized by
$\mathcal N$:
\begin{equation}\label{expfam}
X_i(y)=g\frac{t_i}{y+f^{-1}}g^{-1}.\end{equation} Here $t_i$ are
the standard $2\times 2$ Pauli matrices, with $[t_1,t_2]=t_3$,
etc., $f$ is a non-negative real constant, and $g\in SU(2)$.  For
$f=0$, $\vec X(y)$ identically vanishes; this is the trivial zero
solution of Nahm's equations, which corresponds to the singular
point in $\N$. For all $f\geq 0$, $\vec X(y)$ is regular on the
whole half-line, including $y=0$, and vanishes for $y\to\infty$.

It is not difficult to describe the topology of the manifold $\M$
that parametrizes this family of solutions of Nahm's equations.
$f$ takes values in the half-line $\R_{\geq 0}$, and (since $g$
and $-g$ are equivalent in (\ref{expfam})) $g$ takes values in
$SU(2)/\Z_2=S^3/\Z_2$.  So $\M=S^3/\Z_2\times \R_{\geq 0}$.  But
this is the same as $\R^4/\Z_2$, which  is the same as $\C^2/\Z_2$
and coincides with the nilpotent cone $\mathcal N$ for $SL(2,\C)$.
In fact, one can readily verify that in this family of solutions,
$\CX(0)=X_1(0)+iX_2(0)$ is always nilpotent, and that every
nilpotent element of $\frak{sl}(2,\C)$ equals $\CX(0)$ for
precisely one choice of $g$ (up to sign) and $f$.

Going back to the original problem, for $G=SU(2)$, if $\vec
X_\infty$ is regular, then the moduli space $\M$ of solutions of
Nahm's equations is a smooth manifold that, in a generic complex
structure, is the orbit of a regular semisimple element of
$\frak{sl}(2,\C)$.  But if we turn off the symmetry breaking and
set $\vec X_\infty=0$, then $\M$ becomes the nilpotent cone $\N$.

Starting from $\vec X_\infty=0$, if we turn on $X_{1,\infty}$ and
$X_{2,\infty}$, then $\N$ is deformed and becomes the smooth orbit
of a regular semi-simple element
$\CX_\infty=X_{1,\infty}+iX_{2,\infty}$.  But if we keep
$X_{1,\infty}=X_{2,\infty}=0$ and turn on $X_{3,\infty}$, then the
singularity of $\mathcal N$ is resolved, rather than deformed.

\subsubsection{Analog For Any $G$}\label{omigosh}

The analog for any $G$ is as follows.  The complex Nahm equation
\begin{equation}\label{turto}\frac{\CD \CX}{\CD y}=0\end{equation}
implies that the Casimir invariants of $\CX$ are independent of
$y$. The Casimir invariants of $\CX(0)$ therefore coincide  with
those of $\CX_\infty$. $\CX(0)$ is gauge-invariant (since we only
divide by gauge transformations that equal 1 at $y=0$).  Up to a
complex gauge transformation, the complex Nahm equation has a
unique solution for every choice of $\CX(0)$ that has the same
Casimir invariants as $\CX_\infty$.

If $\CX_\infty$ is regular, then any element of $\frak{g}_\C$ with
the same Casimir invariants is conjugate to $\CX_\infty$.  Hence
the moduli space $\mathcal H$ of solutions of Nahm's equations is
simply the orbit of $\CX_\infty$ in $\frak{g}_\C$.  We denote this
orbit as $\O_{\CX_\infty}$, and we denote as $\bar\O_{\CX_\infty}$
the space of all elements of $\frak g_\C$ with the same Casimir
invariants as $\CX_\infty$.  These two spaces coincide precisely
if $\CX_\infty$ is regular.

Even if $\CX_\infty$ is not regular, it is always possible to find a
regular element $x\in \frak{g}_\C$ with the same Casimir invariants
as $\CX_\infty$.  For instance, generalizing the example for
$SL(2,\C)$, for $G_\C=SL(n,\C)$, if $\CX_\infty=0$, we can take $x$
to be an $n\times n$ matrix with 1's just above the main diagonal
and all other matrix elements zero:
\begin{equation}\label{nork}x=\begin{pmatrix}0& 1 & 0 & \dots & 0\\
                                           0 & 0 & 1 & \dots & 0\\
                                            & & \ddots && \\
                                            0 & 0 & 0 & \dots & 1 \\                                           0 & 0 & 0 & \dots & 0
                                            \end{pmatrix}.\end{equation}
(The subgroup of $SL(n,\C)$ that commutes with $x$ is generated by
$x,x^2,\dots,x^{n-1}$, and so has the same dimension as a maximal
torus.)  The moduli space ${\mathcal H}=\bar\O_{\CX_\infty}$ of
solutions of Nahm's equations is always the {\it closure} of the
orbit $\CO_x$ of $x$ in $\frak{g}_\C$.  The closure is obtained by
adding to $\CO_x$ the orbits of non-regular elements $x'$ that
have the same Casimir invariants as $x$.  In our example with
$SL(2,\C)$, $x$ was a regular nilpotent element and the only
relevant non-regular $x'$ was $x'=0$; in general, finitely many
non-regular orbits appear. The dimension of a regular orbit is
greater than that of any non-regular orbit (since a regular
element, by definition, has a centralizer of the minimum
dimension) and the non-regular orbits $\O_{x'}$ appear as
singularities in ${\mathcal H}$, just as in our example.

Physically, an important special case is the case that symmetry
breaking is absent at $y=\infty$.  This means that $\vec
X_\infty=0=\CX_\infty$, and therefore the Casimir invariants of
$\CX_\infty$ all vanish.  An element $\CX(0)$ of $\frak g_\C$ has
vanishing Casimir invariants if and only if it is nilpotent.
Therefore, in this situation, the moduli space $\mathcal H$ of
solutions of Nahm's equations coincides with the nilpotent cone
$\N$ consisting of all nilpotent elements of $\frak g_\C$.  These
make up finitely many conjugacy classes.

The orbit $\O_x$ of a regular nilpotent element $x$ is a dense
open set in $\N$. $\N$ actually equals $\bar\O_x$, the closure of
$\O_x$; $\N$ has singularities corresponding to non-regular
nilpotent orbits. Symmetry breaking at infinity (by the choice of
$\vec X_\infty$) causes these singularities to be deformed and
resolved; if one chooses $\vec X_\infty$ to break $G$ to its
maximal torus, then the moduli space of vacua becomes smooth.

\subsubsection{More On Nilpotent Orbits}\label{nilporb}

As we have just seen, nilpotent orbits in complex Lie algebras are
important in our subject.  So we pause for a few words on these
orbits.

A nilpotent element of $\frak{sl}(n,\C)$ is simply an $n\times n$
nilpotent matrix. Any $n\times n$ complex matrix can be conjugated
to a Jordan canonical form.  The Jordan canonical form of a
nilpotent matrix $x$ has all matrix elements vanishing except for
1's in some of the entries just above the main diagonal. For some
decomposition $ n=n_1+n_2+\dots +n_k$, with positive integers
$n_i$, where we can assume $n_1\geq n_2\geq\dots\geq n_k$, $x$
takes a block-diagonal form in which the diagonal blocks are
regular nilpotent $n_p\times n_p$ matrices, $1\leq p\leq k$, each
taking precisely the form in eqn. (\ref{nork}).  The off-diagonal
blocks vanish.

An alternative description is useful for generalizing to any group.
Let $\rho:\frak{su}(2)\to\frak g_\C$ be any homomorphism, and as
usual write $t_1,t_2,t_3$ for the images of standard generators of
$\frak{su}(2)$.  The ``raising'' operator $t_+=t_1+it_2$ is then
nilpotent.  Conversely, according to the Jacobson-Morozov Theorem,
every nilpotent element of a complex semi-simple Lie algebra arises
in this way from some $\frak{su}(2)$ embedding.

Let us verify this assertion in the case of $SL(n,\C)$. The Lie
algebra $\frak{su}(2)$ has, up to isomorphism, one irreducible
representation of each positive integer dimension $1,2,3,\dots$.
So up to isomorphism, the embedding $\rho:\frak{su}(2)\to
\frak{sl}(n,\C)$ is determined by a decomposition
$n=n_1+n_2+\dots+n_k$, with positive integers $n_i$ that we can
assume to be non-increasing. Moreover, in an irreducible
$p$-dimensional representation of $\frak{su}(2)$, the raising
operator is a regular nilpotent element, conjugate to the $p\times
p$ case of the matrix described in eqn. (\ref{nork}).  So the two
descriptions agree.

One advantage of the description by $\frak{su}(2)$ embeddings is
that it gives a convenient way to determine the dimension of an
orbit.  Let $t_+$ be a nilpotent element of $\frak g_\C$ that is
the raising operator for some $\frak{su}(2)$ embedding $\rho$, and
let $\O_\rho$ be its orbit.  The dimension of $\O_\rho$ will be
$d-s$, where $d$ is the dimension of $G_\C$ and $s$ is the
dimension of the subgroup $S$ that commutes with $t_+$.  (All
dimensions here are complex dimensions.)  What is $s$?  Let us
decompose $\frak g_\C$ in irreducible representations ${\mathcal
T}_j$ of $\frak{su}(2)$:
\begin{equation}\label{zabel}\frak g_\C=\oplus_{j=1}^s {\mathcal
T}_j.\end{equation} Elements of $\frak g_\C$ that commute with the
raising operator $t_+$ are precisely the highest weight vectors in
the summands ${\mathcal T}_j$.  Each ${\mathcal T}_j$ has a
one-dimensional space of highest weight vectors.  Therefore the
number $s$ of summands in eqn. (\ref{zabel}) is the dimension of
the centralizer $S$ of $t_+$.  The dimension of the orbit of $t_+$
is therefore $d-s$.

Let us check this calculation for the case of a regular nilpotent
element $x$.  This is the case that $\rho:\frak{su}(2)\to
\frak{sl}(n,\C)$ is associated with an irreducible $n$-dimensional
representation of $\frak{su}(2)$.  For this representation, the Lie
algebra $\frak{sl}(n,\C)$ decomposes as a direct sum of
$\frak{su}(2)$ modules of dimensions $3,5,7,\dots,2n-1$.  There are
$n-1$ pieces in all, so $s=n-1$, as we computed before in another
way.

\bigskip\noindent{\it Some More Examples}

It will be helpful to give a few more examples of nilpotent orbits.

As we have already explained, for every simple Lie group $G$, there
is a unique regular nilpotent orbit.  For $G=SU(n)$, it corresponds
to an irreducible embedding $\rho:\frak{su}(2)\to \frak{su}(n)$. The
regular nilpotent orbit is a dense open set in the nilpotent cone
$\mathcal N$.

If $G$ is simply-laced, there is also a unique subregular nilpotent
orbit $\CO'$ -- one whose complex dimension is precisely 2 less than
the dimension of $\mathcal N$.  $\CO'$ therefore appears as a locus
of singularities in $\mathcal N$, and (in keeping with the
hyper-Kahler nature of $\mathcal N$) these are orbifold
singularities $\C^2/\Gamma$, where $\Gamma$ is a finite subgroup of
$SU(2)$.  In fact, $\Gamma$ is the finite subgroup of $SU(2)$ that
corresponds to $G$ in the usual mapping between such subgroups and
simple groups of type A-D-E.

For $G=SU(2)={\rm A}_1$, the subregular nilpotent element is
simply the zero element.  The fact that the nilpotent cone $\N$
has an ${\rm A}_1$ singularity corresponding to the zero element
is a special case of the general relation of subregular nilpotent
orbits to A-D-E singularities.

More generally, for $G=SU(n)$, the subregular nilpotent orbit is
the raising operator $t_+$ of an $SU(2)$ embedding that
corresponds to a decomposition $n=(n-1)+1$.  A computation as
above shows that the centralizer of such a $t_+$  has dimension
$n+1$, which exceeds by 2 the rank $n-1$ of $SU(n)$. This accounts
for the fact that the orbit $\CO'$ is of codimension 2 in the
nilpotent cone.

At the other extreme, the zero element of $\frak g_\C$ is the
unique nilpotent element whose orbit consists of a single point.
There is also a unique nilpotent orbit of smallest positive
dimension  -- usually called the minimal (non-zero) nilpotent
orbit.  For $G=SU(n)$, it corresponds to the decomposition
$n=2+1+1+\dots+1$. A computation as above shows that the
corresponding orbit must have dimension $2n-2.$ In fact, this
orbit consists of $n\times n$ matrices $M$ of rank $1$ with
$M^2=0$. Such a matrix can be written $M^i{}_j=B^iC_j$, where
$\sum_iB^iC_i=0$; this way of writing $M$ is unique modulo $B\to
\lambda B$, $C\to \lambda^{-1}C$.

\subsection{Solutions Of Nahm's Equations With
Poles}\label{solpole}

In section \ref{incpole}, we considered Nahm's equations with a
pole at $y=0$ determined by a homomorphism $\rho:\frak{su}(2)\to
\frak g$.  As we explained there, the solutions of Nahm's
equations, with boundary conditions that $\vec X$  is conjugate at
infinity to a commuting triple $\vec X_\infty$, are parametrized
by a hyper-Kahler manifold $\M_\rho(\vec X)$.

We proceed, following \cite{Kronheimer}, just as in the case of
trivial $\rho$. Setting $\CX=X_1+iX_2$, $\CA=A+iX_3$, two of Nahm's
equations combine to a form familiar from eqn. (\ref{compnahm}):
\begin{equation}\label{comper}\frac{\CD \CX}{\CD
y}=0.\end{equation} Now, however, $\CX$ and $\CA$ are not regular at
$y=0$. Rather, we have
\begin{align}\label{omper} \CX& = \frac{t_1+it_2}{y}+\dots=\frac{t_+}{y}+\dots\\
                           \CA& = \frac{it_3}{y}+\dots,\end{align}
where the ellipses denote regular terms.

As before, Nahm's three real equations modulo real gauge
transformations are equivalent to the complex equation
(\ref{comper}) modulo complex gauge transformations. Now, however,
we cannot use a complex gauge transformation to set $\CA=0$.  The
reason for this is that we are restricted to gauge transformations
that are trivial at $y=0$.  A gauge transformation that would remove
the singularity from $\CA$ would have to have  a singularity at
$y=0$.

We can, however, make a gauge transformation to set $\CA=it_3/y$
everywhere.  Just as in our previous analysis, the gauge
transformation that does this may not commute with $\CX_\infty$ for
$y\to\infty$.  So after setting $\CA=it_3/y$, we should require that
$\CX(y)$ is conjugate to $\CX_\infty$ for $y\to\infty$, not that the
two are equal.

After setting $\CA=it_3/y$, it is straightforward to solve the
complex Nahm equation.  We pick a basis $v_\alpha$ of $\frak g$ of
vectors of definite weight
\begin{equation}\label{basiso}[it_3,v_\alpha]=m_\alpha
v_\alpha,\end{equation} where $m_\alpha\in \Z/2$.  (For example,
$[it_3,t_\pm]=\pm t_\pm$, so we can take $t_\pm$ for two of the
$v_\alpha$.)  Then the complex Nahm equation has the general
solution
\begin{equation}\label{kime}\CX=\sum_\alpha \epsilon_\alpha
\frac{v_\alpha}{y^{m_\alpha}}\end{equation} with coefficients
$\epsilon_\alpha$.  However, we want solutions in which the
singular part at $y=0$ is precisely $t_+/y$.  So we must have
\begin{equation}\label{zime}\CX=\frac{t_+}{y}+\sum_{m_\alpha\leq
0}\epsilon_\alpha\, v_\alpha y^{-m_\alpha}.\end{equation}

This is not the whole story, because the gauge transformation that
sets $\CA=it_3/y$ is not unique.  This form is preserved by a
further gauge transformation generated by
\begin{equation}\label{hito}\phi=\sum_\alpha f_\alpha v_\alpha y^{-m_\alpha},
\end{equation} with
arbitrary coefficients $f_\alpha$. However, since we are supposed
to allow only gauge transformations that vanish at $y=0$, we must
actually restrict the coefficients so that $\phi=\sum_{m_\alpha<0}
f_\alpha v_\alpha y^{-m_\alpha}$.  By a gauge transformation that
shifts $\CX$ by $[\phi,\CX]$ with $\phi$ of this form, we can
remove everything from $\CX$ except the singular term $t_+/y$ and
the terms in which $v_\alpha$ is a lowest weight vector,
annihilated by $t_-$.  So we reduce to
\begin{equation}\label{ime}\CX=\frac{t_+}{y}+\sum_{\alpha\in P_-}
\epsilon_\alpha v_\alpha y^{-m_\alpha},\end{equation} where $P_-$
labels the lowest weight vectors.

The Slodowy slice ${\mathcal S}_{t_+}$ transverse to a nilpotent
orbit ${\mathcal O}_{t_+}$ is defined to be the subspace of $\frak
g$ consisting of elements of the form
\begin{equation}\label{okime} t_++\sum_{\alpha\in P_-}\epsilon_\alpha
v_\alpha,\end{equation} with arbitrary coefficients
$\epsilon_\alpha$. ${\mathcal S}_{t_+}$ meets ${\mathcal O}_{t_+}$
in a single point (the point with all $\epsilon_\alpha=0$) and has
nice or ``transverse'' intersections with all orbits that it
meets.

\def\S{{\mathcal S}}
Clearly, functions $\CX(y)$ of the form given in (\ref{ime}) are
in one-to-one correspondence with points in the Slodowy slice
${\mathcal S}_{t_+}$; the correspondence is made by setting $y=1$
in (\ref{ime}).
 However, the moduli space $\M$ of vacua is not simply the Slodowy
slice.  We must impose the condition that the characteristic
polynomial of $\CX$ coincides with that of $\CX_\infty$.  The
characteristic polynomial of $\CX$ is independent of $y$ because
of the complex Nahm equation, so we can just evaluate this
condition at $y=1$. We learn that $\CX(1)$ takes values in the
intersection of ${\mathcal S}_{t_+}$ with
$\tilde\CO_{\CX_\infty}$, the subspace of $\frak g$ consisting of
elements with the same characteristic polynomial as $\CX_\infty$.
If $\CX_\infty$ is regular, then $\tilde\CO_{\CX_\infty}$ is the
same as $\CO_{\CX_\infty}$, the orbit of $\CX_\infty$.  In
general, it is the closure $\bar\O_x$ of the orbit $\O_x$ of a
regular element $x$ with the same characteristic polynomial as
$\CX_\infty$.

What we learn, then, is that as a complex manifold, the moduli
space $\M_\rho(\vec X)$ of solutions of Nahm's equations with a
pole determined by $\rho$ is the intersection $\S_{t_+}\cap \bar
\CO_x$.  In particular, from this we can determine the dimension
of this space.  If as before $s$ denotes the number of summands
when $\frak g$ is decomposed in representations of $\frak{su}(2)$,
then the dimension of $\S_{t_+}$ is precisely $s$, since each
irreducible representation of $\frak{su}(2)$ has a one-dimensional
space of lowest weight vectors.  Requiring that $\CX(1)$ should
have the same characteristic polynomial as $\CX_\infty$ reduces
the complex dimension by $r$. So the dimension of $\M_\rho(\vec
X)$ is $s-r$.

\subsubsection{Some Examples}\label{examples}

%$\overline \O_{\CX_\infty}$

At one extreme, if $t_+=0$, the corresponding transversal slice
${\mathcal S}_{t_+}$ is all of $\frak g_\C$.  So $\M_{\rho=0}(\vec
X_\infty)$ is simply (if $\CX_\infty$ is regular) the orbit
$\CO(\CX_\infty)$, as before.

At the other extreme, if $t_+$ is a regular nilpotent element, the
Slodowy slice ${\mathcal S}_{t_+}$ has dimension $s=r$, the rank of
$G$. Its intersection with a regular orbit (or the closure of one)
is therefore of dimension zero, and should consist of a finite set
of points. But since the Slodowy slice ${\mathcal S}_{t_+}$ meets
the regular orbit $\O_{t_+}$ in precisely one point (the element
$t_+\in \frak g$), it likewise meets every regular orbit in just one
point.

One can verify this by hand for $SL(n,\C)$. A transversal to the
orbit of the regular nilpotent element $t_+$ given in (\ref{nork})
that is not actually the Slodowy slice, but arises from it by a
different gauge fixing of the gauge invariance (\ref{hito}),
consists of elements of the form
\begin{equation}\label{norko}x=\begin{pmatrix}0& 1 & 0 & \dots & 0\\
                                           0 & 0 & 1 & \dots & 0\\
                                            & & \ddots && \\
                                            0 & 0 & 0 & \dots & 1\\
                                            a_{n} & a_{n-1} & a_{n-2} & \dots & 0
                                            \end{pmatrix},\end{equation}
with coefficients $a_n,a_{n-1},\dots,a_2,0$ in the bottom row. (We
set the lower right entry to zero to ensure that this matrix is in
$\frak{sl}(n,\C)$; in $\frak{gl}(n,\C)$, this element would be
another coefficient $a_1$.)  Every set of values of the Casimir
operators $\Tr\,x^k$, $k=2,\dots,n$ arise precisely once in this
family.  So this transversal slice meets every regular orbit precise
once.

So we learn that if $\rho$ corresponds to a regular nilpotent
orbit, then the moduli space $\M_\rho(\vec X_\infty)$ consists of
only a single point. As we will see, this result is important in
understanding duality of supersymmetric boundary conditions.

For any other $\rho$, $s$ is larger so the relevant moduli space
has a positive dimension.  For example, if $G$ is simply-laced and
$t_+$ is a subregular nilpotent element, then $s-r=2$ and the
moduli space is of dimension 2.  For $\CX_\infty=0$, it equals
$\C^2/\Gamma$, where $\Gamma$ is the finite subgroup of $SU(2)$
related to $G$, and for other $\CX_\infty$, it is a deformation of
$\C^2/\Gamma$. This is explained in \cite{Kronheimer}.

\subsection{Nahm's Equations And Brane Constructions}\label{nahmdisc}

Now we will extend the analysis of Nahm's equations to allow for
discontinuities as well as poles.  Instead of proceeding in an
abstract way, as we have done so far, we consider a specific (and
well known) string theory situation.  This is useful in
understanding the action of duality, though in the present paper
we take only limited steps in that direction.

\begin{figure}
  \begin{center}
    \includegraphics[width=3.5in]{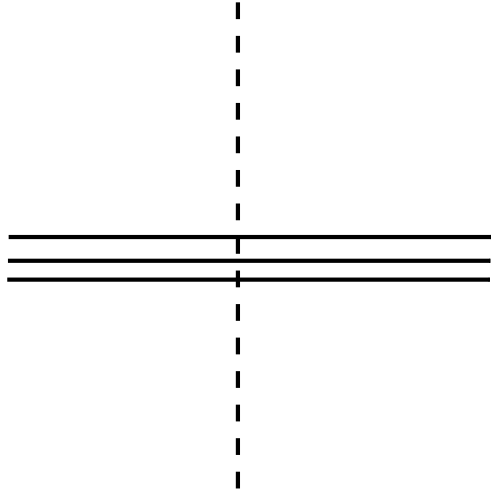}
  \end{center}
\caption{\small Here and later, horizontal solid lines denote
D3-branes whose world-volume is parametrized by $x^0,x^1,x^2,x^3$.
Vertical dotted lines denote D5-branes supported at
$x^3=x^7=x^8=x^9=0.$}
  \label{fig1}
\end{figure}

We consider (fig. \ref{fig1}) a system of $n$ parallel D3-branes,
transversely intersecting a D5-brane.  The D3-branes are
parametrized by $x^0,x^1,x^2,x^3$, and support a four-dimensional
$U(n)$ gauge theory with $\N=4$ supersymmetry; the values of
$x^4,\dots,x^9$ are observed in this theory as scalar fields $\vec
X$ and $\vec Y$.

The D5-brane is supported at $x^3=x^7=x^8=x^9=0$.  The D5-brane
supports a $U(1)$ gauge field.  From the standpoint of the D3-brane
system, which we will focus on, this $U(1)$ can be regarded as a
global symmetry of the D3-brane theory (modulo a caveat noted
below), and the fluctuations in the D5-brane position can be
ignored.

In the D3-brane theory, there is a hypermultiplet $Z$ in the
fundamental representation of $U(n)$, supported at the
intersection with the D5-brane. This intersection is at $x^3=0$;
as usual, we write $y$ for $x^3$.  $Z$ is a ``bifundamental''
hypermultiplet, meaning that it is also charged under the $U(1)$
symmetry coming from the D5-brane.  But the action of this $U(1)$
is the same as that of the center of $U(n)$. Since $U(n)$ is
gauged, this $U(1)$ is not really observed as a global symmetry of
the D3-brane theory. Global symmetries will arise from D5-brane
symmetries when there is more than one D5-brane, as in other cases
that we treat below.

For the same reasons as in the examples that we have already
treated, supersymmetric vacua of the combined system are given by
solutions of Nahm's equations.  However, we must include the
contribution of $Z$ in Nahm's equations.  Essentially the same
derivation\footnote{Because we are now on the full line
$-\infty<y<\infty$, rather than the half-line $y\geq 0$,
integration by parts does not produce a term $\delta(y)\vec X(0)$,
which appears in the previous derivation.} that led to eqn.
(\ref{nypmap}), with $\vec X$ and $\vec Y$ exchanged, shows that
the hyper-Kahler moment map for the combined system consisting of
hypermultiplets  $\vec X$, $A=A_3$, and $Z$ is
\begin{equation}\label{orm}\vec\mu(y)=\frac{D\vec X}{Dy}+\vec
X\times \vec X +\delta(y)\vec \mu^Z,\end{equation} where $\vec\mu^Z$
is the hyper-Kahler moment map for the hypermultiplet $Z$. The
extension of Nahm's equation is therefore
\begin{equation}\label{torm}\frac{D\vec X}{Dy} +\vec
X\times \vec X +\delta(y)\vec \mu^Z=0.\end{equation} The meaning of
the delta function is that $\vec X(y)$ is discontinuous at $y=0$.
The jump $\Delta\vec X$ in crossing $y=0$ obeys
\begin{equation}\label{zorm}\Delta\vec X+\vec\mu^Z=0.\end{equation}

We now want solutions of this extended Nahm equation in which
$\vec X$ approaches one limit $\vec X_{\infty,-}$ for $y\to
-\infty$, and another limit, $\vec X_{\infty,+}$ for $y\to
+\infty$.  Of course, the components of $\vec X_{\infty,-}$
commute with each other, as do the components of $\vec
X_{\infty,+}$.  We want to describe the space ${\mathcal M}(\vec
X_{\infty,-},\vec X_{\infty,+})$ of possible vacua of the combined
system, for specified vacua at the far left and far right.

The usual arguments show that  $\cal M$ is hyper-Kahler. But as in
section \ref{compman}, a useful way to understand $\cal M$ is to
describe it as a complex manifold in one of its complex
structures. Proceeding in the usual way, we introduce the complex
fields $\CX=X_1+iX_2$, $\CA=A+iX_3$, which obey a complex version
of eqn. (\ref{torm}):
\begin{equation}\frac{\CD\CX}{\CD y}+\delta(y)\mu^Z_\C =
0.\end{equation}
 Here $\mu^Z_\C=\mu^Z_1+i\mu^Z_2$ is the complex
moment map of $Z$.  $\cal M$ is the moduli space of solutions of
this equation, with $\CX(y)\to \CX_{\infty,\pm}$ for $y\to
\pm\infty$, and modulo complex gauge transformations that preserve
this asymptotic condition.  As before, the analysis is most
straightforward if $\CX_{\infty,\pm}$ are regular.  (Actually, we
will formulate the argument below in a way that  remains valid as
long as one of the two, say $\CX_{\infty,-}$, is regular; a
singularity develops only when both become non-regular.)

As usual, we can gauge away $\CA$ by a complex gauge transformation
$g(y)$ that does not necessarily preserve the asymptotic condition.
In fact, we can set $g(-\infty)=1$, but then $g(\infty)$ may not
commute with $\CX_{\infty,+}$.  (Of course, we can everywhere
reverse the roles of $+\infty$ and $-\infty$.)  A convenient way to
proceed is to make a gauge transformation with $g(-\infty)=1$ that
sets $\CA=0$ everywhere.  In this gauge, Nahm's equations reduce to
\begin{equation}\label{simpec}\frac{\mathrm{d}\CX}{\mathrm{d}y}+\delta(y)\mu^Z_\C=0,\end{equation}
saying simply that $\CX$ is piecewise constant, with a jump at
$y=0$. Moreover, the boundary condition requires that
$\CX(y)=\CX_{\infty,-}$ for $y<0$. After reducing Nahm's equation to
this form with this boundary condition, we are still free to make a
constant gauge transformation by an element of the group $T_\C$ that
commutes with $\CX_{\infty,-}$.

\def\H{{\mathcal H}}
Now let us count parameters.  A hypermultiplet in the fundamental
representation of $U(n)$ has $2n$ complex parameters.  After
solving (\ref{simpec}), we must impose $n$ complex constraints to
ensure that $\CX(y)$ has the same characteristic polynomial as
$\CX_{\infty,+}$. We also remove $n$ parameters in dividing by the
residual group $T_\C$ of  gauge transformations.  The net effect
is that in this particular example, the moduli space $\M$ is
zero-dimensional.  In fact, it consists of precisely one point, as
we will learn in section \ref{uniqueness}.

\bigskip\noindent{\it Several D5-Branes}

We can apply similar methods to a more general problem with $k$
D5-branes supported at points  $y=y_\alpha$, $\alpha=1,\dots,k$. At
each position $y_\alpha$ is supported a hypermultiplet $Z_\alpha$ in
the fundamental representation of $U(n)$.

Nahm's equation now becomes
\begin{equation}\label{otorm}\frac{D\vec X}{Dy} +\vec
X\times \vec X +\sum_{\alpha=1}^k\delta(y-y_\alpha)\vec
\mu^{Z_\alpha}=0.\end{equation} After gauging $\CA$ to zero and
requiring that $\CX(y)=\CX_{\infty,-}$ for $y<<0$, the complex Nahm
equation becomes
\begin{equation}\label{simptec}\frac{\mathrm{d}\CX}{\mathrm{d}y}+\sum_{\alpha=1}^k\delta(y-y_\alpha)\mu^{Z_\alpha}
_\C=0.\end{equation} So again, in this gauge $\CX$ is piecewise
constant, with jumps at $y=y_\alpha$, $\alpha=1,\dots,k$.

We count parameters as before.  Each fundamental hypermultiplet
$Z_\alpha$ contributes $2n$ parameters. We remove $n$ parameters in
dividing by the residual gauge symmetry, and $n$ more for requiring
that $\CX(y)$ for $y>>0$ has the same characteristic polynomial as
$\CX_{\infty,+}$.  So the total number of parameters is $2n(k-1)$.

\begin{figure}
  \begin{center}
    \includegraphics[width=3.5in]{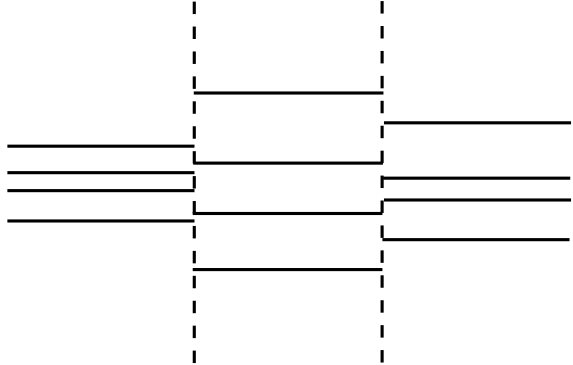}
  \end{center}
  \caption{\small A system of parallel D3-branes interacting with two
parallel D5-branes.  The D3-branes can ``break'' in crossing the
D5-branes.  The position of a D3-brane that connects two D5-branes
is parametrized by the value of a hypermultiplet.}
  \label{fig2}
\end{figure}

So far it does not matter if the points $y_\alpha$ are distinct.
If they are, there is actually a standard way to obtain the
counting of parameters from a brane picture (fig. \ref{fig2}).
Between each pair of successive D5-branes, the $n$ D3-branes can
break away and move freely. The position of a D3-brane is part of
a hypermultiplet, so this gives $n$ hypermultiplets for each pair
of successive D5-branes. With altogether $k$ D5-branes, there are
$k-1$ successive pairs, and so $n(k-1)$ hypermultiplets in all. A
single hypermultiplet corresponds to 2 complex parameters, so
there are $2n(k-1)$ complex parameters to specify the vacuum.

\subsubsection{Uniqueness Of The Vacuum}\label{uniqueness}

Going back to the case of a single D5-brane, we want to show that
for prescribed $\vec X_{\infty,\pm}$, the vacuum is unique. Since
we know that the moduli space $\M$ of vacua is of dimension zero,
it consists of a finite set of points; it suffices to count these
points in a special case.  We will take $\CX_{\infty,-}={\rm
diag}(S_1,S_2,\dots,S_n)$, with all $S_i$ distinct and nonzero,
 while $\CX_{\infty,+}=0$.

To analyze this situation, it helps to describe more explicitly the
moment map of a fundamental hypermultiplet $Z$.  From a complex
point of view, $Z$ consists of $n$ chiral superfields
$B^i,\,i=1,\dots n$ in the fundamental representation of $U(n)$, and
$n$ such superfields $C_j,\,j=1,\dots,n$ in the antifundamental
representation. The complex moment map is the rank 1
matrix\footnote{If $M:V\to V$ is a linear map, we define the rank of
$M$ to be the dimension of the image of $M$, that is, of the
subspace $MV$ of $V$.} $M$ whose matrix elements are
$M^i{}_j=B^iC_j$.  Putting the complex Nahm equation in the form
(\ref{simpec}), and writing $\CX'$ and $\CX''$ for the values of
$\CX$ for $y<0$ and $y>0$, respectively, we have
\begin{equation}\label{tarmac}\CX''=\CX'-M.\end{equation}
Of course, $\CX'=\CX_{\infty,-}$.

Since we are taking $\CX_+=0$, we need $\CX''$ to be nilpotent.  The
group $T_\C=(\C^*)^n$ of diagonal matrices can be used to set all
components $B^i$ (in the basis in which $\CX_{\infty,-}$ is
diagonal) to 1 or 0.  If any of these matrix elements vanishes, it
is impossible for $\CX'-M$ to be nilpotent. For example, for $n=2$,
if $B^1=0$, $B^2=1$, then $\CX'-M$ takes the form
\begin{equation}\label{armac}\begin{pmatrix}S_1 & 0 \\ 0 &
S_2\end{pmatrix}-\begin{pmatrix}0 & 0 \\ C_1 &
C_2\end{pmatrix}.\end{equation} This matrix has $S_1$ as one of its
eigenvalues and is not nilpotent.

So we take
\begin{equation}B=\begin{pmatrix}1\\ 1\\ \vdots\\
1\end{pmatrix}.\end{equation} The condition that $\CX'-M$ is
nilpotent is equivalent to $\det(z-(\CX'-M))=z^n$.  The left hand
side in general equals $z^n+f_{n-1}z^{n-1}+\dots +f_1z+f_0$, and we
must set the coefficients $f_{n-1},\dots,f_0$ to zero. These
coefficients are linear functions of $C_1,\dots,C_n$ because $M$ has
rank 1.  So there is precisely one solution.

\subsubsection{One Extra Brane}\label{zoreq}

\begin{figure}
  \begin{center}
    \includegraphics[width=3.5in]{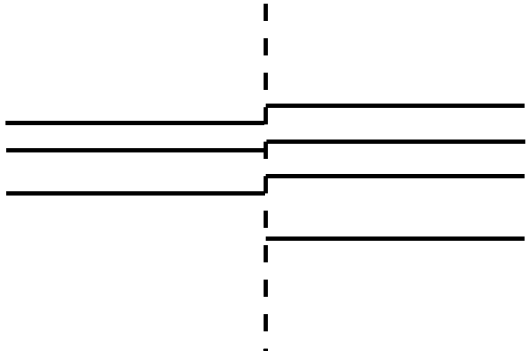}
  \end{center}
  \caption{\small In this example, the number of D3-branes jumps by 1 in crossing
a D5-brane.}
  \label{fig3}
\end{figure}

Our next goal is to describe what happens when there are unequal
numbers of D3-branes on the two sides of a D5-brane.  We begin
with the case that the difference is 1, say $n$ D3-branes for
$y<0$ and $n+1$ for $y>0$ (fig. \ref{fig3}). First we describe
what we claim is the appropriate description of this situation;
then we will try to justify it.

What is  depicted in fig. \ref{fig3} is an example of a
supersymmetric domain wall interpolating between ${\cal N}=4$
super Yang-Mills theories with two different gauge groups -- in
the present case, $U(n)$ for $y<0$ and $U(n+1)$ for $y>0$.  Such
domain walls were discussed in section \ref{domainwalls}. It turns
out that from a field theory point of view, the supersymmetric
domain wall of fig. \ref{fig3} can be described by the
construction of section \ref{secondex} if we take $G=U(n+1)$ and
$G'=U(n)$, and take $H$ to be a copy of $U(n)$ that is a diagonal
product of $G'$ and a $U(n)$ subgroup of $G$. We also set $\tilde
G=U(n)\times U(n+1)$.

For finding supersymmetric vacua, the relevant facts are as
follows. There are no extra matter fields at $y=0$; a
supersymmetric vacuum is to be described by solving Nahm's
equations for $\vec X$. For $y<0$, the gauge group is $U(n)$ so
the components of $\vec X$ are $n\times n$ matrices.  But for
$y\geq 0$, they are $(n+1)\times (n+1)$ matrices. What happens at
$y=0$ is simply that the smaller matrix is embedded as an $n\times
n$ submatrix of the large one; the extra row and column are
arbitrary.

For example, if $n=2$, then $\vec X$ is a $2\times 2$ matrix for
$y<0$:
\begin{equation}\label{mymatrix}\begin{pmatrix} * & * \\ * & *
\end{pmatrix}.\end{equation}
For $y\geq 0$, an extra row and column appear:
\begin{equation}\label{hymatrix}\begin{pmatrix} * & * & \times \\ *
& * & \times \\ \times & \times & \times
\end{pmatrix}.\end{equation}
The upper left block is continuous at $y=0$, and the other matrix
elements are unconstrained at $y=0$.  Nahm's equations determine the
dependence on $y$.

With some care, the recipe stated in the last paragraph can be
extracted from eqn. (\ref{topsy}).   The most relevant part of
eqn. (\ref{topsy}) is $\vec X^+|=D_3\vec X^-|=0$.  As explained in
section \ref{generalization}, eqn. (\ref{topsy}) can be used in
${\cal N}=4$ super Yang-Mills theory with any gauge group $\tilde
G$ and subgroup $H$, if one understands $\Phi^+$ as the projection
of an adjoint-valued field $\Phi$ from $\tilde{\frak g}$ to $\frak
h$, and $\Phi^-$ as the projection to $\frak h^\perp$.  In the
present case, we take $\tilde G=U(n)\times U(n+1)$ and $H$ the
diagonal $U(n)$ subgroup described above.  Then as described in
section \ref{domainwalls}, we ``unfold'' the theory to convert
this boundary condition to a supersymmetric domain wall
interpolating between gauge groups $U(n)$ and $U(n+1)$.  After
unfolding, we arrive at the picture in the last paragraph.

An important detail is that in unfolding, we reverse the sign of
$\vec X$ in one group, say $U(n+1)$.   So the condition that $\vec
X^+|=0$ in the folded theory is equivalent after unfolding to the
statement that the $U(n)$ part of $\vec X$ is continuous at $y=0$.
This is the main claim in (\ref{hymatrix}).

Now we wish to analyze the supersymmetric vacua in this situation.
 As in section \ref{nahmdisc}, we pick commuting triples $\vec X_{\infty,-}$
and $\vec X_{\infty,+}$ to specify choices of vacuum for large
negative and large positive $y$.  We denote as ${\cal M}(\vec
X_{\infty,+},\vec X_{\infty,-})$ the moduli space of vacua in the
full system when the vacua at infinity are fixed. It is the moduli
space of solutions of Nahm's equations (for matrices whose size
jumps as above at $y=0$) with $\vec X(y)\to \vec X_{\infty,\pm}$
for $y\to\pm \infty$. The allowed gauge transformations are by a
function $g(y)$ that is $U(n)$-valued for $y<0$ and
$U(n+1)$-valued for $y>0$. At $y=0$, $g(y)$ takes values in the
subgroup $U(n)$ of $U(n+1)$.

To describe ${\cal M}$, we use again its relation to the complex
Nahm equation
\begin{equation}\label{newc} 0=\frac{\CD\CX}{\CD y}
=\frac{\mathrm{d}\CX}{\mathrm{d}y}+[\CA,\CX],\end{equation} where
$\CX$ and $\CA$ are complex-valued matrices whose size jumps at
$y=0$ as above. As usual, in one of its complex structures, ${\cal
M}$ is the moduli space of solutions of this equation such that
$\CX(y)\to \CX_{\infty,\pm}$ for $y\to \pm\infty$, modulo complex
gauge transformations.  We analyze this problem in the familiar way
by making a gauge transformation with $g(y)\to 1$ for $y\to -\infty$
to set $\CA=0$. The rest of the argument is quite similar to steps
we have already seen. Eqn. (\ref{newc}) implies that $\CX$ is
constant for $y<0$ and hence equals $\CX_{\infty,-}$. At $y=0$, it
acquires $2n+1$ new complex coefficients (from the extra row and
column).  Of these, $n$ can be removed by a gauge transformation
that commutes with $\CX_{\infty,-}$, and $n+1$ are fixed by
requiring that $\CX(y)$ for $y>0$ has the same characteristic
polynomial as $\CX_{\infty,+}$. So counting parameters, we see that
the moduli space ${\cal M}$ is of dimension zero.

In fact, ${\cal M}$ consists of a single point.  The argument for
this closely follows section \ref{uniqueness}.  We assume that
$\CX_{\infty,-}$ is diagonal with distinct and nonzero eigenvalues
and we take $\CX_{\infty,+}=0$.  By a constant gauge
transformation that commutes with $\CX_{\infty,-}$, we can set all
matrix elements in the last column in eqn. (\ref{hymatrix}) to 1
except the bottom one, and then the condition that $\CX(y)$ is
nilpotent for $y>0$ gives $n+1$ linear equations that uniquely
determine the bottom row in eqn. (\ref{hymatrix}).

\bigskip\noindent{\it Comparison To The D3-D5 System}

We now want to show that what has just been described is consistent
with what we know about the D3-D5 system. (See \cite{WeinbergChen}
for a more thorough treatment of similar issues in the context of
monopoles.)

\begin{figure}
  \begin{center}
    \includegraphics[width=2in]{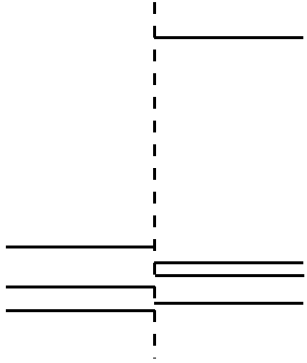}
  \end{center}
 \caption{\small By displacing a D3-brane that is on the right of the D5-brane
very far from the others in the $x^4-x^5-x^6$ directions, we can
reduce to a case with equals numbers of D3-branes on both sides.}
  \label{fig4}
\end{figure}

If one of the D3-branes that are at $y>0$ in fig. \ref{fig3} moves
far away, we reduce (fig. \ref{fig4}) to the case of $n$ D3-branes
meeting a D5-brane. A bifundamental hypermultiplet must appear.
Let us see how this happens.

We suppose that $\CX_{\infty,+}$ has one large eigenvalue, which
we call $W$, and we take $W\to\infty$ keeping all other
eigenvalues of $\CX_{\infty,+}$ fixed. (We also keep
$\CX_{\infty,-}$ fixed.)  For instance, for $n=2$ we have
\begin{equation}\CX_{\infty,+}=\begin{pmatrix} * & * & 0 \\
                                                * & * & 0 \\
                                                0 & 0 &
                                                W\end{pmatrix},\end{equation}
and the matrix elements denoted $*$ will be held fixed while
$W\to\infty$.  We write $\CX'_{\infty,+}$ for the upper left
$n\times n$ block of $\CX_{\infty,+}$.

After gauging $\CA$ to zero and setting $\CX(y)=\CX_{\infty,-}$ for
$y<0$, we are supposed to pick the last row and column in eqn.
(\ref{hymatrix}) so that $\CX(y)$, for $y>0$, is conjugate to
$\CX_{\infty,+}$.  In our example of $n=2$, $\CX_{\infty,-}$ is a
$2\times 2$ matrix that ``grows'' an extra row and column for $y>0$.
 In fact, we pick the last row and column so that for $y>0$
\begin{equation}\label{gelg}\CX(y)=\begin{pmatrix} * & * &  W^{1/2} B^1 \\ *
& * &  W^{1/2} B^2 \\ W^{1/2} C_1 & W^{1/2} C_2 & W
\end{pmatrix},\end{equation}
where the upper left block equals $\CX_{\infty,-}$, and the
coefficients $B^i$, $C_j$, $i,j=1,\dots,n$ are kept fixed for
$W\to\infty$.

Second order perturbation theory shows that for large $W$, one
eigenvalue of $\CX(y)$ equals $W$ and the others are the eigenvalues
of the $n\times n$ matrix
\begin{equation}\label{zelgo} \CX_{\infty,-}-M,\end{equation}
where $M$ has matrix elements $M^i{}_j=B^iC_j$.  Our problem is
now to choose $M$ so that this matrix is conjugate to
$\CX_{\infty,+}$. But this is precisely the problem that we
encountered for the D3-D5 system, with the pair $B^i,C_j$ playing
the role of the bifundamental hypermultiplet.  This shows how the
physics of the D3-D5 intersection follows from our proposal
concerning the asymmetric configuration with an extra D3-brane at
$y>0$.

\bigskip\noindent{\it Flowing In The Opposite Direction}

It is also possible to run this in reverse.  We begin with a D3-D5
system with $n+1$ D3-branes on each side of the D5-brane.  Then we
move one of the D3-branes at $y<0$ very far from the others. We do
this by giving $\CX_{\infty,-}$ one large eigenvalue $W$, while
keeping fixed its other eigenvalues as well as $\CX_{\infty,+}$.
We take $\CX_{\infty,-}={\rm diag}(w_1,w_2,\dots,w_n,W)$, where
$w_1,\dots,w_n$ are the small eigenvalues.  For large $W$, we
should reduce to the problem with $n$ D3-branes at $y<0$ and $n+1$
at $y>0$.

As in eqn. (\ref{tarmac}), we are supposed to satisfy
\begin{equation}\label{lobo}\CX_{\infty,-}-M=\CX'',\end{equation}
where $\CX''$ should be conjugate to $\CX_{\infty,+}$ and in
particular has all eigenvalues fixed as $M\to\infty$.  For this, we
take $M$ of the form
\begin{equation}\label{gelform}M=
\begin{pmatrix} \cdot & \cdot & \times \\ \cdot
& \cdot & \times \\ \times & \times & W+\times
\end{pmatrix}\end{equation}
(illustrated here for $n=2$), where coefficients denoted $\times $
are kept fixed as $W\to\infty$, while coefficients denoted
$~\cdot~$ vanish for $W\to\infty$ and are adjusted so that $M$ is
of rank 1. With this ansatz, the problem of satisfying eqn.
(\ref{lobo}) for large $W$ is equivalent to what we got from
Nahm's equations with matrices that jump in rank in crossing
$y=0$.  The quantities labeled $\times$ in (\ref{gelform}) simply
map to the quantities labeled the same way in (\ref{hymatrix}).

\subsubsection{D3-Branes Ending On A D5-Brane}\label{onesided}

\begin{figure}
  \begin{center}
    \includegraphics[width=3.5in]{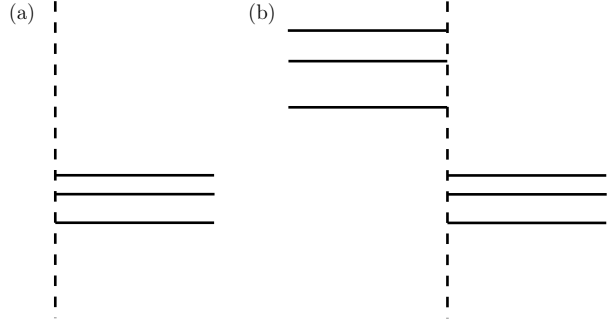}
  \end{center}
  \caption{\small (a) All D3-branes to the right of a D5-brane.
(b) A configuration with equal numbers of D3-branes on both sides
can be reduced to the one-sided configuration in (a) by moving all
D3-branes on one side to large values of $x^{7,8,9}$.}
  \label{fig5}
\end{figure}
When the difference between the numbers of D3-branes on the two
sides of a D5-brane exceeds 1, poles appear in the solutions of
Nahm's equations. To isolate the essential subtleties, we begin
with the extreme case of $n$ D3-branes at $y>0$ and none at $y<0$
(fig. \ref{fig5}a). We approach this starting from the case of $n$
D3-branes on each side, where everything is computable in weakly
coupled string theory, and then we reduce to the case we want by
removing the D3-branes on one side.  To do this (fig.
\ref{fig5}b), we take the eigenvalues of $\vec X_{\infty,-}$ to be
large, while $\vec X_{\infty,+}$ remains small or zero.

In the description by the complex Nahm equations, we use the usual
gauge in which $\CX$ is piecewise constant, equaling $\CX'$ or
$\CX''$ for $y<0$ or $y>0$. $\CX'$ must coincide with
$\CX_{\infty,-}$ (which we assume to be regular), and $\CX''$ must
have the same characteristic polynomial as $\CX_{\infty,+}$. To
achieve the situation depicted in fig. \ref{fig5}b, we take the
eigenvalues of $\CX_{\infty,-}$ to be distinct and large, and we
take $\CX_{\infty,+}=0$.  It follows that $\CX''$ is nilpotent, and
hence its rank is at most $n-1$.

Writing (\ref{tarmac}) in the form $\CX'=\CX''+M$, it says that the
rank $n$ matrix $\CX'$ must be the sum of a rank 1 matrix $M$ and
the matrix $\CX''$.  Hence $\CX''$ must have rank at least $n-1$. In
view of the observation in the last paragraph, this means that
$\CX''$ has rank exactly $n-1$.  Consequently, it is a regular
nilpotent element, conjugate to the matrix in eqn. (\ref{nork}).

Let us suppose that $\vec X_{\infty,+}$ vanishes (and not just its
complex part $\CX_{\infty,+})$, so that $\vec X(y)\to 0$ for
$y\to+\infty$. We know the form of a solution of Nahm's equations
that approaches zero at infinity and for which $\CX$ is a regular
nilpotent.  It is
\begin{equation}\label{pikop}\vec X=\frac{\vec
t}{y+c^{-1}},\end{equation} where $\vec t$ are the generators of an
irreducible $\frak{su}(2)$ subalgebra of $\frak{su}(n)$, and $c$ is
a positive constant.  Any solution on the half-line $y\geq 0$ with
the stated properties has this form.

The constant $c$ depends on the choice of $\CX_{\infty,-}$.  To
reduce to the problem of D3-branes ending on a D5-brane (fig.
\ref{fig5}a), we wish  $\CX_{\infty,-}$ to have very large
eigenvalues. Nahm's equations are invariant under the scaling
$\vec X\to s\vec X$, $y\to s^{-1}y$, for positive $s$.  Under this
operation, we have $c\to sc$. So when we take $s\to\infty$ to send
$\CX_{\infty,-}$ to infinity, we also get $c\to\infty$. The
limiting form of the solution for $y>0$ is then
\begin{equation}\label{pikopo}\vec X=\frac{\vec
t}{y}.\end{equation}

We have learned that the appropriate boundary condition for $n$
D3-branes ending on a D5-brane at $y=0$ is that $\vec X$ must have a
pole at $y=0$ corresponding to an irreducible $\frak{su}(2)$
embedding.  The same reasoning applies for a Dp-D(p+2) system for
any $p$.  While this is a striking and perhaps surprising result, it
has been discovered and explained in the past in several different
ways. The original analysis of the Dp-D(p+2) system and its relation
to Nahm's equations and monopoles \cite{Diaconescu:1996rk} implied
this behavior, in view of the role of such poles in the theory of
monopoles \cite{Nahm:1979yw}.  The pole has an elegant
interpretation in terms of a distortion of the D5-brane by the
``pull'' of the D3-branes \cite{CallanMaldacena}; the different
viewpoints have been related in \cite{Constable:1999ac}.

What happens if we take $\vec X_{\infty,+}$ to be nonzero, and we
keep it fixed while scaling $\vec X_{\infty,-}$ to infinity? As we
learned in our study of the Slodowy slice, for every choice of $\vec
X_{\infty,+}$, there is a unique solution of Nahm's equations on the
half-line $y>0$ that has the singularity of (\ref{pikopo}) for $y\to
0$ and approaches $\vec X_{\infty,+}$ for $y\to\infty$. This is the
behavior that we will get for $y>0$ in the situation just described.

\subsubsection{A More General Case}\label{moreg}

\begin{figure}
  \begin{center}
    \includegraphics[width=3.5in]{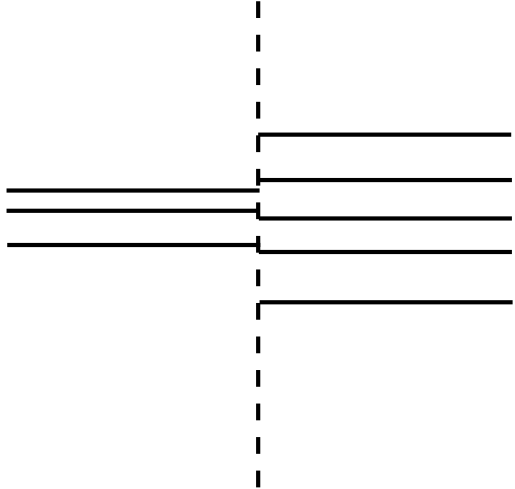}
  \end{center}

  \caption{\small Two extra D3-branes to the right of a D5-brane.}
  \label{fig6}
\end{figure}

Now we consider the general case of a D5-brane with $m$ D3-branes
ending on one side  (fig. \ref{fig6}) and $n>m$ on the other side.
We have already treated the cases that $n=m+1$, or $m=0$. Here we
assume that $n\geq m+2>0$.

It is possible to guess what happens on the following grounds.  We
could simply remove $m$ of the D3-branes by detaching them from the
D5-brane and displacing them in $\vec Y$. This leaves $n-m$
D3-branes on one side of the D5-brane and none on the other side. In
that case, we have just seen that a supersymmetric configuration of
the remaining $n-m$ D3-branes is described by a solution of Nahm's
equations with a pole associated to an irreducible embedding
$\rho:\frak{su}(2)\to \frak{su}(n-m)$. Now if we move the extra
D3-branes back, the simplest possibility is that they do not disturb
this pole.

This suggests that the system is described by a solution of Nahm's
equations with the following properties.  For $y<0$, $\vec X$ is an
$m\times m$ matrix-valued solution of the equations, regular at
$y=0$.  For $y>0$, $\vec X$ is an $n\times n$ matrix-valued
solution.  Near $y=0$, $\vec X$ looks like
\begin{equation}\label{golf}\begin{pmatrix}\vec A& \vec B\\ \vec C &
\vec D\end{pmatrix}\,\end{equation} where the entries are as
follows. $\vec A$ is an $m\times m$ matrix (or rather a trio of such
matrices) and coincides with the limit of $\vec X(y)$ as $y$
approaches zero from below. The lower right hand block is an
$(n-m)\times (n-m)$ block with
\begin{equation}\label{golto}\vec
D=\frac{\vec t}{y}+\dots;\end{equation} here $\vec t$ are generators
of an irreducible embedding $\rho:\frak{su}(2)\to \frak{su}(n-m)$,
and the ellipses are regular terms.  The other blocks $\vec B$ and
$\vec C$ are merely required to be regular for $y\to 0$.

This example is a domain wall of the type described in section
\ref{secondex}, with $G=U(n)$, $G'=U(m)$, a pole in Nahm's equations
that breaks $G\times G'$ to $U(m)\times U(m)$, and $H$ a diagonal
subgroup of $U(m)\times U(m)$.

Notice that if we set $n-m=1$, then  $\vec t=0$ (its components
generate a trivial one-dimensional representation of $\frak{su}(2)$)
so there is no pole in eqn. (\ref{golf}), which actually then
reduces to what we have analyzed in section \ref{zoreq}.

\bigskip\noindent{\it A Useful Trick}

Now let us discuss the solutions of Nahm's equations in this
example. As in the other examples with only a single D5-brane,
once one specifies $\vec X_{\infty,\pm}$, the relevant moduli
space of solutions of Nahm's equation is zero-dimensional. In
order to show this, we need to understand the effects of the pole
at $y=0$. In a one-sided problem that we have already studied in
section \ref{solpole}, the analysis of the pole leads to the
Slodowy slice transversal to a nilpotent orbit. Rather than making
a similar analysis in a new situation, we will use a trick to
reduce to the previous case. The trick in question also has other
applications.

We let $\M_+$ denote the space of $n\times n$ solutions of Nahm's
equations on the half-line $y> 0$, with the form given in eqn.
(\ref{golf}) near $y=0$ and approaching $\vec X_{\infty,+}$ (up to
conjugacy) for $y\to \infty$. Thus, $\vec X$ has a pole at $y=0$
associated with an $\frak{su}(2)$ embedding of rank $n-m$. The
group $U(m)$ acts on $\M_+$, by gauge transformations at $y=0$
that commute with the pole.\footnote{The symmetry group is really
$U(m)\times U(1)$, but the second factor will not be important.
 The moment map for the second factor is $\Tr\,\vec D$.}
 The moment map for the action of $U(m)$ on
$\M_+$ is
\begin{equation}\label{momright}\vec\mu^+=\vec A,\end{equation}
where as in eqn. (\ref{golf}), $\vec A$ is the value at $y=0$ of the
upper left block of $\vec X$.  This formula was obtained in eqn.
(\ref{hypgac}) (except that here we restrict to those global gauge
transformations that commute with the pole).
 The complex dimension of $\M_+$ is $s-n$, where $s$ is the
number of summands when the Lie algebra $\frak{u}(n)$ is
decomposed in irreducible representation of $\frak{su}(2)$
(embedded in $\frak{u}(n)$ via $n=(n-m)+1+1+\dots+1$).  Performing
this computation, we find that
\begin{equation}\label{hobbo}{\rm dim}\,\M_+=m^2+m.\end{equation}

On the other hand, we can solve Nahm's equations in $m\times m$
matrices on the half-line $y\leq 0$.  Now we require the solution
to be regular at $y=0$ and to approach $\vec X_{\infty,-}$ (up to
conjugacy) for $y\to-\infty$. We denote the moduli space of
solutions as $\M_-$. As a complex manifold, it is isomorphic to
the orbit of $\CX_{\infty,-}$ if that element is regular, and in
any event its complex dimension is
\begin{equation}\label{zobbo}{\rm dim}\,\M_-=m^2-m.\end{equation}
The group $U(m)$ acts on $\M_-$ by gauge transformations at $y=0$,
and the hyper-Kahler moment map is
\begin{equation}\label{obbo}\vec\mu\,^-=-\vec X(0)=-\lim_{y\to 0^-}\vec X(y).
\end{equation}
The reason for the minus sign is that we are now solving Nahm's
equations on the half line $y\leq 0$ rather than $y\geq 0$, and
this reverses the sign of the endpoint contribution that results
from integration by parts.

The product $\M_+\times \M_-$ is a hyper-Kahler manifold acted on
by $U(m)$.  Let us take its hyper-Kahler quotient by $U(m)$. This
entails setting to zero the combined moment map
$\vec\mu=\vec\mu^++\vec\mu\,^-$ and dividing by $U(m)$.  Setting
$\vec\mu=0$ means that $\lim_{y\to 0^-}\vec X(0)=\vec A$.  This
means that the two partial solutions on the half-lines $y\leq 0$
and $y\geq 0$ fit together to a solution on the whole line, with
the right singularity at $y=0$ and the right matching condition as
described in eqn. (\ref{golf}) to represent a supersymmetric
vacuum of the full system. Also, in constructing $\M_+$ we have
divided by gauge transformations for $y>0$, and in constructing
$\M_-$ we have divided by gauge transformations for $y<0$.  So
after also dividing by $U(m)$ to construct the hyper-Kahler
quotient of $\M_+\times\M_-$ by $U(m)$, we have divided by all
gauge transformations on the line.

The upshot is that the desired moduli space $\M$ of supersymmetric
vacua of the combined system is the hyper-Kahler quotient
 of $\M_+\times \M_-$ by $U(m)$, often denoted $(\M_+\times \M_-)/\negthinspace
 /\negthinspace /U(m)$.  Taking
the hyper-Kahler quotient by a $w$-dimensional group reduces the
complex dimension by $2w$.  Since the dimension of $U(m)$ is
$m^2$, we see, using (\ref{hobbo}) and (\ref{zobbo}), that $\M$ is
zero-dimensional.

\subsection{Pole Of General Type}\label{genpole}

We are now ready to consider a much more general problem. We
consider supersymmetric boundary conditions of D5-type in $U(n)$
gauge theory.  From a field theory point of view, such a boundary
condition can be constructed for any choice of a homomorphism
$\rho:\frak{su}(2)\to \frak u(n)$. In section \ref{onesided}, we
explained that the case that $\rho$ is an irreducible
$\frak{su}(2)$ embedding corresponds to D3-branes ending on a
D5-brane.  This makes it relatively straightforward to understand
the $S$-dual of this boundary condition.

As a basis for understanding $S$-duality in general, we would like
to find a D-brane construction of the boundary condition
associated to an arbitrary $\rho$. At first sight, this may appear
difficult. A general $\rho$ is specified by a decomposition
$n=n_1+n_2+\dots + n_k$, where the $n_i$ are positive integers and
we can assume that $n_1\geq n_2\geq \dots\geq n_k$. How can we
encode this information in terms of D-branes?

Roughly speaking, we do this by letting the $n$ D3-branes end on
$k$ different D5-branes -- with $n_i$ D3-branes ending on the
$i^{th}$ D5-brane, for $i=1,\dots,k$.  However, it is not clear
what this is suppose to mean if all the D5-branes are located at
$y=0$.  In that case, how do we make sense of the question of
which D3-brane ends on which D5-brane?

\begin{figure}
  \begin{center}
    \includegraphics[width=3.5in]{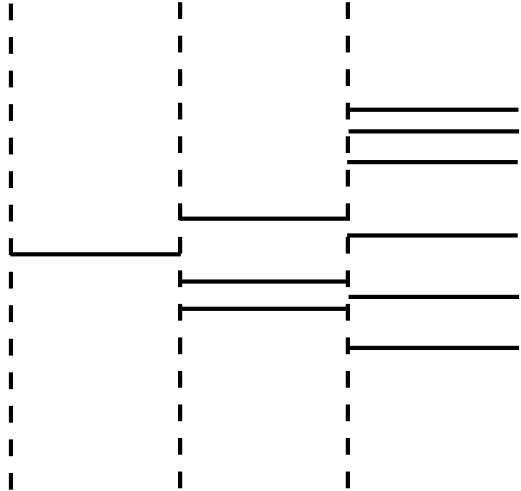}
  \end{center}

  \caption{\small To the right of this picture, there are six D3-branes.
Reading from right to left, three of them end on the first
D5-brane, two end on the second, and one ends on the third and
last.}
  \label{fig7}
\end{figure}

To make sense of it, we displace the D5-branes from each other, as
in fig. \ref{fig7}.  Thus we consider a system with D5-branes at
points $y_\alpha$, $\alpha=1,\dots,k$; we assume that $\tilde
n_\alpha$ D3-branes end on the $\alpha^{th}$ D5-brane.  Thus the
total number of D3-branes is reduced by $\tilde n_\alpha$ when one
crosses the $\alpha^{th}$ D5-brane from right to left.  The
numbers $\tilde n_\alpha$ will equal the $n_i$ up to permutation
-- but it will be crucial, as we will see, to choose the right
permutation.

From section \ref{nahmdisc}, we know what field theory construction
corresponds to the configuration of fig. \ref{fig7}. Supersymmetric
vacua, for example, are described by a solution of Nahm's equations
for matrices $\vec X(y)$ whose rank jumps whenever $y=y_\alpha$ for
some $\alpha$, and which have a pole for $y\to y_\alpha^+$ whenever
$\tilde n_\alpha\geq 2$.  Moreover, fig. \ref{fig7} is useful
because it is described in terms of branes; this will be our
starting point in a separate paper in analyzing its $S$-dual.

Our hypothesis then, is that for some choice of the $\tilde
n_\alpha$, which will equal the $n_i$ up to permutation, the brane
configuration of fig. \ref{fig7} is equivalent, in the limit that
all $y_\alpha\to 0$, to a field theory construction based on a
corresponding $\frak{su}(2)$ embedding.  The $\frak{su}(2)$
embedding, of course, is the one associated with the decomposition
$n=n_1+n_2+\dots + n_k$. In section \ref{ancomp}, we justify this
claim by analyzing the moduli space of supersymmetric vacua.  But
first, we determine exactly how the $\tilde n_\alpha$ must be
related to the $n_i$.

\begin{figure}
  \begin{center}
    \includegraphics[width=4.5in]{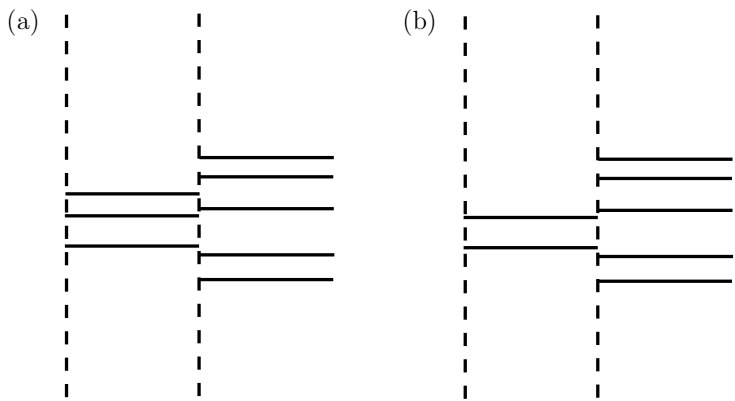}
  \end{center}

  \caption{\small To the right of the picture, there are five parallel D3-branes.
Reading from right to left, in (a), two of them end on the first
D5-brane and three on the second, while in (b), the numbers are
reversed.  This is the special case $n=5$, $n_1=3$, $n_2=2$ of a
decomposition $n=n_1+n_2$.}
  \label{fig8}
\end{figure}

To explain the issue, we first consider the example $k=2$.  Thus,
there are precisely two D5-branes; $n_1$ end on one and $n_2$ on
the other.  There are two possible arrangements (fig. \ref{fig8}),
depending on whether the D5-brane on which the larger number of
branes end is on the right or the left.

In section \ref{ancomp}, we will use Nahm's equations to describe
the moduli space $\M$ of supersymmetric vacua (for a given limit
$\vec X_\infty$ at infinity) in this situation.  But for now, we
count the parameters directly from the brane diagram.

Each D3-brane that is free to move between two D5-branes
contributes one hyper-Kahler modulus or two complex moduli.  In
fig. \ref{fig8}a, there are $n_1$ such branes and in fig.
\ref{fig8}b, there are $n_2$ such branes. (The figure is drawn for
$n_1=3$, $n_2=2$.) So the number of complex moduli is $2n_1$ in
one case and $2n_2$ in the other.

Let us compare this to a boundary condition in  Yang-Mills theory
on a half-space  given by a solution of Nahm's equation with a
pole at $y=0$.  We suppose that the pole is determined by a
homomorphism $\rho:\frak{su}(2)\to\frak{u}(n)$ associated with a
decomposition $n=n_1+n_2$.  The moduli space $\M$ of vacua,
according to section \ref{solpole}, has complex dimension $s-n$,
where $s$ is the number of summands when the Lie algebra
$\frak{u}(n)$ is decomposed as a direct sum of irreducible
representations of $\frak{su}(2)$. Assuming that $n_1\geq n_2$,
one finds that $s=n+2n_2$.  The complex dimension of $\M$ is
therefore $2n_2$.

This shows that the configuration of fig. \ref{fig8}b, but not that
of fig. \ref{fig8}a, may as $y_1,y_2\to 0$ approach the conformally
invariant boundary condition determined by $\rho$. We believe this
to be the case.  We are not certain what is the limit for
$y_1,y_2\to 0$ of the configuration of fig. \ref{fig8}a, but we
believe that it may be that the $n_1-n_2$ extra hypermultiplets
simply decouple in this limit.

\begin{figure}
  \begin{center}
    \includegraphics[width=3.5in]{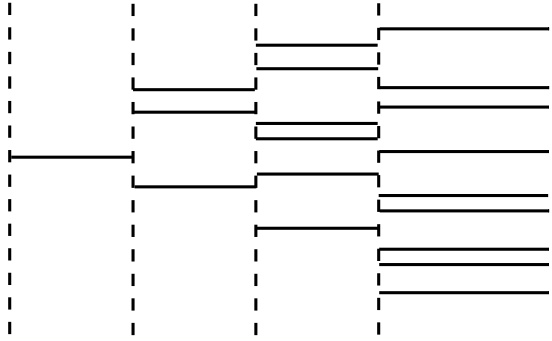}
  \end{center}
  \caption{\small Reading from right to left, the numbers of D3-branes ending on
successive D5-branes are 3,3,2, and 1. This is a non-increasing
sequence of numbers, so this configuration has a nice limit when
the D5-branes become coincident.}
  \label{fig9}
\end{figure}

In section \ref{ancomp}, we will confirm the hypothesis about fig.
\ref{fig8}b by analyzing $\M$ as a complex manifold.  For now,
however, we just explain the counting for the case that $\rho$ is
an $\frak{su}(2)$ embedding associated with a general
decomposition $n=n_1+n_2+\dots+n_k$, with $n_1\geq
n_2\geq\dots\geq n_k$.  The number of irreducible $\frak{su}(2)$
modules in the decomposition of $\frak{u}(n)$ is
$s=n+2\sum_{i<j}n_j=n+2\sum_{j=1}^k(j-1)n_j$. The hyper-Kahler
dimension of $\M$ is therefore
\begin{equation}\label{dorkky}(s-n)/2=\sum_{j=1}^k(j-1)n_j.\end{equation}
We stress that this is the dimension of $\M$ if and only if the
$n_i$ are labeled in non-ascending order.

Eqn. (\ref{dorkky}) agrees with the number of parameters suggested
by the brane diagram if and only if, as one approaches the boundary
from the bulk, the number of D3-branes ending on one D5-brane is at
least as great as the number ending on the next one (fig.
\ref{fig9}). Indeed, the number of hyper-Kahler parameters suggested
by the brane picture is $\sum_{j=1}^{k-1} b_j$, where $b_j$ is the
number of D3-branes between the $j^{th}$ and $j+1^{th}$ D5-brane. If
$\tilde n_j$ D3-branes end on the $j^{th}$ D5-brane (counting them
from right to left), then  $b_j=\sum_{s=j+1}^k \tilde n_s$, so
$\sum_{j=1}^{k-1}b_j=\sum_{j=1}^k (j-1)\tilde n_j$.  Assuming that
the numbers $\tilde n_j$ are supposed to be a permutation of the
$n_j$, this number coincides with (\ref{dorkky}) if and only if
$n_j=\tilde n_j$, so that the $\tilde n_j$ are non-ascending.

\begin{figure}
  \begin{center}
    \includegraphics[width=5in]{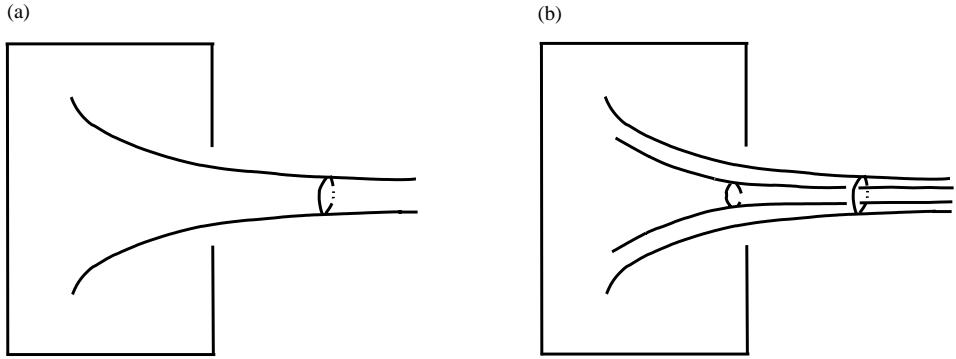}
  \end{center}

  \caption{\small (a) D3-branes ending on a D5-brane can be represented by a
``spike.'' (b) The condition that the numbers of D3-branes ending
on successive D5-branes are non-increasing (from right to left)
ensures that one spike can fit inside the next. }
  \label{fig10}
\end{figure}

This has a heuristic explanation if one thinks of the D3-branes
ending on a D5-brane as creating a ``spike'' in a D5-brane
\cite{CallanMaldacena}.  The condition of fig. 9 then makes it
possible for the various spikes to avoid intersecting each other
(fig. \ref{fig10}).

\subsubsection{Analysis Of $\M$ As A Complex
Manifold}\label{ancomp}

\begin{figure}
  \begin{center}
    \includegraphics[width=4in]{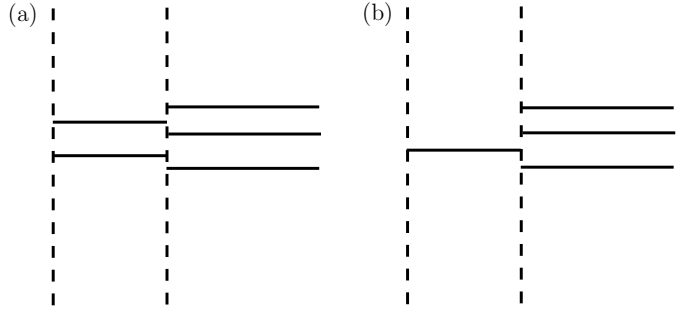}
  \end{center}
  \caption{\small These two configurations correspond to two different problems in
$U(3)$ gauge theory, as described in the text.  The configuration
of (b) has a straightforward limit as the two NS5-branes approach
each other, and that of (a) does not.}
  \label{fig11}
\end{figure}

We are now going to compare the complex manifolds associated with
the brane diagrams of figs. \ref{fig9}a and \ref{fig9}b to the
answer coming from the corresponding $\frak{su}(2)$ embedding.  To
make it easy to write explicit formulas, we will just describe the
case $n=3$, $n_1=2$, $n_2=1$ (fig. \ref{fig11}).  The general case
is similar. We want to compare three complex manifolds:

(1) ${\cal M}$ parametrizes supersymmetric vacua in $U(3)$ gauge
theory on the half-space $y\geq 0$, with a pole at $y=0$ given by
the $\frak{su}(2)$ embedding (associated with the decomposition
$3=2+1$), and with $\vec X(y)\to\vec X_\infty$ for $y\to\infty$.

(2) ${\cal M}'$ parametrizes supersymmetric vacua in the brane
picture of fig. \ref{fig11}a; one D3-brane ends on a D5-brane at
$y=y_1>0$ and the other two continue to $y=0$.

(3) ${\cal M}''$ parametrizes supersymmetric vacua corresponding
to fig. \ref{fig11}b; now two D3-branes end at $y=y_1$ and the
other two continue to $y=0$.

We can describe case (1) using eqn. (\ref{ime}).  If we choose
$\rho$ to map $\frak{su}(2)$ to matrices supported in the upper
left $2\times 2$ block of a $3\times 3$ matrix (so that in
particular the pole in $\CX$ lives in that block), then the
general form of a solution of the complex Nahm equations in case
(1) is
\begin{equation}\label{gurcase}\CX=
\begin{pmatrix} a & y^{-1} & 0 \\
by & a    & cy^{1/2}\\
dy^{1/2} & 0 & e\end{pmatrix},\end{equation} where $a,b,c,d$ and
$e$ are complex parameters.  After fixing three parameters to
specify the characteristic polynomial of $\CX$, we see that $\M$
is two-dimensional.

In case (2), on the interval $0\leq y\leq y_1$, we solve Nahm's
equations via $2\times 2$ matrices with a pole at $y=0$.  The
general allowed form is again given by eqn. (\ref{ime}):
\begin{equation}\label{urcase}\CX=\begin{pmatrix} a & y^{-1} \\
                                                  by & a
\end{pmatrix}.\end{equation}
So far so good: this agrees with the upper left block in eqn.
(\ref{gurcase}).  However, when we cross $y=y_1$, the matrix
simply grows a new row and column with no restriction on the new
matrix elements:
\begin{equation}\label{turcase}\CX=\begin{pmatrix} a & y^{-1}& c \\
                                                  by & a& d\\ e &
f & g
\end{pmatrix}.\end{equation}
Now there are seven complex parameters, and after adjusting three
to fix the characteristic polynomial of $\CX$, we find that $\M'$
has complex dimension four and hyper-Kahler dimension two.  This
agrees with what we would expect from the brane picture in fig.
\ref{fig11}a, and shows that $\M'$ cannot coincide with $\M$.

In case (3), on the interval $0\leq y\leq y_1$, we solve  Nahm's
equations with $1\times 1$ matrices. In particular, in the usual
gauge, $\CX$ is simply a complex constant $e$.  Upon crossing
$y=y_1$, two new rows and columns appear, and there is a pole at
$y=y_1$ in the new $2\times 2$ block.  The general allowed form is
\begin{equation}\label{zurcase}\CX=
\begin{pmatrix} a & (y-y_1)^{-1} & 0 \\
b(y-y_1) & a    & c(y-y_1)^{1/2}\\
d(y-y_1)^{1/2} & 0 & e\end{pmatrix}.\end{equation} The parameters
correspond to those in (\ref{gurcase}) in an obvious way.

So the moduli space of vacua derived from the brane picture of
fig. \ref{fig11}b agrees with the one associated with the
embedding $\rho:\frak{su}(2)\to \frak{u}(3)$.  This makes it
reasonable to expect that for $y_1\to 0$, the model derived from
the brane picture converges to $\N=4$ super Yang-Mills theory with
the superconformal boundary condition derived from $\rho$.  This
will be our starting point elsewhere in studying duality.

\subsection{Moduli Space Of Vacua With More General Boundary
Conditions}\label{nahmmore}

Most of what we have done so far is to analyze Nahm's equations in
the presence of a boundary condition associated to a homomorphism
$\rho:\frak{su}(2)\to\frak g$.  The moduli space $\M_\rho$ (or
$\cal M_\rho(\vec X_\infty)$) has turned out to be a Slodowy slice
transverse to the raising operator $t_+$ associated to $\rho$,
intersected with the closure of an orbit. A group $F$ that is the
commutant of $\rho$ in $G$ acts as a group of symmetries of
$\M_\rho$. The hyper-Kahler moment map for the action of $F$ is
\begin{equation}\label{tolgor}\vec\mu=\vec
X(0)_{\f},\end{equation} that is, the projection of $\vec X(0)$ to
$\f$.  The derivation of this statement is exactly the same as the
derivation of eqn. (\ref{hypgac}), except that here we consider
only gauge transformations that are $F$-valued at $y=0$, so we
project the formula to $\f$.

In section \ref{combining}, we described more general
supersymmetric boundary conditions in which, roughly speaking,
after picking $\rho$, we gauge a subgroup $H$ of $F$ and couple it
to boundary degrees of freedom with $H$ symmetry.  Our next goal
is to describe the moduli space of supersymmetric vacua in this
more general context.

First we describe the effect of gauging $H$ without adding
boundary variables.  Away from $y=0$, supersymmetry still requires
that $\vec X$ should obey Nahm's equations. According to
(\ref{newpars}), the boundary condition requires $\vec X(0)^+=\vec
v$, where $\vec v$ is a triple of elements of the center of $\frak
h$. (We omit the $\vec\mu^Z$ term as we are not yet including
boundary variables.)  We also must divide by the action of $H$,
since this is now part of the gauge group. As the gauge symmetry
has been reduced to $H$ at the boundary, $\vec X(0)^+$ is just the
projection of $\vec X(0)$ from $\f$ to $\frak h$. So, according to
(\ref{tolgor}), $\vec X^+(0)$ is a moment map for the action of
$H$ on $\M_\rho$.  But as usual, we are free to add central
elements to the moment map, and so $\vec X^+(0)-\vec v$ is an
equally good moment map. The combined operation of setting $\vec
X^+(0)=\vec v$ and dividing by $H$ is therefore a hyper-Kahler
quotient. Thus the moduli space $\M_{\rho,H}$ of solutions of
Nahm's equations for the boundary condition associated to a
general $H$ is the same as the hyper-Kahler quotient by $H$ of
$\M_\rho$:
\begin{equation}\label{kiops}\M_{\rho,H}=\M_\rho/\neg/\neg/
H.\end{equation} The hyper-Kahler quotient is taken for a
specified value of the FI constants $\vec v$.

\def\H{\mathcal H}
There is no problem to add a boundary theory $B$ with $H$ action.
$B$ has its own moduli space of vacua, say $\H$, also a
hyper-Kahler manifold with $H$ action. If $\vec\mu_B$ is the
moment map for the action of $H$ on $\H$, then the incorporation
of the boundary variables has the effect of replacing the boundary
condition $\vec X(0)^+=0$ by $\vec X(0)^++\vec \mu_B=0$.  This was
 demonstrated  in eqn. (\ref{ympap}).

So when boundary variables are added, we construct the moduli
space of vacua by beginning with $\M_\rho\times \H$, imposing the
boundary condition $\vec X(0)^++\vec \mu_B=\vec v$, and dividing
by $H$. On the other hand, $\vec X(0)^++\vec \mu_B-\vec v$ is a
moment map for the action of $H$ on $\M_\rho\times \H$, so setting
this to zero and dividing by $H$ amounts to a hyper-Kahler
quotient. The moduli space of vacua of the combined system is
therefore
\begin{equation}\label{comvac}\M_{\rho,H,B}=(\M_\rho(\vec X_\infty)\times
\H)/\neg/\neg/ H.\end{equation}

\subsection{Including $\vec Y_\infty$}\label{incy}

We have written explicitly the dependence of $\M_\rho$ on $\vec
X_\infty$ in eqn. (\ref{comvac}) to emphasize that this analysis
does incorporate the value of $\vec X$ at infinity.  However,
throughout this section, we have taken $\vec Y_\infty=0$.  It is
now time to incorporate $\vec Y_\infty$.  In doing so, we assume
that $\vec X_\infty$ and $\vec Y_\infty$ taken together are
regular and break $G$ to a maximal torus $T$.

Supersymmetry requires that in bulk  $\vec Y$ must obey
(\ref{ildo})
\begin{equation}\label{zildox}\frac{D\vec Y}{Dy}=[\vec Y,\vec
Y]=[\vec Y,\vec X]=0,\end{equation} and thus, the components of
$\vec Y$ generate symmetries of the solution of Nahm's equations. We
must supplement this with additional information associated with the
boundary conditions.

As summarized in section \ref{combining}, the most general
half-BPS boundary conditions depends on the choice of a triple
$(\rho,H,B)$, where $\rho:\frak{su}(2)\to \frak g$ is a
homomorphism, $H$ is a subgroup of $G$ that commutes with $\rho$,
and $B$ is a boundary theory with $H$ symmetry.  To explain the
main points most directly, we first suppose that $\rho$ and $B$
are trivial.

The boundary condition on $\vec Y$ was described in eqn.
(\ref{newpars}).  We decompose $\frak g=\frak g^+\oplus \frak
g^-$, where $\frak g^+=\frak h$ and $\frak g^-$ is the
orthocomplement.  Then we pick elements $\vec w\in \frak g^-$ that
commute with each other and with $\frak h$, and require
\begin{equation}\label{oko}\vec Y^-(0)=\vec w.\end{equation}
Equivalently, we require
\begin{equation}\label{zokio}\vec Y(0)=\vec w~{\rm mod}~\frak
h.\end{equation}

This equation plus the covariant constancy of $\vec Y$, which is
part of eqn. (\ref{zildox}), says that $\vec Y_\infty$ must be
conjugate to $\vec w$ mod $\frak h$.  If this is not the case,
then the moduli space of supersymmetric vacua is empty.  For most
choices of $H$, that is the situation for a generic choice of
$\vec Y_\infty$ and $\vec w$.  For example, suppose that $H$ is
trivial and $\vec w=0$. Then the condition is that $\vec Y_\infty$
must be conjugate to 0, that is, it must vanish.  Otherwise, there
are no supersymmetric vacua.

If $\vec Y_\infty$ is conjugate to $\vec w$, this may be so in
inequivalent ways. For example, suppose that $G=SU(4)$ and
$H=SU(2)$, consisting of matrices of the form
\begin{equation}\label{pkio}\begin{pmatrix} 1 & 0 & 0 & 0\\
                                            0 & 1 & 0 & 0 \\
                                            0 & 0 & * & * \\
                                            0 & 0 & * & *
\end{pmatrix}.\end{equation}
Let $\vec w={\rm diag}(\vec a,-\vec a,0,0)$ (so $\vec w\in \frak
g^-$, and its components commute with each other and with $H$),
$\vec X={\rm diag}(\vec x_1,\vec x_2,\vec x_3,\vec x_4)$, $\vec
Y={\rm diag}(\vec y_1,\vec y_2,\vec y_3,\vec y_4)$. Conjugating
$\vec Y$ to equal $\vec w$ mod $\frak h$ means finding $i$ and $j$
such that $\vec y_i=\vec a$ and $\vec y_j=-\vec a$. Generically
this cannot be done (and the moduli space of supersymmetric vacua
is empty), but it can happen that there is more than one way to do
this. (This occurs if the $\vec y_i$ are pairwise equal, but $\vec
x_i$ is generic enough that the collection $\vec X,\vec Y$ is
regular.) When that is the case, each choice leads potentially to
a component of the moduli space of vacua.

Making a particular choice of how to conjugate $\vec Y_\infty$ to
be in the form \begin{equation}\label{gafter}\vec Y_\infty = \vec
w+h,\end{equation} with $h\in H$, let us describe the associated
component of the moduli space. It is convenient to think of $H$ as
a fixed subgroup of $G$, and $\vec w$ and $\vec Y_\infty$ as fixed
elements of $\frak g$, rather than all this being given up to
conjugacy.

Now in solving Nahm's equations, we have worked in a gauge with
$A_3=0$. Since $\vec Y$ is covariantly constant, it is actually
constant in this gauge.  According to (\ref{zildox}), the solution
of Nahm's equations takes values in the subgroup $G_{\vec Y}$ of
$G$ that commutes with $\vec Y$.  This has the important
consequence that for all $y$, $\vec X^+(y)$ takes values in
$H_{\vec Y}$, the subgroup of $H$ that commutes with $\vec Y$.
This is so even before we specialize to the boundary, $y=0$.

In section \ref{nahmmore}, we found that the effect of having $H$
non-trivial is that, after constructing the moduli space of
solutions of Nahm's equations, we must take the hyper-Kahler
quotient by the action of $H$.  There are two related reasons that
this is not the right thing to do when $\vec Y_\infty\not=0$.

First, a key part of the hyper-Kahler quotient was to set to zero
the hyper-Kahler moment map $\vec X^+(0)-\vec v$. (We recall that
$\vec v$ are constants valued in the center of $\frak h$.) In the
present context,  part of $\vec X^+(y)$ already vanishes before
setting $y=0$, namely the part that does not commute with $\vec
Y$.  It only makes sense at the boundary to add a condition on the
projection of $\vec X^+$ to $\frak h_{\vec Y}$ (the centralizer of
$\vec Y$ in $\frak h$). So we may as well regard the constraint
$\vec X^+(0)-\vec v=0$ as an equation in $\frak h_{\vec Y}$, the
Lie algebra of $H_{\vec Y}$. This is the moment map for the action
of $H_{\vec Y}$, not the action of $H$.

Second, the other part of the hyper-Kahler quotient is to divide
by $H$.  But, with $\vec Y\not=0$ and equal to a specified
constant, there is no $H$ sysmmetry, so we cannot divide by $H$.
We can divide only by $H_{\vec Y}$.

The conclusion from each of the last two paragraphs is that what
we want is a hyper-Kahler quotient by $H_{\vec Y}$.  After making
a choice $\alpha$ of how to put $\vec Y_\infty$ in the form
(\ref{gafter}), we should  construct the corresponding moduli
space $\M^\alpha_{\vec Y}(\vec X_\infty)$ of $G_{\vec Y}$-valued
solutions of Nahm's equations in which $\vec X$ is conjugate at
infinity (by an element of $G_{\vec Y}$) to $\vec X_\infty$, and
take its hyper-Kahler quotient by $H_{\vec Y}$. After summing over
$\alpha$, we get the moduli space of vacua:
\begin{equation}\label{gofoer} \M_{H,\vec Y}=\bigcup_\alpha \M^\alpha_{\vec
Y}/\neg/\neg/H_{\vec Y}.\end{equation}

This discussion is not changed in an essential way by including a
homomorphism $\rho:\frak{su}(2)\to\frak g$ or a boundary CFT with
$H$ symmetry.  The boundary CFT just gives another factor $\H$
(its moduli space of vacua) that must be included in the
hyper-Kahler quotient. The effect of $\rho$ is just that, as
usual, in solving Nahm's equations we must require $\vec X(y)$ to
have a pole at $y=0$.  Eqn. (\ref{gafter}) in any case implies
that $\vec Y$ commutes with $\rho$.

\subsection{Duality: First Steps}\label{firststeps}

Though we defer a serious study of how duality acts on half-BPS
boundary conditions to a subsequent paper, we make here some
preliminary observations that may help place in context some of
the constructions that we have described.

A basic question is to ask what is the $S$-dual of the simplest
Neumann boundary conditions, described in section \ref{boundcon}.
With these boundary conditions, the vacuum is uniquely determined
if one specifies the value of $\vec X$ at infinity.  For fixed
$\vec X_\infty$, the moduli space of supersymmetric vacua consists
of only a single point.

The duality transformation $S:\tau\to -1/\tau$ transforms the
unbroken supersymmetries of ${\cal N}=4$ super Yang-Mills in a
non-trivial fashion, which is described for example in eqn. (2.25)
of \cite{KW}.  For simplicity, let us specialize to the case that
$\tau$ is imaginary (or in other words the case that the $\theta$
angle vanishes).  The transformation of the unbroken supersymmetries
is then
\begin{equation}\label{transfor} \varepsilon\to
\frac{1-\Gamma_{0123}}{\sqrt 2}\varepsilon,\end{equation} and this
has the effect of exchanging the supersymmetry preserved with
Neumann boundary conditions with the supersymmetry preserved by
Dirichlet conditions.  (The generalization of (\ref{transfor}) to
$\theta\not=0$ is given in eqn. (\ref{kurlo}).)

One might think that the dual of Neumann boundary conditions would
be Dirichlet boundary conditions.  This, however, cannot be the
case, because Dirichlet boundary conditions lead to a non-trivial
moduli space of solutions of Nahm's equations, which has no analog
for the Neumann case.  The dual of Neumann boundary conditions must
be a boundary condition which preserves the same supersymmetry as
Dirichlet, but which does not lead to a non-trivial moduli space.

We have learned that for each choice of $\rho:
\frak{su}(2)\to\frak{g}$, we can generalize Dirichlet boundary
conditions to a more general boundary condition that preserves the
same supersymmetry.  The resulting moduli space of solutions of
Nahm's equations is a Slodowy slice associated to $\rho$, and is
trivial if and only if $\rho$ is the principal $\frak{su}(2)$
embedding.  So this, rather than naive Dirichlet boundary
conditions, is the natural candidate for the $S$-dual of Neumann
boundary conditions.

For the case of $G=U(N)$, this proposal can be confirmed by
considering the D3-NS5 and D3-D5 systems.\footnote{For the other
classical groups $SO(N)$ and $Sp(N)$, a similar argument can be
given by combining the branes with an orientifold threeplane.} For
$N$ D3-branes ending on an NS5-brane, we get $U(N)$ gauge theory
with Neumann boundary conditions.  The $S$-dual consists of $N$
D3-branes ending on a D5-brane.   As we have learned, this
corresponds not to naive Dirichlet boundary conditions but to
Dirichlet boundary conditions modified with the principal
embedding $\rho:\frak{su}(2)\to\frak{g}$.  (This fact seems to
underlie many occurrences of the principal $\frak{su}(2)$
embedding in the geometric Langlands program.)

A converse question is to ask what is the $S$-dual of ordinary
Dirichlet boundary conditions (with $\rho=0$).  The answer
involves Neumann boundary conditions modified by coupling to a
certain boundary superconformal field theory.  To elucidate the
nature of this theory, and to answer analogous questions for other
boundary conditions described in section \ref{boundcon}, will be
our goal in a separate paper.

\subsection{Other Moduli Spaces Of Solutions Of Nahm's
Equations}\label{othernahm}

\def\CG{\mathcal G}
In our study of Nahm's equations so far, the goal has been to
describe the moduli space of vacua of gauge theory on a half-space
$y\geq 0$, with BPS boundary conditions at $y=0$ and specified
values of $\vec X$ and $\vec Y$ at $y=\infty$. Here we will briefly
describe some other related spaces of solutions of Nahm's equations.
(Apart from their intrinsic interest, these are relevant to a more
detailed study of $S$-duality of boundary conditions that will
appear elsewhere.)

\subsubsection{Hyper-Kahler Analog Of A Lie Group}\label{hypan}

The most basic of these \cite{Kronheimer} is the moduli space of
solutions of Nahm's equations on a finite interval $0\leq y\leq
\ell$.  We consider the gauge-invariant form of Nahm's equations
\begin{equation}\label{onruf}\frac{D\vec X}{Dy}+\vec X\times \vec
X=0\end{equation} for a pair $\vec X,A$, modulo gauge
transformations that equal 1 at both $y=0$ and $y=\ell$.  By the
usual reasoning, the moduli space, which we will call $\CG_\ell$, is
a hyper-Kahler manifold. (A simple scaling argument shows that the
$\ell$-dependence of the hyper-Kahler metric of $\CG_\ell$ is a
simple factor of $1/\ell$. The same is true of the generalizations
introduced below.) Moreover, the group $G\times G$ acts on
$\CG_\ell$. One copy of $G$ acts by gauge transformations at $y=0$
and the second copy acts by gauge transformations at $y=\ell$. We
write $G_L$ and $G_R$ for $G$ acting on the left or right, that is
at $y=0$ or at $y=\ell$.

We can calculate the moment map $\vec\mu_L$ and $\vec\mu_R$ for the
left and right action of $G$ as in eqn. (\ref{hypgac}), leading to
\begin{align}\label{ygac}\notag \vec\mu_L& =\vec X(0) \\ \vec\mu_R&
=-\vec X(\ell).\end{align} (The minus sign in the second line
comes in integration by parts.) As an example of the use of this
formula, let us compute the hyper-Kahler quotient
$\CG/\neg/\neg/G_R$.  We do this by setting $\vec\mu_R=0$ and
dividing by $G_R$.  Since $\vec\mu_R=\vec X(\ell)$ and Nahm's
equations are of first order in $\vec X$, a solution with
$\vec\mu_R=0$ has $\vec X$ identically zero.  Dividing by $G_R$,
after already dividing by gauge transformations that are 1 at
$y=0,\ell$, means that we divide by all gauge transformations that
are 1 at $y=0$.  This enables us to set $A=0$ in a unique fashion.
So the hyper-Kahler quotient of $\CG$ by $G_R$ -- or likewise its
hyper-Kahler quotient by $G_L$ -- is a single point.

As a second example, pick two positive numbers $\ell$ and $\ell'$,
and consider the group $G$ acting on the right on $\CG_\ell$ and
on the left on $\CG_{\ell'}$.  We claim that the hyper-Kahler
quotient, which we abbreviate as $\CG_\ell\times_G\CG_{\ell'}$, is
simply $\CG_{\ell+\ell'}$.  To get this result, we think of
$\CG_\ell$ as the moduli space of pairs $\vec X,A$ that obey
Nahm's equations on the interval $[0,\ell]$, and $\CG_{\ell'}$ as
the moduli space of pairs $\vec X',A'$ that obey Nahm's equations
on the interval $[\ell,\ell+\ell']$.  In each case we divide by
gauge transformations that are 1 on the boundary of the interval.
To compute the hyper-Kahler quotient by the diagonal product of
the right action of $G$ on the first factor and the left action on
the second factor, we set to zero the moment map, which is
$\hat\mu=-\vec X(\ell)+\vec X'(\ell)$, and then divide by gauge
transformations acting on all fields at $y=L$.   Once we set
$\hat\mu=0$ and divide by gauge transformations at $y=L$, the
quantities $\vec X, A$ and $\vec X',A'$ fit together to a single
solution of Nahm's equations on the full interval
$[0,\ell+\ell']$, modulo gauge transformations that are 1 on the
boundary.  Hence
\begin{equation}\label{corny}\CG_\ell\times_G\CG_{\ell'}=\CG_{\ell+\ell'}.\end{equation}

In any one of its complex structures, $\CG_\ell$ is isomorphic to
$T^*G_\C$, the cotangent bundle of the complex Lie group $G_\C$  (in
particular, as a complex manifold, it is independent of $\ell$). To
see this, as usual we introduce the variables $\CX=X_1+iX_2$ and
$\CA=A+iX_3$. In one of its complex structures, $\CG_\ell$ is
equivalent to the moduli space of solutions of the complex Nahm
equation
\begin{equation}\label{zorno}\frac{{\cal D}\CX}{{\cal
D}y}=0\end{equation} modulo complex-valued gauge transformations
that equal 1 at $y=0,\ell$.    The gauge-invariant data
characterizing this solution is $\CX(0)$ and the ``Wilson line'' or
holonomy
\begin{equation}\label{worno} g =P\exp\left(-\int_0^\ell
\CA\right).\end{equation} (We need not include $\CX(\ell)$ since the
complex Nahm equation implies that it coincides with
$g\CX(0)g^{-1}$.)  Here $g$ takes values in $G_\C$ and we can
consider $\CX(0)$ to take values in the cotangent bundle of $G_\C$
at the point $g$.  They are subject to no additional restrictions,
so we can identify $\CG_\ell$ holomorphically, in any one of its
complex structures, with $T^*G_\C$.

The subgroup of $G\times G$ leaving fixed a given point in $\CG$
is always a subgroup of $G$, with a diagonal embedding in $G\times
G$. In fact, a symmetry of a solution of Nahm's equations must be
generated by a covariantly constant gauge parameter, which is
determined by its restriction to $y=0$. Any solution with the full
$G$ symmetry has $\vec X=0$.  The gauge-invariant data contained
in the solution is then the $G$-valued holonomy
$P\exp(-\int_0^\ell A)$. This may be any element of $G$, so
solutions with $\vec X=0$ furnish a copy of $G$ embedded in
$\CG_\ell$, and these are the solutions for which the unbroken
symmetry is maximal. By identifying a solution of Nahm's equations
with the initial values $\vec X(0)$ plus the holonomy
$P\exp\left(-\int_0^\ell A\right)$, one can show that as a
manifold with $G\times G$ action, $\CG$ is equivalent to $G\times
\frak g^3$, with a natural action of $G\times G$.

\subsubsection{Hyper-Kahler Analog Of A Homogeneous
Space}\label{hypanhom}

\def\CT{{\mathcal T}}
The space $\CG_\ell$ is in a sense the hyper-Kahler analog of a
Lie group.  There is also \cite{Biel} a hyper-Kahler analog of a
homogeneous space. For this, we  solve Nahm's equations on the
interval $[0,\ell)$, modulo gauge transformations that are 1 on
the boundary, but now we pick a homomorphism $\rho:\frak{su}(2)\to
\frak g$ and we require that $\vec X$ should have a pole of type
$\rho$ at $y=\ell$:
\begin{equation}\label{terf}\vec X\sim \frac{\vec
t}{y-\ell}.\end{equation} (As usual, $\vec t$ is the image under
$\rho$ of a standard set of $\frak{su}(2)$ generators.) We call
the moduli space $\CT^\rho_\ell$. It is a hyper-Kahler manifold,
as usual. The group $G$ acts by gauge transformations at $y=0$,
with moment map
\begin{equation}\label{oterf} \vec \mu =\vec X(0).\end{equation}
The hyper-Kahler quotient $\CT^\rho_\ell/\neg/\neg/G$ is empty,
because (for non-zero $\rho$) it is impossible to solve Nahm's
equations with the initial condition $\vec X(0)=0$ and with the
polar behavior (\ref{terf}) at $y=\ell$.

We denote as $\CG_\ell\times_G\CT^\rho_{\ell'}$ the hyper-Kahler
quotient of $\CG_\ell\times\CT^\rho_{\ell'}$ by $G$, with $G$ acting
on the right on the first factor and as just stated on the second
factor.  The same steps that led to (\ref{corny}) give
\begin{equation}\label{borny}\CG_\ell\times_G\CT^\rho_{\ell'}=\CT^\rho_{\ell+\ell'}.\end{equation}

If $\rho$ has a nontrivial centralizer $H\subset G$, then
$\CT^\rho_\ell$ admits an action of $H$, by gauge transformations
at $y=\ell$, commuting with the action of $G$.  The moment map is
$\vec\mu_H=\vec X(\ell)_{\frak h}$, where $\vec X(\ell)_{\frak h}$
is the projection of $\vec X(\ell)$ to $\frak h$.  The unbroken
subgroup of $G\times H$ at any point in $\CT^\rho_\ell$ is a
subgroup of $H$, with a diagonal embedding in $H\times H\subset
G\times H$. To see this, observe that a symmetry that leaves fixed
a given solution of Nahm's equations is generated by a gauge
parameter that is covariantly constant, and whose restriction to
$y=\ell$ must commute with the Nahm pole.  A solution whose
unbroken symmetry is actually $H$ can be obtained by setting $A=0$
and $\vec X=\vec t/(y-\ell)$.

Just like $\CG_\ell$, $\CT^\rho_\ell$ can be described explicitly as
a complex manifold in any one of its complex structures.  It is
parametrized by a pair $(g,\eta)$, where $g\in G_\C$ and $\eta\in
\frak g$ is a lowest weight vector with respect to $\rho$ (in other
words, $[\rho(t_-),\eta]=0$).  We cannot quite define $g$ as the
holonomy operator (\ref{worno}); this holonomy does not converge,
since $\CA$ has a pole at $y=\ell$.  Instead, we define $g$ as a
regularized version of the holonomy:
\begin{equation}\label{vexo} g=\lim_{\delta\to
0}\left[(-\delta)^{it_3}P\exp\left(-\int_0^{\ell-\delta}\CA\right)\right].\end{equation}
Similarly, $\eta$ is defined as a regularized version of
$\CX(\ell)$.  (It is simpler to use $\CX(\ell)$ rather than
$\CX(0)$, since the conditions that it obeys are more simply
stated.) In a gauge in which $\CA=it_3/(y-\ell)$ in a neighborhood
of $y=\ell$, the solution for $\CX$ is given essentially by
(\ref{ime}):
\begin{equation}\label{zimea}\CX(y)=\frac{t_+}{y-\ell}+\sum_{\alpha}
\epsilon_\alpha v_\alpha (y-\ell)^{-m_\alpha},\end{equation} where
$\epsilon_\alpha$ are complex constants and the $v_\alpha$ are  a
basis of lowest weight vectors with $[it_3,v_\alpha]=m_\alpha
v_\alpha$, $m_\alpha\leq 0$.  So we define
\begin{equation}\label{rime}\eta=(-\delta)^{it_3}\bigl(\CX(\ell-\delta)+t_+/\delta\bigr)
(-\delta)^{-it_3},\end{equation} which is independent of $\delta$
for small $\delta$.

\def\CS{{\mathcal S}}
We can go one step farther and define $\CS^{\rho,\rho'}_\ell$ to be
the moduli space of solutions of Nahm's equations on the interval
$[0,\ell]$ with poles of type $\rho$ and $\rho'$, respectively, at
the two endpoints.  (These are the boundary conditions used by Nahm
in the original work \cite{Nahm:1979yw} relating Nahm's equations to
BPS monopoles.) This more general moduli space can be constructed
from the ones that we have already considered as a hyper-Kahler
quotient:
\begin{equation}\label{zext}\CS^{\rho,\rho'}_{\ell+\ell'}=\left(\CT^\rho_\ell\times\CT^{\rho'}_{\ell'}\right)
/\neg/\neg/G.\end{equation} This can be shown by following the
derivation of (\ref{corny}).

\subsubsection{Including An NS5-Brane}\label{incvice}

Our last topic is to consider what happens to Nahm's equations in
the presence of an NS5-brane.\footnote{To preserve the same
supersymmetry as that of D3-branes that span directions 0123 and
D5-branes that span directions 012456, the NS5-brane should span
directions 012789.}

We suppose that the NS5-brane is located at $y=0$.  We assume that
there are $n$ D3-branes ending on this NS5-brane on its left, and
$m$ on its right.  The low energy physics is well known.  For $y<0$,
there is a $U(n)$ gauge theory with $\N=4$ supersymmetry.  For
$y>0$, the gauge group is $U(m)$.  At $y=0$, there is a
bifundamental hypermultiplet of $U(n)\times U(m)$.  We write $Z$ for
the space parametrized by the bifundamental hypermultiplet and
$\vec\mu_L^Z$, $\vec\mu^Z_R$ for the moment maps for the action on
$Z$ of $U(n)$ and $U(m)$, respectively.

Similarly, we write $\vec X_L$ and $\vec X_R$ for the fields $\vec
X$ for $y<0$ and $y>0$, respectively.  Like $\vec\mu_L^Z$ and
$\vec\mu_R^Z$, they take values  in the adjoint representations of
$U(n)$ and $U(m)$, respectively.   In a supersymmetric
configuration, $\vec X_L$ and $\vec X_R$ must obey Nahm's equations
away from $y=0$.  The appropriate boundary conditions at $y=0$ are
special cases of (\ref{ympap}):
\begin{align}\label{zympap}\notag -\vec X_L(0)+\vec\mu_L & = 0\\
                             \vec X_R(0)+\vec\mu_R & = 0.\end{align}
(The minus sign in the first line comes from integrating by parts in
determining the boundary contribution to the moment maps.)

To get some insight, we look at the space of solutions of Nahm's
equations as a complex manifold in one of its complex structures. We
introduce $\CX_L=X_{L,1+i2}$, $\CX_R=X_{R,1+i2}$.  Also, from the
point of view of one complex structure, the bifundamental
hypermultiplet is equivalent to a pair $A,B$ where $A$ is an
$n\times m$ matrix and $B$ is an $m\times n$ matrix.  The complex
moment maps are $\mu_{\C,L}=AB$, $\mu_{\C,R}=-BA$, and the boundary
conditions are therefore
\begin{align}\label{gympap}\notag \CX_L(0)& = AB\\
                                  \CX_R(0)&=BA. \end{align}
Of course, Nahm's equations imply that $\CX_L(y)$ and $\CX_R(y)$ are
conjugate for all $y$ to $\CX_L(0)$ and $\CX_R(0)$.

 It follows from (\ref{gympap}) that the nonzero eigenvalues of $\CX_L$ and
$\CX_R$ are the same.  If $n>m$, then $\CX_L$ is at most of rank
$n$. If $n=m$, then $\CX_L$ and $\CX_R$ have the same characteristic
polynomials and are conjugate if they are regular, but in general
not otherwise.

We briefly conclude with some examples (which will be useful
elsewhere). For $n=2$ and $m=1$, eqn. (\ref{gympap}) says that
$\CX_L(0)$ can be any $2\times 2$ matrix of rank 1, and that
$\CX_R(0)=\Tr\,\CX_L(0)$. Now let us embed this problem in a
larger one.  We assume that we want to solve Nahm's equations on
the interval $(-\ell,\ell]$, by $2\times 2$ matrices for $y<0$,
$1\times 1$ matrices for $y>0$, and with an NS5-brane at $y=0$.
Also, let us ask for a regular Nahm pole at $y=-\ell$ and
Dirichlet boundary conditions at $y=\ell$.  All nonzero conjugacy
classes arise in the Slodowy slice transverse to a regular Nahm
pole, so these boundary conditions allow $\CX_L(0)$ to be any
nonzero matrix, and in particular any matrix of rank 1.  Thus, a
solution with the indicated boundary conditions does exist.  Since
$\CX_L(0)$ must be nonzero, the hypermultiplets $A$ and $B$ are
likewise nonzero. The group $U(1)$ acts on the space of solutions,
by gauge transformations at $y=\ell$. Because $A$ and $B$ are
nonzero, a solution with these boundary conditions cannot be
$U(1)$-invariant.

Finally, let us consider $n=m=2$, with the same boundary
conditions on the interval $(-\ell,\ell]$, still with a regular
Nahm pole at $y=-\ell$, Dirichlet boundary conditions at $y=\ell$,
and a fivebrane at $y=0$. The group that acts on the space of
solutions by gauge transformations at $y=\ell$ is now $G=U(2)$.
It is possible to find a solution with these boundary conditions
that is invariant under a non-central subgroup of $U(2)$
consisting of matrices of the form ${\rm diag}(1,*)$.  To do this,
simply embed the $m=1$ solution of the last paragraph in the $m=2$
problem.

\def\Bbb{\mathbb}
\def\Tr{{\rm Tr}}
\def\16{{\bf 16}}
\def\1{{\bf 1}}
\def\2{{\bf 2}}
\def\3{{\bf 3}}
\def\4{{\bf 4}}
\def\bar{\overline}
\def\tilde{\widetilde}
\def\R{{\Bbb{R}}}\def\Z{{\Bbb{Z}}}
\def\N{{\mathcal N}}
\def\hat{\widehat}

\section{Supersymmetry Without Lorentz Invariance}\label{without}

What were described in section \ref{boundcon} were Lorentz-invariant
half-BPS boundary conditions.  Here we will discuss what happens if
the requirement of Lorentz invariance is dropped.  By the Lorentz
group in this context we mean $SO(1,2)$, the group of Lorentz
transformations that act trivially on $y=x^3$.  As we will see, it
is possible to break Lorentz invariance but still preserve eight
supersymmetries.

Though it is possible to explain this purely in field theory, and we
will do so, we will introduce the subject by describing brane
constructions that give significant examples. We will simply deform
the usual D3-D5 and D3-NS5 systems by turning on a flux on the
fivebrane, in a way that ``rotates'' the unbroken supersymmetries
while preserving their number.  The deformation breaks both the
$SO(1,2)$ Lorentz symmetry and the $SU(2)_X$ $R$-symmetry, but
leaves $SU(2)_Y$ and translation symmetry in the $012$ directions.

\subsection{Deforming The D3-D5 System}\label{deformdthree}
\def\CMF{\cmmib F}
We start with the D3-D5 system in Type IIB superstring theory. This
theory in $\R^{1,9}$ has 32 supersymmetries, consisting of two
copies $\varepsilon_L$ and $\varepsilon_R$ of the $\bf {16}$ of
$SO(1,9)$.  Here $\varepsilon_L$ and $\varepsilon_R$ arise
respectively from left- and right-moving excitations on the string
worldsheet.  Now as usual we introduce four-dimensional $U(N)$ gauge
theory by considering $N$ D3-branes with worldvolume extending in
the directions 0123.  Half of the supersymmetry is broken; the
unbroken supersymmetries obey
\begin{equation}\label{d3d5} \varepsilon_R =
\Gamma_{0123}\varepsilon_L=-B_0\varepsilon_L.\end{equation} (The
$B_i$ were defined in (\ref{gomot}).) Then we introduce a D5-brane
extending in directions 012456. This again reduces the
supersymmetry by a factor of two; in the absence of any flux, the
unbroken supersymmetries obey
$\varepsilon_R=\Gamma_{012456}\varepsilon_L=-B_0B_1\varepsilon_L$.

The D5-brane supports a $U(1)$ gauge field, whose curvature we
will call  $\cmmib F$ and measure in string units. We take $\cmmib
F$ to be a two-form with constant coefficients on the D5-brane
worldvolume, preserving translation invariance but breaking
Lorentz invariance. The condition for unbroken supersymmetry due
to the presence of the D5-brane is deformed to
\begin{equation}\label{d5}\varepsilon_R=-\exp(\Gamma^{IJ}\CMF_{IJ}/4)B_0B_1
\varepsilon_L.\end{equation}

Generically, the two conditions (\ref{d3d5}) and (\ref{d5}) are
inconsistent and there are no unbroken supersymmetries. Indeed, we
can combine the two equations into
\begin{align}\label{jutz} \varepsilon_L &=-B_0wB_0B_1\varepsilon_L
\end{align}
where $w=\exp(\Gamma^{IJ}\CMF_{IJ}/4)$.  We can think of
$W=-B_0wB_0B_1$ as an element of $Spin(1,6)$, the double cover of
$SO(1,6)$, the Lorentz group acting on directions 0123456.  For
generic $\CMF$, 1 is not an eigenvalue of $W$ (acting on spinors)
and there are no unbroken supersymmetries.  For $\CMF=0$, $W=B_1$,
which is the lift to ${\rm Spin}(1,6)$ of the $SO(1,6)$ element
${\rm diag}(1,1,1,-1,-1,-1,-1)$. Eqn. (\ref{jutz}) then has an
eight-dimensional space of solutions.  In general, for $W$ to
preserve one-half of the supersymmetries, it must belong to an
$SU(2)$ subgroup of ${\rm Spin}(1,6)$ (embedded via $SU(2)\subset
SU(2)\times SU(2)={\rm Spin}(4)\subset {\rm Spin}(1,6)$).  On the
other hand, the explicit form of $W$ shows that it anticommutes
with $\Gamma_3$, so one of its eigenvalues in the $\bf 7$ of
$SO(1,6)$ is $-1$.  So $W$ must be conjugate to ${\rm
diag}(1,1,1,-1,-1,-1,-1)$.  In particular, $W^2=1$, which is
equivalent to
\begin{equation}\label{zoff}B_1wB_1=w^{-1}.\end{equation} That
equation holds if
\begin{equation}\label{toff}B_1\CMF =-\CMF B_1.\end{equation}
Conversely, (\ref{toff}) gives a component of solutions of
(\ref{zoff}), namely the component of solutions that come by
deformation from $\CMF=0$.  We will only consider this component.

For $\CMF$ of the form found in the last paragraph, $B_0 w
=w^{-1}B_0$, so the condition (\ref{jutz}) for unbroken
supersymmetry simplifies to $\varepsilon_L=
w^{-1}B_1\varepsilon_L$. If we introduce a natural square root of
$w$ by
\begin{equation}\label{dolf} h
=\exp(\Gamma^{IJ}\CMF_{IJ}/8),\end{equation} the condition becomes
\begin{equation} \label{defol}h^{-1} B_1 h \varepsilon_L =
\varepsilon_L
\end{equation} The matrix $h$ is an element of $SO(1,5)$ (acting on directions
012456), and the eight-dimensional subspace of unbroken
supercharges is conjugate under $h^{-1}$ to the standard space of
unbroken supersymmetries of the D3-D5 system.

\subsubsection{Field Theory Interpretation}

We will now reinterpret purely in field theory terms the brane
construction just described.  One advantage of this is that the
field theory description is valid for arbitrary gauge group, not
just for gauge groups (such as $U(N)$) that are conveniently
realized by branes.

So far we could be considering intersecting branes or D3-branes
ending on a D5-brane.  We now focus on that latter case.

At $\CMF=0$, the appropriate boundary condition for $N$ D3-branes
ending on a D5-brane was described in section \ref{onesided}. $\vec
Y$ obeys Dirichlet boundary conditions, as do the three-dimensional
gauge fields $A_\mu$.  However, the scalar fields $X^a$ do not obey
simple Dirichlet or Neumann boundary conditions, but have a pole at
the boundary:
\begin{equation}\label{apole}X^a\sim\frac{t^a}{y} +\dots.
\end{equation}
Here $y=x^3$ vanishes at the boundary and $t^a$ are the images of
standard $\frak{su}(2)$ generators for a principal embedding
$\rho:\frak{su}(2) \to \frak{g}$. The unbroken supersymmetries obey
\begin{equation}\label{kiop}B_1\varepsilon=\varepsilon.\end{equation}

When we turn on $\CMF$, the D5-brane is still located at $\vec
Y=0$, so there is no change in the Dirichlet boundary conditions
for $\vec Y$. However, the boundary conditions on $A_\mu$ and
$\vec X$ do change.

The boundary condition (\ref{apole}) is supersymmetric because the
polar behavior of the $X^a$ is consistent with Nahm's equations
\begin{equation}\label{cnahm}\frac{DX^a}{Dy}+\frac{1}{2}\epsilon^{abc}[X_b,X_c]=0.
\end{equation}
This equation is consistent with supersymmetry because Nahm's
equations are the dimensional reduction of the selfdual Yang-Mills
equations, which are of course compatible with supersymmetry. Here
we are considering the selfdual Yang-Mills equations in the $3456$
plane, even though the covariant derivatives $D_a=\partial_a+A_a$
in the $456$ direction have been replaced by matrices $X^a$.

The selfdual Yang-Mills equations in any four-dimensional plane
preserve the same amount of supersymmetry.  Therefore, we can make
an $SO(1,5)$ rotation of the polar behavior that is assumed in
(\ref{cnahm}).  (Of course, we use the $SO(1,5)$ that fixes the 3
direction and acts on 012456.)  We take three orthonormal linear
combinations of the $012456$ directions, and postulate that the
corresponding fields $C^i$ (which are orthonormal linear
combinations of $A_\mu$, $\mu=0,1,2$ and $X^a,$ $a=4,5,6$) have a
pole $C^i\sim t^i/y$.  This preserves supersymmetry, just like the
special case of eqn. (\ref{apole}), since it is a special solution
of the selfdual Yang-Mills equations (or a dimensional reduction
thereof) in a certain four-dimensional subspace.

We thus get a family of boundary conditions that are all
associated with a principal embedding
$\rho:\frak{su}(2)\to\frak{g}$.  They are obtained by making an
$SO(1,5)$ rotation of the pole at $y=0$, even though $SO(1,5)$ is
not a symmetry of the theory.  The unbroken supersymmetry is
obtained by making the same $SO(1,5)$ rotation from the $3456$
plane to the appropriate four-plane.\footnote{The relevant
rotation group is $SO(1,5)$, not $SO(1,6)$, because we do not
rotate the $y=x^3$ direction.  This direction is distinguished by
the fact that the boundary is at $y=0$.}  Thus, we can immediately
characterize the supersymmetry left unbroken by a boundary
condition of this type. For some element $h\in SO(1,5)$, the
unbroken supersymmetries obey the rotated version of (\ref{kiop}),
namely
\begin{equation}\label{ystrict} h^{-1}B_1h\varepsilon=\varepsilon\end{equation}
or
\begin{equation}\label{zstrict}
B_1h\varepsilon=h\varepsilon.\end{equation} This coincides with
the condition (\ref{defol}) that we found for the D3-D5 system (in
the present field theory approach, we denote $\varepsilon_L$
simply as $\varepsilon$).

It is convenient to also introduce the six-dimensional chirality
operator $\Gamma'=\Gamma_{012456}$ and make a chiral decomposition
$\varepsilon=\varepsilon_++\varepsilon_-$, where
\begin{equation}\label{zorf}\Gamma'\varepsilon_\pm=\pm\varepsilon_\pm.
\end{equation}
Since $B_1$ anticommutes with $\Gamma'$, eqn. (\ref{zstrict}) is
equivalent to \begin{equation}\label{bstrict}
h\varepsilon_-=B_1h\varepsilon_+.
\end{equation}
Because of (\ref{zorf}), we can replace $B_1=\Gamma_{3456}$ by
$\Gamma^*=\Gamma_{0123}$ and write
\begin{equation}\label{zombo}h\varepsilon_-=-\Gamma^* h \varepsilon_+.
\end{equation}
This will be useful in section \ref{anexample}.

The condition (\ref{zombo}) is invariant under $h\to qh$ for $q\in
Q=SO(1,2)\times SO(3)_X$ (which commutes with $\Gamma^*$).  So we
can think of $h$ as taking values in the nine-dimensional space
$\tilde Z_9=Q\backslash SO(1,5)$. (Of course, nine is also the
number of components of $\CMF$, given that it has one index of type
012 and one of type 456.)  Thus, this construction gives a
nine-dimensional family $\tilde Z_9$ of half-BPS boundary conditions
that are associated with the principal embedding
$\rho:\frak{su}(2)\to\frak{g}$.   This generalizes what we found
from the D3-D5 system for $G=U(N)$.

\subsection{Rotating The D3-NS5 System}

Similarly it is possible to ``rotate'' the unbroken supersymmetry
of the D3-NS5 system.  This is particularly simple if the
four-dimensional $\theta$-angle vanishes.

We first rewrite (\ref{d5}) in the form
\begin{equation}\label{ploy}\begin{pmatrix}\varepsilon_R\\
\varepsilon_L\end{pmatrix} =
-\exp\left(\frac{1}{4}\Gamma^{IJ}\CMF_{IJ}\begin{pmatrix} 1 & 0 \\
0 & -1
\end{pmatrix}\right) B_0B_1\begin{pmatrix} 0 & 1\\ 1& 0
\end{pmatrix}\begin{pmatrix}\varepsilon_R\\
\varepsilon_L\end{pmatrix}. \end{equation} At $\theta=0$, the
duality transformation $S:\tau\to -1/\tau$ maps a D5-brane to an
NS5-brane and transforms $\varepsilon_R, $ $\varepsilon_L$ to
\begin{equation}\label{supertrans}\begin{pmatrix}\varepsilon_R'\\
\varepsilon_L'\end{pmatrix}=\frac{1}{\sqrt 2}\begin{pmatrix} 1 &
1\\ -1 & 1\end{pmatrix}
\begin{pmatrix}\varepsilon_R\\
\varepsilon_L\end{pmatrix}.\end{equation} The supersymmetry
condition in the presence of an NS5-brane can be deduced from
(\ref{ploy}) and is
\begin{equation}\label{ovoy}\begin{pmatrix}\varepsilon_R'\\
\varepsilon_L'\end{pmatrix}=-\exp\left(-\frac{1}{4}\Gamma^{IJ}\CMF_{IJ}
\begin{pmatrix} 0 & 1\\ 1& 0
\end{pmatrix}\right) B_0B_1    \begin{pmatrix} 1 & 0 \\ 0 & -1
\end{pmatrix}                  \begin{pmatrix}\varepsilon_R'\\
\varepsilon_L'\end{pmatrix}\end{equation}  If also D3-branes are
present, we must supplement this with
\begin{equation}\label{gof}\varepsilon'_R=-B_0\varepsilon'_L,\end{equation}
which follows from (\ref{d3d5}). With a little algebra, one can
eliminate $\varepsilon_R'$ and obtain the condition on
$\varepsilon'_L$:
\begin{equation}\label{zof}
\left(
1-\cosh(\Gamma^{IJ}\CMF_{IJ}/4)B_2+\sinh(\Gamma^{IJ}\CMF_{IJ}/4)B_1\right)
\varepsilon'_L=0.
\end{equation}
(In deriving this, note that $\Gamma^{IJ}\CMF_{IJ}$ commutes with
$B_2$, and anticommutes with $B_0$ and $B_1$.)

As preparation for interpreting this result in field theory, we
again make a chiral decomposition
$\varepsilon'_L=\varepsilon_++\varepsilon_-$, where
\begin{equation}\label{okolt}
\Gamma'\varepsilon_\pm = \pm \varepsilon_\pm.\end{equation}  (In
field theory, we omit the primes and the subscript $L$ and denote
the supersymmetry generator simply as $\varepsilon$.)  We can use
(\ref{zof}) to solve for $\varepsilon_-$ in terms of
$\varepsilon_+$:
\begin{equation}\label{xsendo}\varepsilon_-=\frac{1}
{1-\cosh(\Gamma^{IJ}\CMF_{IJ}/4)B_2}
\sinh(\Gamma^{IJ}\CMF_{IJ}/4)B_1\varepsilon_+. \end{equation}

We can make a small simplification as follows.  We have
$B_2\varepsilon_+=-\varepsilon_+$ (since
$\Gamma'\varepsilon_+=\varepsilon_+$ and $\varepsilon_+$ has
positive ten-dimensional chirality).  Also $B_2$ commutes with
$\Gamma^{IJ}\CMF_{IJ}$ and anticommutes with $B_1$.  Using these
facts, one can omit $B_2$ in (\ref{xsendo}), which becomes
\begin{equation}\label{sendo}\varepsilon_-=\frac{1}
{1-\cosh(\Gamma^{IJ}\CMF_{IJ}/4)}
\sinh(\Gamma^{IJ}\CMF_{IJ}/4)B_1\varepsilon_+. \end{equation}

Expanding this in powers of $\CMF$, the first term is
\begin{equation}\label{endo}
\varepsilon_-=\frac{1}{4}\Gamma^{IJ}\CMF_{IJ}B_1\varepsilon_+.\end{equation}
Using the fact that explicitly $B_1=\Gamma_{3456}$, and that $\CMF$
has one index of type 012 and one of type 456, we see we can write
this as
\begin{equation}\label{zendo}\varepsilon_-=\sum_{I<J<K}\Gamma^{IJK}q_{IJK}\Gamma_3
\varepsilon_+,\end{equation} where here the indices $I,J,K$ take
values $012456$, and $q_{IJK}$ is a third rank antisymmetric
tensor that depends on $\CMF$.

It is convenient to regard
$q=\sum_{I<J<K}q_{IJK}{\mathrm{d}}x^I\wedge {\mathrm{d}}x^J\wedge
{\mathrm{d}}x^K$ as a three-form on $\R^{1,5}$ with constant
coefficients. It is not immediately obvious that $q$ is selfdual
or anti-selfdual,\footnote{We define an antisymmetric tensor
$\epsilon_{IJKLMN}$ with $\epsilon^{012456}=1$.
 Indices $I,J,K$ will take values $0,1,2,4,5,6$.
 Self-duality for a third rank antisymmetric tensor $q$ means that
$q^{IJK}=\epsilon^{IJKLMN}q_{LMN}/3!$.  In Lorentz signature in
six dimensions, a third rank real antisymmetric tensor can be
selfdual or anti-selfdual. For example, with this definition, the
three-form $-{\mathrm{d}}x^0\wedge {\mathrm{d}}x^1\wedge
{\mathrm{d}}x^2+{\mathrm{d}}x^4\wedge {\mathrm{d}}x^5\wedge
{\mathrm{d}}x^6$ is selfdual. } but in fact, because
$\Gamma'\varepsilon_+=\varepsilon_+$, the anti-selfdual part of
$q$ does not contribute, and hence we can project $q$ to its
selfdual part.  Let $\star_{012}$ and $\star_{456}$ be the Hodge
$\star$ operators in the 012 and 456 directions.  We can pick
conventions so that $\star_{012}^2=\star_{456}^2=1$,
$\star_{012}\star_{456}=\star_{456}\star_{012}$; the
six-dimensional $\star$ operator is
$\star=\star_{012}\star_{456}$.  The relation between $q$ and
$\CMF$ in linear order is
\begin{equation}\label{medox}
q=\frac{(\star_{012}+\star_{456})\CMF}{8}\end{equation} and here $q$
is selfdual.

Though this analysis has been only to linear order in $\CMF$, in
fact, (\ref{sendo}) is precisely equivalent to (\ref{zendo}), with
the selfdual three-form $q$ in general a nonlinear function of
$\CMF$. To see this, we observe that gamma matrices
$\Gamma_7,\Gamma_8,\Gamma_9$ are absent in (\ref{sendo}) and
$\Gamma_3$ appears only as a linear factor in $B_1$ multiplying
$\varepsilon_+$.  So eqn. (\ref{sendo}) takes the form
$\varepsilon_-=\Omega\Gamma_3\varepsilon_+$, where $\Omega $ is
constructed from gamma matrices $\Gamma^I$, with $I$ ranging over
$012456$.  $\Omega$ must be of odd order in the $\Gamma^I$, since
it must reverse the six-dimensional chirality; and because
$\varepsilon_+$ obeys (\ref{okolt}), we can reduce to the case
$\Omega=\Gamma^IS_I+\sum_{I<J<K}\Gamma^{IJK}q_{IJK}$, with a
one-form $S$ and selfdual three-form $q$.  Moreover, the one-form
is absent for a reason that will be explained in section
\ref{general}.

Thus, the unbroken supersymmetry can be characterized by a
selfdual three-form in six dimensions.  However, the construction
as described so far does not lead to the most general selfdual
three-form.  Indeed, as in section \ref{deformdthree}, $\CMF$
depends on only nine parameters, but a selfdual three-form (with
constant coefficients) in $\R^{1,5}$ depends on ten parameters.
The missing parameter is  the four-dimensional $\theta$ angle,
which preserves half of the supersymmetry (and actually preserves
Lorentz invariance).  It is absent from the above formulas because
we obtained them starting with $S$-duality from the D3-D5 system
at $\theta=0$. This tenth parameter will be included in section
\ref{deformns} as well as section \ref{general}.

If we restrict to $\theta=0$, we get a nine-parameter family $Z_9$
of half-BPS (but not Lorentz-invariant) deformations of the D3-NS5
system.  They are $S$-dual to the corresponding nine-parameter
family  $\tilde Z_9$ of deformations of the D3-D5 system, described
in section \ref{deformdthree}, in the sense that the strong coupling
limit of one is the weak coupling limit of the other.

The reason that we have not seen a tenth parameter for the D3-D5
system is that $S$-duality becomes more complicated when
$\theta\not=0$; it does not simply exchange weak and strong
coupling.  As soon as $\theta\not=0$, the $S$-dual of a strongly
coupled D3-NS5 system is no longer a weakly coupled D3-D5 system.

\subsubsection{Realization In Field Theory}\label{deformns}

We will now re-examine the deformation of the D3-NS5 system just
described from the point of view of field theory.  As usual, one
advantage of this is that the discussion is valid for any gauge
group.

D3-branes ending on a single NS5-brane without any flux are governed
by Neumann boundary conditions for the vector multiplet $A_\mu$ and
$\vec X$ and Dirichlet boundary conditions for $\vec Y$. This was
described in section \ref{boundcon}.  Can we modify these boundary
conditions in a way that depends on a selfdual or anti-selfdual
third rank tensor and preserves the half-BPS property? In fact, in a
special case this has essentially been done in section
\ref{boundcon}.

In that analysis, a deformation was considered from Neumann boundary
conditions for gauge fields, which assert that $F_{3\lambda}=0$ on
the boundary for $\lambda=0,1,2$, to a more general boundary
condition with three-dimensional Lorentz invariance:
\begin{equation}\label{pmor}\epsilon_{\lambda\mu\nu}F^{3\lambda}+\gamma
F_{\mu\nu}=0.\end{equation} The physical meaning of the term
linear in $\gamma$ was explained in eqn. (\ref{takesf}).  It
corresponds to adding to the action a term proportional to $\int
\Tr \,F\wedge F$, or equivalently, after integrating by parts to
convert this to a surface term, it corresponds to adding a
boundary interaction
\begin{equation}\label{plor}-\frac{\gamma}{2e^2}\int_{\partial M}
\mathrm{d}^3x\,
\epsilon^{\mu\nu\lambda}\,\Tr\,\left(A_\mu\partial_\nu
A_\lambda+\frac{2}{3}A_\mu A_\nu A_\lambda\right).\end{equation}
Here $M$ is spacetime, and $\partial M$ is its boundary at $y=0$.
The interaction that we have added is of dimension three and
therefore preserves conformal invariance.

$SO(1,2)$ invariance allows us to add one more
conformally-invariant interaction constructed from bosons.  This
is
\begin{equation}\label{lor}-\frac{u}{3e^2}\int_{\partial
M}\mathrm{d}^3x\,\epsilon_{abc}\Tr\,X^a[X^b,X^c].\end{equation} If
we do add this interaction, then Neumann boundary conditions for $X$
are modified to
\begin{equation}\label{zmor}\frac{DX_a}{Dy}+\frac{u}{2}\epsilon_{abc}[X_b,X_c]
=0.\end{equation}

If we are willing to relax $SO(1,2)$ invariance, we can add
additional bosonic interactions that preserve global scale
invariance. Define quantities $Z^I$, $I=0,1,2,4,5,6$, as follows.
For $I=4,5,6$, set $Z^I=X^a$.  And for $I=0,1,2$, set $Z^I$ equal
to the covariant derivative $D_I=\partial_I+A_I$.  Then we can add
to the action a dimenstion three term that we loosely describe as
\begin{equation}\label{zxor}\int_{\partial M}
\mathrm{d}^3x\,q_{IJK}\Tr\,Z^I[Z^J,Z^K],\end{equation} where $q$ is
an arbitrary third-rank antisymmetric tensor.   The special case
involving the component $q_{012}$ corresponds to the Chern-Simons
interaction in (\ref{plor}), and the case involving $q_{456}$
corresponds to the Lorentz-invariant coupling in (\ref{lor}). Other
components of $q$ give couplings that violate Lorentz invariance;
they are schematically of the form $\Tr\,X^aF_{\mu\nu}$ or $\Tr\,
X^aD_\mu X^b$, with $\mu,\nu=0,1,2$, $a,b=4,5,6$.

Now the question arises of whether the bosonic interaction
(\ref{zxor}) can be completed to a supersymmetric theory by
suitably modifying the fermion boundary conditions (or
equivalently, by adding boundary interactions bilinear in
fermions).  If so, will a constraint come in related to
selfduality or anti-selfduality? We would expect this from the
discussion of (\ref{zendo}).

Happily, we do not really need to do a new calculation. For the
Lorentz-invariant case, with $q_{012}$ and $q_{456}$ the only
non-zero matrix elements of $q$, a  half-BPS boundary condition
was constructed in section \ref{basicboundary}. The quantities
$\gamma$ and $u$ were not independent but were parametrized by
\begin{equation}\label{paramz}
\gamma=-\frac{2a}{1-a^2}, ~~u=-\frac{2a}{1+a^2}.\end{equation}
NS5-brane and NS5-antibrane boundary conditions correspond to
$a=\infty$ and $a=0$. Expanding to first order in $1/a$ near
$a=\infty$ or to first order in $a$ near $a=0$, we have
$\gamma=\mp u$, which corresponds to the expected selfduality or
anti-selfduality of the tensor $q$.  The condition $\gamma=\mp u$
means that the three-form $q$ is
\begin{equation}\label{aramz} q=u(\mp {\mathrm{d}}x^0\wedge {\mathrm{d}}x^1\wedge {\mathrm{d}}x^2 +
{\mathrm{d}}x^4\wedge {\mathrm{d}}x^5\wedge
{\mathrm{d}}x^6)\end{equation} and so is Lorentz-invariant.   Note
that this particular three-form cannot be expressed in terms of
$\CMF$ as in (\ref{medox}), so we are here indeed describing the
tenth parameter that was missing in that derivation.

\subsubsection{Canonical Form Of $q$}\label{canonical}

One might think that the supersymmetry of the construction of
section \ref{boundcon} that we have just reviewed is only a
special case.  But in a certain sense it is actually generic. Let
us count the number of parameters of a general selfdual three-form
that, by an $SO(1,5)$ transformation, can be put in the form of
(\ref{aramz}). One parameter, namely $u$, is visible in
(\ref{aramz}).  We must also allow 9 more parameters generated by
$SO(1,5)$ transformations. ($SO(1,5)$ has dimension 15; its
subgroup that leaves $q$ fixed is $SO(1,2)\times SO(3)$, of
dimension 6; the difference is 9.) This gives a total of $1+9=10$
parameters.  But 10 is the dimension of the space of selfdual or
anti-selfdual three-forms, so a generic such form is of this type.

The half-BPS boundary condition derived from D3-branes ending on
an NS5-brane actually has a direct analog in $6+1$-dimensional
super Yang-Mills theory. In string theory, this can be understood
by replacing the D3-branes ending on an NS5-brane by D6-branes
which end on the NS5-brane.\footnote{Since the NS5-brane, which is
supposed to provide the boundary, has a six-dimensional
world-volume, we cannot make a construction like this above $6+1$
dimensions.  This can also be understood from a field theory point
of view; the Dirichlet boundary conditions on the three scalar
fields $Y^p$ do not have analogs if one or more of those scalars
is replaced by covariant derivatives in extra dimensions.} From a
field theory point of view, we simply allow all fields to depend
on three more coordinates $x^4,x^5,x^6$, and replace the three
scalar fields $X^a$ with covariant derivatives $D_a+A_a$ in the
$x^4,x^5,x^6$ directions. This substitution makes sense because
$X^a$ enters the ${\cal N}=4$ super Yang-Mills Lagrangian only via
its commutators with other fields and with covariant derivatives.

The boundary condition for $D6$-branes ending on an NS5-brane has
$SO(1,5)$ symmetry. So after lifting the D3-NS5 system to $6+1$
dimensions, and making the deformation involving the three-form in
eqn. (\ref{aramz}), we can make an $SO(1,5)$ rotation. Then we can
reduce back to $3+1$ dimensions, taking the fields to be once again
independent of $x^4,x^5,x^6$, and turning the covariant derivatives
$D_a+A_a$ back into scalar fields $X^a$.

What we gain by the detour through $6+1$ dimensions is the
knowledge that we can, in effect, make an $SO(1,5)$ transformation
of the deformed boundary conditions even though $SO(1,5)$ is not a
symmetry of the theory. Hence, without any need for further
computation, there is a half-BPS boundary condition in which
(\ref{aramz}) is replaced by a general selfdual three-form.

This construction gives a ten-dimensional family $Z_{10}$ of
half-BPS boundary conditions.  The Neumann boundary conditions of
the D3-NS5 system, with any number of D3-branes and a single
NS5-brane, represent a point in $Z_{10}$.  The generic point
represents a half-BPS but not Lorentz-invariant deformation. At a
generic point, the unbroken supersymmetry is described by
\begin{equation}\label{dex}\varepsilon_-=\sum_{I<J<K}\Gamma^{IJK}q_{IJK}\Gamma_3
\varepsilon_+\end{equation} for a selfdual three-form $q$. As we
explain in section \ref{general}, the family $Z_{10}$ also
contains points ``at infinity'' that cannot be described in this
way.  (These include points describing D3-branes ending on an NS
anti-fivebrane.)

Only the sublocus $Z_9$ describes deformations that have a simple
$S$-duality relationship to the analogous family $\tilde Z_9$ of
deformations of the D3-D5 system.  For a given $q$, how can we
determine if the corresponding deformation of the D3-NS5 system lies
in $Z_9$?  One necessary and sufficient criterion is that it must be
possible to parametrize $q$
 via (\ref{sendo}) in terms of a two-form $\CMF$.  An equivalent
 criterion
is that the space of unbroken supersymmetries, characterized by
(\ref{dex}), must transform under $S:\tau\to-1/\tau$ into a space of
supersymmetries that can be characterized in terms of the analogous
formula (\ref{ystrict}) of the D3-D5 system.

To use the last-mentioned criterion, we need to know how the space
of unbroken supersymmetries transforms under duality. This can be
deduced from string theory formulas presented earlier, but can
also be understood purely in four-dimensional terms. In general,
under a duality transformation that transforms the coupling
parameter $\tau$ by $\tau\to (a\tau+b)/(c\tau+d)$, the
supersymmetry generators $\varepsilon$ transform by
\begin{equation}\label{kurlo}\varepsilon\to
\left(\frac{|c\tau+d|}{c\tau+d}\right)^{-i\Gamma^*/2}\varepsilon,\end{equation}
with $\Gamma^*=\Gamma_{0123}$.  (For example, see \cite{KW}, eqn.
2.25.) For the transformation $S:\tau\to -1/\tau$, with $\theta=0$
so that $\tau$ is on the imaginary axis, this becomes
\begin{equation}\label{urlo}\varepsilon\to\frac{1-\Gamma^*}{\sqrt
2}\varepsilon.\end{equation}

\subsection{An Example}\label{anexample}

Now we are going to consider an example: we will take a boundary
condition representing a point in $Z_{10}$, and show that it
actually represents a point in $Z_9$, and so is $S$-dual to a D3-D5
boundary condition with a pole.

As explained in section (\ref{canonical}),  a generic selfdual
three-form $q$ can be put in the canonical form of equation
(\ref{aramz}).  But it is not true that every selfdual three-form
can be put in this form.  A counterexample can be written
\begin{equation}\label{qcan}q= \frac{1}{4}({\mathrm{d}}x^0+{\mathrm{d}}x^4)\wedge ({\mathrm{d}}x^1\wedge
{\mathrm{d}}x^5+{\mathrm{d}}x^2\wedge
{\mathrm{d}}x^6).\end{equation} This three-form cannot be put in
the form (\ref{aramz}) by an $SO(1,5)$ transformation, because
when that is done, $|u|$ is an invariant. However, $q$ can be
rescaled by a Lorentz boost in the $04$ plane, and hence cannot be
characterized by any nonzero invariant.

A deformation of the D3-NS5 system associated with this choice of
$q$ appears in the gauge theory approach to geometric
Langlands.\footnote{In eqn. (12.31) of \cite{KW}, boundary
conditions are given for a gauge theory description of the
``canonical coisotropic brane.'' These boundary conditions can be
obtained by perturbing $\N=4$ super Yang-Mills by a boundary
interaction associated with a three-form, as in (\ref{zxor}).  The
necessary three-form is the one indicated in eqn. (\ref{qcan}). To
see this, one must take into account a clash in notation between
the present paper and \cite{KW}.  The boundary direction that we
call $x^3$ is called $x^1$ in \cite{KW}, and the directions that
we label $012456$ are $023567$ in \cite{KW}.}  The fact that the
$S$-dual of this particular boundary condition is associated with
a point in $\tilde Z_9$, and thus is associated with a principal
embedding $\rho:\frak{su}(2)\to\frak{g}$, is important in
geometric Langlands, and has until now been mysterious from the
gauge theory point of view.

We can relate this particular deformation of the D3-NS5 system to
a D3-D5 deformation using either of the two approaches mentioned
at the end of section \ref{canonical}.  First, we can show
directly that with $q$ as above, the deformed NS5 supersymmetry
relation
$\varepsilon_-=\sum_{I<J<K}\Gamma^{IJK}q_{IJK}\Gamma_3\varepsilon_+$
is equivalent to the deformed D5 relation (\ref{sendo}), with
$\CMF$ a multiple of ${\mathrm{d}}x^1\wedge
{\mathrm{d}}x^6-{\mathrm{d}}x^2\wedge {\mathrm{d}}x^5$. (The
precise multiple is determined below by another method.) The
evaluation of (\ref{sendo}) is simple for $\CMF$ of this form
because $M=\Gamma_{16}-\Gamma_{25}$ obeys $M^3=-4M$, reflecting
the fact that it is a generator of an $SU(2)$ subgroup of
$Spin(1,5)$.

Alternatively, we can proceed by analyzing the unbroken
supersymmetries. As usual, we write the generator of an unbroken
supersymmetry as $\varepsilon=\varepsilon_++\varepsilon_-$, where
\begin{equation}\label{olf}\Gamma'\varepsilon_\pm=\pm\varepsilon_\pm,\end{equation}
and moreover
\begin{equation}\label{polf}
\varepsilon_-=\sum_{I<J<K}q_{IJK}\Gamma^{3IJK}\varepsilon_+
=\frac{1}{4}\Gamma_3(-\Gamma_0+\Gamma_4)(\Gamma_{15}+\Gamma_{26})\varepsilon_+.\end{equation}
According to (\ref{urlo}), the duality transformation $S:\tau\to
-1/\tau$ maps $\varepsilon$ to $\tilde\varepsilon = \frac{1}{\sqrt
2} (1-\Gamma^*)\varepsilon$, or equivalently
$\varepsilon=\frac{1}{\sqrt 2}(1+\Gamma^*)\tilde\varepsilon$.
Since $\Gamma^*$ anticommutes with $\Gamma'$, it exchanges
$\varepsilon_\pm$, so this becomes
\begin{align}\label{unco}\notag \varepsilon_+ & = \frac{1}{\sqrt
2}\left(\tilde \varepsilon_++\Gamma^* \tilde \varepsilon_-\right) \\
 \varepsilon_- & = \frac{1}{\sqrt 2}\left(\tilde \varepsilon_-+\Gamma^*\tilde
 \varepsilon_+\right).\end{align}
If we set
$M=\frac{1}{4}\Gamma_3(-\Gamma_0+\Gamma_4)(\Gamma_{15}+\Gamma_{26})$,
so that eqn. (\ref{polf}) reads $\varepsilon_-=M\varepsilon_+$,
then the $S$-dual version is $\tilde \varepsilon_-+\Gamma^*\tilde
\varepsilon_+=M(\tilde \varepsilon_++\Gamma^*\tilde
\varepsilon_-)$, or
\begin{equation}\label{sdualv} (1-M\Gamma^*)\tilde
\varepsilon_-=-\Gamma^*(1+\Gamma^*M)\tilde
\varepsilon_+.\end{equation} Upon evaluating $M\Gamma^*\tilde
\varepsilon_-$ and $\Gamma^*M\tilde \varepsilon_+$, using
(\ref{olf}), we find after some gamma matrix algebra that
(\ref{sdualv}) is equivalent to
\begin{equation}\label{udaulv}P\tilde \varepsilon_-=-\Gamma^*P\tilde
\varepsilon_+,\end{equation} where
\begin{equation}\label{pmo}P=1+\frac{\Gamma_{16}-\Gamma_{25}}{2}.
\end{equation}

If $P$ were an element of $SO(1,5)$, this relation would be in the
desired form (\ref{zombo}), showing that the $S$-dual of the
boundary condition that we started with does represent a point in
$\tilde Z_9$.  It is actually not true that $P$ is an element of
$SO(1,5)$. However, both $P$ and $\Gamma^*$ commute with
\begin{equation}\label{gervo} T=
\frac{1-\Gamma_{1256}}{2}+\frac{1+\Gamma_{1256}}{2\sqrt
2},\end{equation} so eqn. (\ref{pmo}) is equivalent to
\begin{equation}\label{wervo} TP\tilde\varepsilon_-=-\Gamma^* TP\tilde
\varepsilon_+.\end{equation} Here\footnote{To obtain the second
equality in (\ref{usu}), observe that both sides equal 1 when
acting on spinors $\psi$ with $\Gamma_{1256}\psi=-\psi$, and then
evaluate the two sides assuming $\Gamma_{1256}\psi=\psi$.}
\begin{equation}\label{usu} TP=\left(\frac{1-\Gamma_{1256}}{2}+
\frac{1+\Gamma_{1256}}{2\sqrt
2}\right)\left(1+\frac{\Gamma_{16}-\Gamma_{25}}{2}\right)=
\exp\left(\frac{\pi}{8}(\Gamma_{16}-\Gamma_{25})\right)
\end{equation}
is an element of $SO(1,5)$, in fact an element of the
one-parameter subgroup of $SO(1,5)$ generated by
$\Gamma_{16}-\Gamma_{25}$. So eqn. (\ref{wervo}) is of the form of
(\ref{zombo}).

\def\N{{\cal N}}
\def\S{{\cal S}}
\subsection{General Formulation}\label{general}

Until this point, we have relied upon explicit constructions using
either branes or field theory.  Here, we will study conceivable
half-BPS boundary conditions from a more general point of view. This
will give a clearer understanding of some things that we originally
described by hand.

First of all, if one has a boundary at $x^3=0$, then regardless of
the nature of the boundary condition, there is no translation
invariance in the $x^3$ direction.  Hence if $\varepsilon$ and
$\tilde \varepsilon$ are two generators of supersymmetries that
remain valid in the presence of the boundary, we must have
\begin{equation}\label{momor}\bar\varepsilon\Gamma_3\tilde\varepsilon=0.
\end{equation}
For a half-BPS boundary condition, 8 of the possible 16
supersymmetries are unbroken.  We can interpret the condition
(\ref{momor}) as follows.  Let $V_{16}$ be the 16-dimensional real
vector space (the irreducible positive chirality spinor
representation of $SO(1,9)$) in which $\varepsilon$ takes values.
The expression
$(\varepsilon,\tilde\varepsilon)=\bar\varepsilon\Gamma_3\tilde\varepsilon$
defines a non-degenerate quadratic form on this space, of
signature $(8,8)$.  The condition (\ref{momor}) asserts that
$\varepsilon$ takes values in a subspace  $T\subset V_{16}$ such
that the quadratic form vanishes when restricted to $T$.  A
maximal subspace with this property is eight-dimensional, and the
half-BPS condition asserts precisely that $T$ is maximal.

Regardless of what we pick $T$ to be, the condition that
$\varepsilon,\tilde\varepsilon\in T$ does not suffice to set
$\bar\varepsilon \Gamma^\mu\tilde \varepsilon=0$ for any value of
$\mu$ other than 3.  Hence, a half-BPS boundary condition, though
not necessarily Lorentz-invariant, is invariant under translations
in the $0,1,$ and $ 2$ directions.

Let ${\cal S}$ be the space of all eight-dimensional null
subspaces of $V_{16}$.  Every half-BPS boundary condition
determines a point in ${\cal S}$.  ${\cal S}$ is a homogeneous
space for a group $H=SO(1,8)$ that formally rotates the
coordinates $x^I$, $I\not=3$. $SO(1,8)$ is not really a symmetry
group of $\N=4$ super Yang-Mills theory; only its subgroup
$SO(1,2)\times SO(6)$ is a group of symmetries. ($SO(1,2)$ is the
Lorentz group that acts on the $0,1$ and $2$ directions, and
$SO(6)$ is the group of $R$-symmetries.) But the action of
$SO(1,8)$ on $\S$ will be useful in the following analysis of
half-BPS boundary conditions that lack $SO(1,2)$ symmetry.

We make a preliminary simplification along the following lines. We
will only consider half-BPS boundary conditions that  can be
obtained by marginal (scale-invariant) deformation of a
Lorentz-invariant one. As explained in section
\ref{basicboundary}, any $SO(1,2)$-invariant half-BPS boundary
condition has Dirichlet boundary conditions on precisely three of
the scalars, denoted there as $\vec Y$.  Since the only fields in
$\N=4$ super Yang-Mills theory of conformal dimension 1 are the
scalar fields $\vec X$ and $\vec Y$, the only possible marginal
deformation of the Dirichlet boundary condition $\vec Y|=0$ is to
rotate $\vec Y$ to a linear combination of $\vec X$ and $\vec Y$.
Making such a rotation does not give anything essentially new, so
we will stick with $\vec Y|=0$.

Furthermore, we will consider only half-BPS boundary conditions that
are invariant under the group $SO(3)_Y$ that rotates $\vec Y$. It is
now useful to decompose the space $V_{16}$ under the action of
$SO(1,5)\times SO(3)_Y$, where $SO(1,5)$, which rotates the
directions $012456$, is the subgroup of $SO(1,8)$ that commutes with
$SO(3)_Y$. We can decompose $V_{16}$ as $W_8\otimes W_2$, where
$SO(3)_Y$ acts on $W_2$ in the spinor representation, and $SO(1,5)$
likewise acts on $W_8$ in the spinor representation. (Both $SO(1,5)$
chiralities are included in $W_8$.) For $\varepsilon=\mu\otimes
\nu$, $\tilde\varepsilon=\tilde\mu\otimes\tilde\nu$, we can
decompose the inner product as
\begin{equation}\label{zolm}(\varepsilon,\tilde\varepsilon)=\langle
\mu,\tilde\mu\rangle \,\langle \nu,\tilde\nu\rangle',\end{equation}
where $\langle ~,~\rangle$ is an inner product on $W_8$ and
$\langle~,~\rangle'$ is one on $W_2$.  The second inner product
$\langle~,~\rangle'$  is antisymmetric (the spinor representation of
$SO(3)_Y$ admits only an antisymmetric inner product), so
$\langle~,~\rangle$ is also antisymmetric.\footnote{This does not
follow  from $SO(1,5)$ invariance alone; since $W_8$ is the direct
sum of the two spinor representations of $SO(1,5)$ of opposite
chirality, it admits both a symmetric and an antisymmetric invariant
inner product.  This is clear from the group theory described
below.} The decomposition $V_{16}=W_8\otimes W_2$ is obviously
similar to a decomposition made in section \ref{basicboundary}, but
here we make this decomposition using a different subgroup of
$SO(1,8)$.

Now let us return to the eight-dimensional null subspace $T\subset
V_{16}$ that parametrizes the supersymmetries left unbroken by a
half-BPS boundary condition.  If the boundary condition is to be
$SO(3)_Y$-invariant, we must have $T=U\otimes W_2$, where $U$ is a
four-dimensional null subspace of $W_8$.

\def\4{{\bf 4}}
\def\one{{\bf 1}}
\def\6{{\bf 6}}
\def\fif{{\bf 15}}
\def\10{{\bf 10}}
It will help to know something about such null subspaces.  For
this, we need some $SO(1,5)$ group theory.  Let us write $\4$ and
$\4'$ for the positive and negative chirality spinor
representations of $SO(1,5)$.  Thus, we have $W_8\cong
W_{\4}\oplus W_{\4'}$, where $W_\4$ and $W_{\4'}$ transform,
respectively, in the representations $\4$ and $\4'$. We denote the
trivial representation, the vector representation, and the second
rank antisymmetric tensor representation of $SO(1,5)$ as
$\one,\6,$ and $\fif$, respectively. Finally, the third rank
tensor representation has dimension $6\cdot 5\cdot 4/3!=20$, but
decomposes as a direct sum of two representations $\10$ and $\10'$
that consist respectively of anti-selfdual and selfdual third rank
tensors.

The tensor products of  spinor representations of $SO(1,5)$
decompose as follows:
\begin{align}\label{dormo}\notag \4\otimes \4 \,\,& = \6_A\oplus \10_S \\
             \notag       \4'\otimes \4' & = \6_A\oplus \10'_S\\
                          \4\otimes \4'& = \one\oplus
                          \fif.\end{align}
The subscripts $A$ and $S$ refer respectively to the antisymmetric
and symmetric parts.  For example, the first line means that the
antisymmetric part of $\4\otimes \4$ is $\6$ and the symmetric
part is $\10$.

From (\ref{dormo}), we see that an invariant inner product between
two spinors must pair a $\4$ and a $\4'$.  So $W_\4$ and $W_{\4'}$
are two examples of null subspaces of $V_8$.  It is not hard to
describe half-BPS boundary conditions associated with these
subspaces.  The condition that $\mu\in W_4$ is that
$\Gamma'\mu=\mu$, where we set
$\Gamma'=\Gamma_{012456}=\epsilon^{IJKLMN}\Gamma_{IJKLMN}/6!$. We
can write the condition on $\mu$ in terms of
$\varepsilon=\mu\otimes\nu$ as $\Gamma'\varepsilon=\varepsilon$.
Equivalently, since $\varepsilon$ has positive chirality for
$SO(1,9)$, and thus obeys $\Gamma_{012\dots
9}\varepsilon=\varepsilon$, the condition is
\begin{equation}\label{zormo}B_2\varepsilon=-\varepsilon,\end{equation}
where (as in eqn. (\ref{gomot})) $B_2=\Gamma_{3789}$. Likewise, the
condition $\mu\in W_{\4'}$ corresponds to
\begin{equation}\label{noxto}B_2\varepsilon=\varepsilon.\end{equation}

As we know by now, many different boundary conditions  preserve the
supersymmetry of (\ref{zormo}) or (\ref{noxto}).
 As explained in section \ref{interp},
 a particularly simple example arises for a system of D3-branes
ending on an NS5-brane (or NS5-antibrane; the two choices
correspond to the two possible conditions
$B_2\varepsilon=\pm\varepsilon$). This corresponds to Neumann
boundary conditions for gauge fields and for the scalar fields
$\vec X$, extended to the fermions in a supersymmetric fashion.

Now let us describe what a generic choice of $U$ would look like.
We write $\mu=\eta\oplus \zeta$, $\eta\in W_\4$, $\zeta\in
W_{\4'}$. Thus
\begin{equation}\label{torpe}\Gamma' \eta = \eta,~~\Gamma' \zeta =
-\zeta.\end{equation} For a suitable choice of basis, the inner
product of $\mu$ with $\tilde\mu=\tilde \eta\oplus \tilde \zeta$ is
\begin{equation}\label{yorm}\langle\mu,\tilde\mu\rangle =
\sum_{a=1}^4\left(\eta^a\tilde \zeta_{ a}-\zeta_a\tilde \eta^a
\right).\end{equation} Since this inner product on $W_8$ is
antisymmetric, a convenient way to proceed is to think of $\mu,
\eta, $ and $\zeta$ as fermionic variables, and then the inner
product can be described via a quadratic function of $\mu$:
\begin{equation}\label{zormy}F(\mu)=\sum_{a=1}^4\eta^a\zeta_a.\end{equation}
A subspace $U\subset W_8$ is null if $F(\mu)=0$ for $\mu\in U$. The
simplification here is that there is no need to mention a second
spinor $\tilde \mu$.

In this formulation, it is straightforward to describe the generic
four-dimensional null subspace $U$.   A generic four-dimensional
subspace of $W_8$ can be defined by a condition
\begin{equation}\label{zomr} \zeta_a=\sum_b\, f_{ab}\eta^b  \end{equation}
for some tensor $f_{ab}$.  In order for this equation to imply
that $0=F(\mu)=\sum_a\eta^a\zeta_a$, the condition we need is that
$f$ should be symmetric, $f_{ab}=f_{ba}$.  Here $f$ transforms as
the symmetric product $\4'\otimes \4'$, that is, like a selfdual
three-form $q$ (with constant coefficients) on $\R^{1,5}$. For
$q=\sum_{I<J<K}q_{IJK}{\mathrm{d}}x^I\wedge {\mathrm{d}}x^J\wedge
{\mathrm{d}}x^K$ (with $I,J,K$ taking values in $012456$), we can
write (\ref{zomr}) in terms of gamma matrices in the
form\footnote{If $q$ is anti-selfdual, then as $\Gamma'\eta=\eta$,
we have $q_{IJK}\Gamma^{IJK}\eta=0$, giving another explanation
for why in (\ref{pomr}), $q$ is selfdual.}
\begin{equation}\label{pomr}
\zeta=\sum_{I<J<K}q_{IJK}\Gamma^{IJK}\eta.\end{equation}   In
terms of the supersymmetry generator $\varepsilon$, which we
decompose as $\varepsilon=\varepsilon_++\varepsilon_-$ where
$\Gamma'\varepsilon_\pm=\pm\varepsilon_\pm$, the condition is
\begin{equation}\label{tomr}
\varepsilon_-=\sum_{I<J<K}q_{IJK}\Gamma^{IJK}\Gamma^3\varepsilon_+.\end{equation}
This condition is familiar from (\ref{zendo}), whose structure is
hopefully now more clear.

Each choice of $q$ gives a maximal null subspace $U$, but not
every such subspace arises this way.  The ones that so arise are
precisely those that have trivial intersection with $W_{\4'}$, or
in other words contain no vector with $\eta=0$.  Conversely, every
maximal null subspace whose intersection with $W_\4$ is trivial
can be defined by an equation
\begin{equation}\label{bromr} \eta^a=\sum_b\, g^{ab}\zeta_b,\end{equation}
where again $g^{ab}$ is symmetric.  Thus, $g^{ab}$ transforms as
an anti-selfdual three-form  $\tilde q$ on $\R^{1,5}$.  As in
(\ref{pomr}), we can equivalently write
\begin{equation}\label{bomr}
\eta=\sum_{I<J<K}\tilde q_{IJK}\Gamma^{IJK}\zeta.\end{equation}

For NS5-brane boundary conditions, $q$ vanishes; so for a small
perturbation of those boundary conditions, $q$ is small.  When $q$
is small, (\ref{pomr}) is a good description.  Close to the
NS5-antibrane case, $\tilde q$ is small and (\ref{bomr}) is more
useful.

\bigskip
Research of DG supported in part by DOE Grant DE-FG02-90ER40542.
Research of EW supported in part by NSF contract PHY-0503584. We
would like to thank E. Weinberg for helpful comments about Nahm's
equations.

\bibliography{stringsclass}{}

\providecommand{\href}[2]{#2}\begingroup\raggedright\begin{thebibliography}{10}

\bibitem{Nahm:1979yw}
W.~Nahm, {\it {A Simple Formalism for the BPS Monopole}},  {\em Phys. Lett.}
  {\bf B90} (1980) 413.
%%CITATION = PHLTA,B90,413;%%

\bibitem{WeinbergYi}
E.~J. Weinberg and P.~Yi, {\it {Magnetic Monopole Dynamics, Supersymmetry, and
  Duality}},  {\em Phys. Rept.} {\bf 438} (2007) 65--236
  [\href{http://arXiv.org/abs/hep-th/0609055}{{\tt hep-th/0609055}}].
%%CITATION = HEP-TH/0609055;%%

\bibitem{Diaconescu:1996rk}
D.-E. Diaconescu, {\it {D-Branes, Monopoles and Nahm Equations}},  {\em Nucl.
  Phys.} {\bf B503} (1997) 220--238
  [\href{http://arXiv.org/abs/hep-th/9608163}{{\tt hep-th/9608163}}].
%%CITATION = HEP-TH/9608163;%%

\bibitem{Tsimpis:1998zh}
D.~Tsimpis, {\it {Nahm Equations and Boundary Conditions}},  {\em Phys. Lett.}
  {\bf B433} (1998) 287--290 [\href{http://arXiv.org/abs/hep-th/9804081}{{\tt
  hep-th/9804081}}].
%%CITATION = HEP-TH/9804081;%%

\bibitem{Kapustin:1998pb}
A.~Kapustin and S.~Sethi, {\it {The Higgs Branch of Impurity Theories}},  {\em
  Adv. Theor. Math. Phys.} {\bf 2} (1998) 571--591
  [\href{http://arXiv.org/abs/hep-th/9804027}{{\tt hep-th/9804027}}].
%%CITATION = HEP-TH/9804027;%%

\bibitem{DeWolfe:2001pq}
O.~DeWolfe, D.~Z. Freedman and H.~Ooguri, {\it Holography and defect conformal
  field theories},  {\em Phys. Rev.} {\bf D66} (2002) 025009
  [\href{http://arXiv.org/abs/hep-th/0111135}{{\tt hep-th/0111135}}].
%%CITATION = HEP-TH/0111135;%%

\bibitem{Erdmenger:2002ex}
J.~Erdmenger, Z.~Guralnik and I.~Kirsch, {\it {Four-Dimensional Superconformal
  Theories with Interacting Boundaries or Defects}},  {\em Phys. Rev.} {\bf
  D66} (2002) 025020 [\href{http://arXiv.org/abs/hep-th/0203020}{{\tt
  hep-th/0203020}}].
%%CITATION = HEP-TH/0203020;%%

\bibitem{Constable:2002xt}
N.~R. Constable, J.~Erdmenger, Z.~Guralnik and I.~Kirsch, {\it {Intersecting
  D3-Branes and Holography}},  {\em Phys. Rev.} {\bf D68} (2003) 106007
  [\href{http://arXiv.org/abs/hep-th/0211222}{{\tt hep-th/0211222}}].
%%CITATION = HEP-TH/0211222;%%

\bibitem{Brink:1976bc}
L.~Brink, J.~H. Schwarz and J.~Scherk, {\it {Supersymmetric Yang-Mills
  Theories}},  {\em Nucl. Phys.} {\bf B121} (1977) 77.
%%CITATION = NUPHA,B121,77;%%

\bibitem{GaWi}
D.~Gaiotto and E.~Witten, {\it {Janus Configurations, Chern-Simons Couplings,
  And The Theta-Angle in N=4 Super Yang-Mills Theory}},
  \href{http://arXiv.org/abs/0804.2907}{{\tt 0804.2907}}.
%%CITATION = 0804.2907;%%

\bibitem{DEG}
E.~D'Hoker, J.~Estes and M.~Gutperle, {\it {Interface Yang-Mills,
  Supersymmetry, and Janus}},  {\em Nucl. Phys.} {\bf B753} (2006) 16
  [\href{http://arXiv.org/abs/hep-th/0603013}{{\tt hep-th/0603013}}].
%%CITATION = HEP-TH/0603013;%%

\bibitem{Constable:1999ac}
N.~R. Constable, R.~C. Myers and O.~Tafjord, {\it {The Noncommutative Bion
  Core}},  {\em Phys. Rev.} {\bf D61} (2000) 106009
  [\href{http://arXiv.org/abs/hep-th/9911136}{{\tt hep-th/9911136}}].
%%CITATION = HEP-TH/9911136;%%

\bibitem{Kronheimer}
P.~B. Kronheimer, {\it {Instantons and the Geometry of the Nilpotent Variety}},
   {\em J. Diff. Geom.} {\bf 32} (1990) 473--490.
%%CITATION = JDGEA,32,473;%%

\bibitem{Kronheimer2}
P.~B. Kronheimer, {\it {A Hyper-Kahlerian Structure On Coadjoint Orbits Of A
  Complex Lie Group}},  {\em J. London Math. Soc.} {\bf 42} (1990) 193--208.

\bibitem{Biel}
R.~Bielawski, {\em { Lie Groups, Nahm's Equations and Hyperkaehler Manifolds}}.
\newblock Universit¨atsverlag G¨ottingen (Gottingen, 2007).

\bibitem{Abiel}
M.~F. Atiyah and R.~Bielawski, {\it {Nahm's Equations, Configuration Spaces and
  Flag Manifolds}},  {\em Bull. Braz. Math. Soc. (N.S.)} {\bf 33} (2002)
  157--176.

\bibitem{WeinbergChen}
X.~Chen and E.~J. Weinberg, {\it {ADHMN Boundary Conditions From Removing
  Monopoles}},  {\em Phys. Rev.} {\bf D67} (2003) 065020
  [\href{http://arXiv.org/abs/hep-th/0212328}{{\tt hep-th/0212328}}].
%%CITATION = HEP-TH/0212328;%%

\bibitem{CallanMaldacena}
C.~G. Callan and J.~M. Maldacena, {\it {Brane Dynamics from the Born-Infeld
  Action}},  {\em Nucl. Phys.} {\bf B513} (1998) 198--212
  [\href{http://arXiv.org/abs/hep-th/9708147}{{\tt hep-th/9708147}}].
%%CITATION = HEP-TH/9708147;%%

\bibitem{KW}
A.~Kapustin and E.~Witten, {\it {Electric-Magnetic Duality and the Geometric
  Langlands Program}},  {\em Comm. Number Theory and Physics} {\bf 1} (2007)
  1--236 [\href{http://arXiv.org/abs/hep-th/0604151}{{\tt hep-th/0604151}}].
%%CITATION = HEP-TH/0604151;%%

\end{thebibliography}\endgroup
\bibliographystyle{JHEP-2}
\end{document}